\documentclass[12pt,a4paper]{article}
\usepackage[margin=0.6in]{geometry}

\renewcommand{\title}[1]{\vbox{\center\LARGE{#1}}\vspace{5mm}}
\renewcommand{\author}[1]{\vbox{\center#1}\vspace{5mm}}
\newcommand{\address}[1]{\vbox{\center\em#1}}
\newcommand{\email}[1]{\vbox{\center\tt#1}\vspace{5mm}}

\renewcommand{\date}[1]{\vbox{\center#1}}

\usepackage{amssymb}
\usepackage{amsmath,bm}
\usepackage{amssymb}
\usepackage{graphicx}
\usepackage{amsfonts}         
\usepackage{fancybox}   

\usepackage{enumitem}

\usepackage{slashed}

\usepackage{epstopdf}
\usepackage[usenames,dvipsnames]{xcolor}
\usepackage[pagebackref, bookmarks={false}, pdfauthor={SM}, pdftitle={}]{hyperref}
\hypersetup{colorlinks=true, linkcolor=darkblue, citecolor=red, filecolor=OliveGreen, urlcolor=blue, filebordercolor={.8 .8 1}, urlbordercolor={.8 .8 0}}
\usepackage{soul}
\setstcolor{Red}

\usepackage{wrapfig}

\definecolor{jazzberryjam}{rgb}{0.65, 0.04, 0.37}
\definecolor{lust}{rgb}{0.9, 0.13, 0.13}
\definecolor{sandybrown}{rgb}{0.96, 0.64, 0.38}
\definecolor{mountainmeadow}{rgb}{0.19, 0.73, 0.56}
\definecolor{glaucous}{rgb}{0.38, 0.51, 0.71}
\definecolor{chromeyellow}{rgb}{1.0, 0.65, 0.0}
\definecolor{emerald}{rgb}{0.31, 0.78, 0.47}
\definecolor{deepsaffron}{rgb}{1.0, 0.6, 0.2}
\definecolor{darkgreen}{rgb}{0,0.4,0}
\definecolor{darkred}{rgb}{0.4,0,0}
\definecolor{darkblue}{rgb}{0,0,0.4}
\definecolor{lightblue}{rgb}{.6,.6,0.9}

\definecolor{uglybrown}{rgb}{0.8,  0.7,  0.5}

\definecolor{palatinatepurple}{rgb}{0.41, 0.16, 0.38}
\definecolor{celebrationcolor}{rgb}{0.75,  0.0,  0.9}

 \usepackage{mdframed}

\usepackage{framed}
\definecolor{shadecolor}{rgb}{0.90,0.90,0.90}

\usepackage{tkz-euclide}

\usepackage{makeidx}
\usepackage{cite}
\usepackage{bm}
\usepackage{geometry}
\usepackage{braket}
\usepackage[margin=20pt,small]{caption}

\usepackage[toc]{appendix}

\usepackage{tikz}
\usetikzlibrary{calc,decorations.markings,arrows.meta,shapes.misc,decorations.pathmorphing,calc,bending}

\usetikzlibrary{arrows.meta,shapes.misc,decorations.pathmorphing,calc,bending}

\tikzset{
  branch point/.style={cross out,draw=black,fill=none,minimum size=2*(#1-\pgflinewidth),inner sep=0pt,outer sep=0pt}, 
  branch point/.default=5
}
\tikzset{
  branch cut/.style={
    decorate,decoration=snake,
    to path={
      (\tikztostart) -- (\tikztotarget) \tikztonodes
    },
    }
  }

\input epsf

\newcommand{\nn}{\nonumber}
\newlength{\extraspace}
\newlength{\extraspaces}
\def\be{\begin{equation}}
\def\ee{\end{equation}}

\newcommand{\bea}{\begin{eqnarray}}
\newcommand{\eea}{\end{eqnarray}}

\def\Tr{{{\rm Tr~ }}}
\def\tr{{\rm tr}}

\def\II{\relax{I\kern-.10em I}}

\def\IB{\relax{\rm I\kern-.18em B}}

\def\ID{\relax{\rm I\kern-.18em D}}
\def\IE{\relax{\rm I\kern-.18em E}}
\def\IF{\relax{\rm I\kern-.18em F}}
\def\IG{\relax\hbox{$\inbar\kern-.3em{\rm G}$}}
\def\IGa{\relax\hbox{${\rm I}\kern-.18em\Gamma$}}
\def\IH{\relax{\rm I\kern-.18em H}}
\def\II{\relax{\rm I\kern-.18em I}}
\def\IK{\relax{\rm I\kern-.18em K}}

\def\inbar{\,\vrule height1.5ex width.4pt depth0pt}

\def\lp10{\ell_p^{10}}
\def\lp11{\ell_p^{11}}
\def\R11{R_{11}}

\def\frac#1#2{{#1 \over #2}}

\newdimen\tableauside\tableauside=1.0ex
\newdimen\tableaurule\tableaurule=0.4pt
\newdimen\tableaustep
\def\phantomhrule#1{\hbox{\vbox to0pt{\hrule height\tableaurule width#1\vss}}}
\def\phantomvrule#1{\vbox{\hbox to0pt{\vrule width\tableaurule height#1\hss}}}
\def\sqr{\vbox{%
  \phantomhrule\tableaustep
  \hbox{\phantomvrule\tableaustep\kern\tableaustep\phantomvrule\tableaustep}%
  \hbox{\vbox{\phantomhrule\tableauside}\kern-\tableaurule}}}
\def\squares#1{\hbox{\count0=#1\noindent\loop\sqr
  \advance\count0 by-1 \ifnum\count0>0\repeat}}
\def\tableau#1{\vcenter{\offinterlineskip
  \tableaustep=\tableauside\advance\tableaustep by-\tableaurule
  \kern\normallineskip\hbox
    {\kern\normallineskip\vbox
      {\gettableau#1 0 }%
     \kern\normallineskip\kern\tableaurule}%
  \kern\normallineskip\kern\tableaurule}}
\def\gettableau#1 {\ifnum#1=0\let\next=\null\else
  \squares{#1}\let\next=\gettableau\fi\next}

 \def\eqnn#1{\xdef #1{(\secsym\the\meqno)}\writedef{#1\leftbracket#1}%
 \global\advance\meqno by1\wrlabeL#1}
 \def\eqna#1{\xdef #1##1{\hbox{$(\secsym\the\meqno##1)$}}
 \writedef{#1\numbersign1\leftbracket#1{\numbersign1}}%
 \global\advance\meqno by1\wrlabeL{#1$\{\}$}}
 \def\eqn#1#2{\xdef #1{(\secsym\the\meqno)}\writedef{#1\leftbracket#1}%
 \global\advance\meqno by1$$#2\eqno#1\eqlabeL#1$$}

\global\newcount\itemno \global\itemno=0

\def\itemaut#1{\global\advance\itemno by1\noindent\item{\the\itemno.}#1}

\def\({\left(}
\def\){\right)}

\def\lsim{\mathrel{\mathstrut\smash{\ooalign{\raise2.5pt\hbox{$<$}\cr\lower2.5pt\hbox{$\sim$}}}}}
\def\gsim{\mathrel{\mathstrut\smash{\ooalign{\raise2.5pt\hbox{$>$}\cr\lower2.5pt\hbox{$\sim$}}}}}

\def\overleftrightarrow#1{\vbox{\ialign{##\crcr
     $\leftrightarrow$\crcr\noalign{\kern-0pt\nointerlineskip}
     $\hfil\displaystyle{#1}\hfil$\crcr}}}
     
     \def\overleftarrow#1{\vbox{\ialign{##\crcr
     $\leftarrow$\crcr\noalign{\kern-0pt\nointerlineskip}
     $\hfil\displaystyle{#1}\hfil$\crcr}}}

\usepackage[yyyymmdd,hhmmss]{datetime}
\newif{\ifeq}           
\eqtrue                 
\newcounter{lecturecounter}

\usepackage{placeins}  
\usepackage{flafter}   

\usepackage[utf8]{inputenc}
\usepackage{comment}
\usepackage{float, xcolor}
\usepackage{physics}
\usepackage{ragged2e}
\usepackage{booktabs}
\usepackage{latexsym}
\usepackage{mathtools}

\usepackage{subcaption}
\usepackage{graphicx}
\usepackage{cleveref}
\usepackage{amsmath}

\numberwithin{equation}{section}
\definecolor{darkgreen}{rgb}{0,0.5,0}
\definecolor{darkblue}{rgb}{0,0,0.6}
\usepackage[format=plain,
            labelfont={bf},
            textfont={it}]{caption}
\usepackage{caption}
\def \be {\begin{equation}}
\def \ee {\end{equation}}
\def \bea {\begin{eqnarray}}
\def \eea {\end{eqnarray}}
\newcommand{\eq}[1]{(\ref{#1})}

\hypersetup{
citecolor=red,
colorlinks=true,
filecolor=red,
linkcolor=blue,
linktocpage=true,
urlcolor=darkgreen
}

\begin{document}

\title{Entanglement Harvesting and Quantum Discord of Alpha Vacua in de Sitter Space}

\author{Feng-Li Lin$^\alpha$ and  Sayid Mondal$^\beta$}

\address{ \vspace{0.0cm}
	{Department of Physics, National Taiwan Normal University, Taipei, 11677, Taiwan}
	
}

\email{$^\alpha$\href{mailto: fengli.lin@gmail.com}{ fengli.lin@gmail.com},  $^\beta$\href{mailto:sayid.mondal@gmail.com}{sayid.mondal@gmail.com}}

\abstract{ The CPT invariant vacuum states of a scalar field in de Sitter space, called $\alpha$-vacua, are not unique. We explore the $\alpha$-vacua from the quantum information perspective by a pair of static Unruh-DeWitt (UDW) detectors coupled to a scalar field with either monopole or dipole coupling, which are in time-like zero separation or space-like antipodal separation. The analytical form of the reduced final state of the UDW detector is derived. We study the entanglement harvesting and quantum discord of the reduced state, which characterize the quantum entanglement and quantum correlation of the underlying $\alpha$-vacua, respectively. Our results imply that the quantum entanglement gravitated by de Sitter gravity behaves quite differently for time-like and space-like separations. It experiences ``sudden death" for the former and grows for the latter as the measuring time or the value of $\alpha$ increases. This demonstrates the nonlocal nature of quantum entanglement. For the quantum discord, we find no ``sudden death" behavior, and it experiences superhorizon suppression, which explains the superhorizon decoherence in the inflationary universe scenario. Overall, the time-like or space-like quantum entanglement and correlation behave differently on their dependence of $\alpha$, measuring time and spectral gaps, with details discussed in this work. 
}

\tableofcontents


\section{Introduction}

De Sitter space is intriguing in many different ways. On the one hand, it is one of the simplest spacetimes because it is maximally symmetric. On the other hand, it is a highly dynamic spacetime adopted to explain inflationary universe scenarios. The existence of the eternal Hubble horizon initiated the primordial curvature fluctuations, which caused the signature of cosmic microwave background (CMB) in today's observational universe \cite{gibbons1983very}. Besides, in the static patch, the existence of a cosmic event horizon mimics the one associated with black holes, such as the thermal properties related to Hawking-like radiation. These thermal properties are also encoded in the vacuum Wightman function of a quantum field in de Sitter space and manifest in the infinite number of imaginary double poles in its spectral representation. Furthermore, the full isometry of the de Sitter space also allows a variety of the vacua states of a quantum field, the so-called $\alpha$-vacua \cite{Mottola:1984ar, PhysRevD.32.3136,  PhysRevD.98.065014, PhysRevD.65.104039}, which has nontrivial UV and IR properties such as the acausal correlations.

Therefore, it is worth exploring further the physical properties of vacua of de Sitter space. Among many properties, the quantum information perspective has been studied less for de Sitter space. Quantum fluctuations of a relativistic field carry both energy and quantum fluctuation, and it is natural to ask how the Hubble horizon affects long-range quantum correlations such as quantum entanglement or non-classical mutual information. The answers to these questions can help understand the quantum information perspective of the primordial fluctuations and their imprints in CMB. Moreover, the zero-point energy fluctuations lead to the famous cosmological constant problem, which could result in the de Sitter space. This has puzzled theoretical physicists for decades, and the resolution still seems elusive. As information and energy are twin partners of a physical entity such as quantum fields, the insight about the quantum information of de Sitter space may help to uncover the mysterious cosmological constant in the long run. This further motivates the study of quantum information problems in de Sitter space.

Due to the almost infinite degrees of freedom, directly studying the quantum information of a relativistic quantum field is a formidable task. Despite that, many important results have been obtained in the past few decades. For example, the entanglement entropy of vacuum states between adjacent regions is shown to obey the area-law \cite{Srednicki:1993im, Callan:1994py, Holzhey:1994we}. Also, the Reeh-Schlieder theorem \cite{Schlieder1965SomeRA, Witten:2018zxz, Sanders:2008gs} showed that the entanglement of a quantum field exists among regions separated at all scales. Introducing new tools such as holographic principle or its manifest via AdS/CFT duality \cite{Maldacena:1997re} helps to explore the quantum information of nontrivial quantum field theories by their dual descriptions \cite{Ryu:2006bv, Chen:2021lnq}. On the other hand, a more conservative way is to explore the quantum information of a relativistic quantum field by some elementary probe, by which we can watch or monitor the change in quantum information due to its interaction with the environmental quantum field. A typical probe for such a purpose is the atomic or qubit-like Unruh-DeWit (UDW) detector \cite{PhysRevD.14.870, DeWitt:1980hx}. 
Via environmental scalar as a quantum channel, the distant UDW detectors can build up long-range entangled states violating Bell's inequality; this confirms the implication of the Reeh-Schlieder theorem \cite{summers1985vacuum, Summers:1987ze, VALENTINI1991321, Reznik:2002fz, PhysRevA.71.042104}.

Extending the above discussions from the Minkowski spacetime to the curved spacetime yields more interesting phenomena because gravity force, even as a classical entity, can cause nontrivial effects on the vacuum states. For example, the self-force may induce radiation damping for the UDW detector via the fluctuation-dissipation theorem \cite{Wilson-Gerow:2024ljx} and induce decoherence \cite{Danielson:2021egj, Danielson:2022sga, Danielson:2022tdw, Dhanuka:2022ggi, Gralla:2023oya}, reflecting either the thermal nature of the Rindler vacuum for an accelerating detector or of the cosmic horizon or the event horizon of a black hole. Inspired by these results and the aforementioned motivation to explore the quantum information perspective of de Sitter space, this paper will consider a pair of static UDW detectors coupled to the scalar field in the de Sitter space. By evolving the total system to arrive at its final state in the far future, we will see how the strong gravity affects the quantum information content of the UDW detectors via the coupling to the vacuum channel of the environmental scalar.  By tracing out the scalar field part of the total final state, we obtain the reduced density matrix of the UDW detectors in a mixed state. Examining this reduced mixed state, we can read out the effect of de Sitter gravity on its quantum information content and uncover the gravity effect on the quantum fluctuations of the scalar field.

Two characteristics of quantum information considered in this paper for the reduced final are entanglement harvesting \cite{Salton:2014jaa, Martin-Martinez:2015eoa, PhysRevD.93.044001, Henderson:2017yuv, Kukita:2017etu, Koga:2019fqh, Perche:2022ykt, Mendez-Avalos:2022obb} and quantum discord \cite{ollivier2001introducing, Henderson:2001wrr,Barman:2021kwg,Barman:2021bbw,Barman:2022xht}. The first explores the time scale and energy gap dependences of the quantum entanglement generated by strong gravity and then harvested by the UDW detectors. The second examines the amount of non-classical quantum correlation generated by the gravity of de Sitter space. These two quantum information characteristics are related but not equivalent. For example, we will see the ``sudden death" behavior for the entanglement harvesting at the superhorizon scale or large energy gap, but not for the quantum correlation. These quantum informational quantities, in general, are nonlocal and will depend on the separation scale between two UDW detectors. However, the general separation will be quite difficult for the analytical calculations. To avoid technical involvement, we will consider two extreme cases: the zero and antipodal separations, representing the time-like and space-like separations, respectively.  For the same reason, we will only consider a conformally coupled scalar, of which the corresponding Green functions of vacuum states take the simpler form. With this technical simplification, we can obtain the analytical form of the reduced density matrix of the final state for the UDW detectors in the saddle point approximation, which is formally valid for large measuring time scales. This will avoid the numerical inaccuracy arising from the peculiar $i\epsilon$ prescription when evaluating the quantum informational quantities of the reduced states. Based on these results, we will study the dependence of quantum information on the measuring time scale and energy gaps of the UDW detectors and examine how they behave at the superhorizon scale. Besides, we consider both monopole and dipole couplings of UDW detectors to the environmental scalar. In some cases, we will see the essential differences between two different ways of coupling. Our results serve as the prototype for the quantum correlation between two different regions separated by either time-like or space-like distances in de Sitter space and can shed new insight into the quantum information perspective of the primordial fluctuations or quantum gravity. 

The rest of the paper is organized as follows. In the next section, we will briefly review the basics of the UDW detector, concurrence, and discord, scalar vacuum states in de Sitter space, and finally derive and classify some issues of the spectral representations of Wightman Green functional of de Sitter $\alpha$-vacua. In section \ref{sec3}, we obtain the matrix elements of the reduced density of a pair of UDW detectors in terms of the spectral density of Wightman functions in the saddle point approximation. Based on the analytical results of section \ref{sec3}, in section \ref{sec4} and \ref{sec5}, we present the numerical plots respectively for entanglement harvesting and quantum discord to demonstrate their dependence on measuring time scale, energy gaps and the variety of vacua. We finally conclude our paper in section \ref{sec6}.

\section{Brief review on UDW detectors, the quantum characteristics of $X$-states and $\alpha$-vacua of de Sitter space}

In this section, we sketch the basics of UDW detectors,  quantum characteristics of the resultant final $X$-states of their reduced dynamics, and $\alpha$-vacua of a scalar field in de Sitter space. At the same time, we set up the notations for our constructions and calculations. Despite this being a review section, the quantum discord for the resultant $X$-states of the UDW detectors is obtained here for the first time. Besides, in the last part of this section, we also give the analytical forms for the spectral representations of Wightman functions probed by two static UDW detectors with zero or antipodal separation.

\subsection{Probing environmental vacuum state by UDW detectors}
The UDW detector \cite{PhysRevD.14.870, DeWitt:1980hx} is a particle detector that serves as a local probe of the quantum field. For simplicity, we will consider a two-level system (qubit) for such a particle detector with its worldline trajectory denoted as  $x(\tau)$ parameterized by its own proper time $\tau$. This qubit-type UDW detector is characterized by the energy gap $\Omega$ separating the ground $|0\rangle$ from the excited state $|1\rangle$, and by its {monopole} interaction Hamiltonian with the environmental scalar field $\phi(x)$, which in the interaction picture is expressed as
\begin{equation}\label{Hamil_int_mono}
H^{(0)}(\tau)=g \;  \chi(\tau)\left(e^{i \Omega \tau} |1\rangle\langle 0| +e^{-i \Omega \tau} |0\rangle\langle 1|\right) \otimes \phi\big(x(\tau)\big)\;.
\end{equation}
Here, the superscript $(0)$ in $H^{(0)}$ refers to monopole coupling, and we also introduce a coupling constant $g$ to tune the interaction strength and the window function $\chi(\tau)$ to characterize the duration of interaction for the UDW detector coupled to $\phi(x)$. In this work, we will use the typical Gaussian-type window function,  
\begin{equation}\label{chi_w}
\chi(\tau)=\frac{1}{(2 \pi)^{1 / 4}} \mathrm{e}^{-\frac{\tau^2}{4 T^2}}\;,
\end{equation}
where the normalization is chosen so that $\int_{-\infty}^{\infty} d\tau [\chi(\tau)]^2 = T$, and the parameter $T$ can be considered as the interaction time scale or its inverse as the resolution energy scale for measuring the field fluctuation by the UDW detector.

In this work, we like to study the quantum information perspective of the quantum field fluctuations in its vacuum states probed by a pair of qubit-type UDW detectors. Compared to the scheme of using just a single UDW detector, we can explore the quantum entanglement or quantum correlation between the detectors, which we label as $A$ and $B$. The associated quantities for each detector will be indicated by the subscripts $A$ and $B$. To characterize the strong gravitational effect on the relativistic quantum information, we will prepare an initial state that contains minimal content of quantum information and then examine the influence of gravity through quantum evolution by the interaction Hamiltonian \eq{Hamil_int_mono}. The simplest initial state for such purpose is the direct product of the vacuum states for all the involved partners, i.e., 
\be 
|\Psi\rangle_i:=|0\rangle_A \otimes |0\rangle_B \otimes |\Lambda\rangle\;,
\ee
where $|0\rangle_{A,B}$ and $|\Lambda\rangle$ are the corresponding vacuum states of UDW detectors and the scalar field.  We then evolve the initial state $|\Psi\rangle_i$, prepared at the coordinate time $t=-\infty$ to the final total state $|\Psi\rangle_i$ at $t=\infty$, which in the interaction picture can be expressed as
\begin{equation}
|\Psi\rangle_f = \mathcal{T} \exp\left[-i \int_{-\infty}^{\infty} dt \left[\frac{d\tau_A}{dt} H^{(0)}_A\left(\tau_A\right) + \frac{d\tau_B}{dt} H^{(0)}_B\left(\tau_B\right) \right] \right] |\Psi\rangle_i \;,
\end{equation}
where $\mathcal{T}$ is the time ordering operator, $\frac{d\tau_{A, B}}{dt}$ are the corresponding boost factors from the comoving frames to the laboratory frame. In this work, we will study only the static UDW detectors, for which we can simply choose $\tau_{A,B}=t$.

The final total state $|\Psi\rangle_f$ encodes all the gravitational effects on the quantum fluctuations and quantum information of the coupled system. However, due to the complication of the quantum field by its infinite number of degrees of freedom, it is not easy to examine it directly. As the quantum evolution entangles the UDW detectors and the fields, we can then glimpse the imprint effect of gravity on the quantum field fluctuations by using the UDW detectors as the probe. This implies we can examine the reduced final state of the UDW detector by taking the partial trace of $|\Psi\rangle_f\langle \Psi|$ over $\phi$. Up to ${\cal O}(g^2)$ this results in the reduced density matrix $\rho_{AB}$, which takes the famous form of $X$-states \cite{yu2005evolution, rau2009algebraic} \footnote{Bell states and Werner states \cite{PhysRevA.40.4277} are special examples of the $X$-states.}, and is dictated by the Wightman function of $\phi$, which we denote by $W_{\Lambda}(x, x^{\prime}):=\left\langle \Lambda \left|\phi(x) \phi\left(x^{\prime}\right)\right| \Lambda \right\rangle$. In the basis of $|i\rangle_A \otimes |j\rangle_B=\{|00\rangle,|01\rangle,|10\rangle,|11\rangle\}$, it is explicitly given by \cite{PhysRevD.93.044001, Martin-Martinez:2015eoa, Henderson:2017yuv}
\begin{equation}\label{dtec_den_mat}
\rho_{A B}=\left(\begin{array}{cccc}
1-P_A-P_B & 0 & 0 & X \\
0 & P_B & C & 0 \\
0 & C^* & P_A & 0 \\
X^* & 0 & 0 & 0
\end{array}\right)+\mathcal{O}\left(g^4\right),
\end{equation}
where
\begin{align}
&P_D = g^2 \int_{-\infty}^{\infty} d\tau_D \int_{-\infty}^{\infty} d \tau_D' \, e^{-i \Omega_D \left(\tau_D-\tau_D'\right)} W_{\chi\Lambda}\big(x_D(\tau_D),x_D(\tau_D')\big)   \quad \mbox{for} \quad D \in \{A,B\},
\label{PJ} \\
&C= g^2 \int_{-\infty}^{\infty}  d\tau_A \int_{-\infty}^{\infty}  d\tau_B   \,  e^{- i \left( \Omega_A \tau_A - \Omega_B \tau_B \right)} W_{\chi\Lambda}\big(x_A(\tau_A),x_B(\tau_B) \big) \label{defC}, \\
&X =-g^2  \int_{-\infty}^{\infty}   d\tau_A  \int_{-\infty}^{\infty}   d\tau_B \,   e^{-i\left( \Omega_A  \tau_A + \Omega_B  \tau_B \right)}
\Big[ \theta\big[t_B(\tau_B)-t_A(\tau_A)\big] W_{\chi\Lambda}\big(x_A(\tau_A),x_B(\tau_B) \big) \nn
\\
& \qquad \qquad \qquad + \theta\big[t_A(\tau_A)-t_B(\tau_B)\big] W_{\chi\Lambda}\big(x_B(\tau_B),x_A(\tau_A) \big) \Big] \;,
\label{defX}
\end{align}
with $\theta[t]$ denoting the Heaviside step function of the coordinate time $t$, and the windowed Wightman function is defined by 
\be\label{win_G}
W_{\chi\Lambda}(x_F(\tau_F),x_G(\tau_G)):=\chi(\tau_F) \chi(\tau_G) \; \langle \Lambda| \phi(x_F(t_F)) \phi(x_G(t_G))|\Lambda \rangle \;, \quad \mbox{for} \quad F,G \in \{A,B\}\;.
\ee
Here, $P_{D=A, B}$, by definition, is the transition probability for a single UDW detector, which yields the transition rate $\sim P_D/T$ for characterizing the Unruh effect for a constantly accelerating UDW detector \cite{PhysRevD.14.870}. This can be verified by the fact that $\rho_A=\tr_B \rho_{AB}={\rm diag}(1-P_A, P_A) + \mathcal{O}(g^2)$. The eigenvalues of $\rho_{AB}$ are given by \cite{ali2010quantum, Koga:2018the} (up to $\mathcal{O}(g^2)$),
\be
\lambda_{0,1,\pm}=0, \quad 1- P_A- P_B, \quad \frac{1}{2} \Big( P_A+ P_B \pm \sqrt{(P_A-P_B)^2+ 4 |C|^2} \Big)\;.
\ee
Note that the positive condition for $\lambda_-$ requires 
\be
\frac{P_A P_B}{|C|^2}\ge 1\;.
\ee
This condition is usually satisfied for generic vacuum states of $\phi$. Furthermore, if the two UDW detectors are identical, i.e.,  $\Omega_A=\Omega_B$ (and also  $\chi_A=\chi_B$), then $P_A=P_B:=P$. In this case, the eigenvalues of $\rho_{AB}$ are $0, 1-2 P, P+|C|, P-|C|$. Moreover, if two UDW detectors are also placed at the same spatial location, then it is straightforward to see $C=P$ \footnote{$C=P$ is the case only if the switching functions, i.e., $\chi(\tau)$ are the same for both UDW detectors as assumed in this work. Otherwise, $C$ will not equal $P$ as seen from the defining equations \eq{PJ} and \eq{defC}.}, so that $\rho_{AB}$ becomes a rank-$2$ matrix. 

It is also straightforward to extend the above consideration into the dipole interactions with the interaction Hamiltonian taking this form \cite{Wilson-Gerow:2024ljx}
\be\label{dipole_H}
H^{(2)}(\tau)=g \;  \chi^{\mu}(\tau)\left(e^{i \Omega \tau} |1\rangle\langle 0| +e^{-i \Omega \tau} |0\rangle\langle 1|\right) \otimes \Phi_{\mu}\big(x(\tau)\big)\;,
\ee 
where  the superscript $(2)$ in $H^{(2)}$ refers to dipole coupling, and $\Phi_{\mu}$ can be either $\partial_{\mu}\phi$ for a scalar-dipole interaction, or $F_{0 \mu}$ for an electric-dipole interaction with $F_{\mu\nu}$ the Maxwell's field strength.  We introduce the local tetrad $e^I_{\mu}$ and its inverse along the worldline to define the local vector field $\Phi_I:=\Phi_{\mu} e^{\mu}_I$ and the local vector window function $\chi^I:=\chi^{\mu} e^I_{\mu}$ which will be again chosen to be Gaussian-type. Then, the reduced density matrix for the pair of UDW detectors up to ${\cal O}(g^2)$ will take the same $X$-state form as in \eq{dtec_den_mat}. However, the windowed Wightman function defined in \eq{win_G} and used in \eq{PJ}-\eq{defX} will be replaced by 
\be\label{dipole_W}
{\cal W}_{\chi\Lambda}\big(x_F(\tau_F), x_G(\tau_G)\big):=\chi^I(\tau_F) \chi^J(\tau_G) \; \langle \Lambda| \Phi_I(x_F(t_F)) \Phi_J(x_G(t_G)) |\Lambda\rangle\;, \quad F,G \in \{A,B\}\;.
\ee
We will provide the explicit forms of the above windowed Green functions later.

\subsection{Characterizing quantumness: concurrence and discord}

Given a  $\rho_{AB}$, we can characterize the quantum information by some entanglement measure to characterize the entanglement harvested by the UDW detectors from the vacuum fluctuations of the scalar field. A common quantity for evaluating the entanglement of a mixed state is concurrence \cite{PhysRevLett.80.2245}, which we will adopt. The concurrence for the $X$-state of \eq{dtec_den_mat} up to ${\cal O}(g^2)$ can be found to be \cite{PhysRevD.93.044001,Henderson:2017yuv},
\begin{equation}\label{def_concurrence}
\mathcal{C}\left(\rho_{A B}\right)=2 \max \left[0,|X|-\sqrt{P_A P_B}\right] \;.
\end{equation}
The concurrence is an entanglement monotone, so its value quantifies the amount of quantum entanglement.  Since we start with a pair of UDW detectors in their unentangled product of ground states, any entanglement quantified by the nonzero concurrence for the reduced final state can be seen as harvesting from the entanglement of the environmental scalar gravitated by the de Sitter space during the time evolution. Thus, this quantity can be coined as entanglement harvesting \cite{Salton:2014jaa, Martin-Martinez:2015eoa, PhysRevD.93.044001, Henderson:2017yuv, Kukita:2017etu, Perche:2022ykt, Mendez-Avalos:2022obb}. By studying the scale dependence of the entanglement harvesting, we can uncover the gravity effect of de Sitter space on the generation of entanglement in the scalar vacua at different length/time scales. 
 
Another quantity adopted in this paper for characterizing the quantum coherence of a $X$-state is the quantum discord \cite{ollivier2001introducing, Henderson:2001wrr}, which measures the difference between two natural extensions of classical mutual information between $A$ and $B$. Thus, the quantum discord is used to characterize the quantumness of correlations between subsystems, which does not necessarily involve quantum entanglement. In classical information theory, the mutual information $I(A,B):=S(A)+S(B)-S(AB)=S(A)-S(A||B)$ where $S(A)$, $S(B)$ and $S(AB)$ are the Shannon entropy of the subsystem $A$ and $B$, and total system AB, respectively; and $S(A||B)=S(AB)-S(B)$ is the relative entropy. A natural extension to quantum mutual information is to replace the Shannon entropy with the von Neumann entropy, e.g., $S(A)=-\Tr_A \rho_A \ln \rho_A$. Alternatively, one can define the quantum mutual information through its operational meaning, i.e., obtaining the information about $A$ by observing $B$. Introduce the projective measurement basis $\{B_k\}$ for performing measurements on subsystem $B$, one can define alternative quantum mutual information other than $I(A,B)$ by
\be
J(A,B)=S(A)- {\rm min}_{\{ B_k \}} \sum_k p_k S(A|| B_k)\;,
\ee
where $p_k:=\Tr_B(B_k \rho_{AB})$. The measurement destroys quantum correlation so that $J(A,B)$ quantifies the classical correlation. Thus, the quantum discord quantifies the pure quantum correlation by defining it as 
\be 
D(A,B):= I(A,B)- J(A,B) = {\rm min}_{\{ B_k \}} \sum_k p_k S(A|| B_k) - S(A||B) \ge 0 \;.
\ee
The subadditivity of entanglement entropy ensures the last inequality. The quantum discord vanishes when $\rho_{AB}$ is in the pointer states, i.e., environment-induced superselection \cite{ollivier2001introducing} such that $\rho_{AB}=\sum_k B_k \rho_{AB} B_k$.

Evaluation of quantum discord for generic $X$-state was studied extensively \cite{ali2010quantum, chen2011quantum, PhysRevA.88.014302, yurischev2015quantum}. Following the procedure outlined in \cite{yurischev2015quantum} to evaluate the quantum discord for $\rho_{AB}$ of \eq{dtec_den_mat}, we find that the conditioned entropy $\sum_k p_k S(A|| B_k)$ is independent of the choices of measurement basis ${\{ B_k \}}$ up to ${\cal O}(g^2)$ so that there is no need for minimization to obtain quantum discord\footnote{The measurement-basis independence of conditioned entropy $\sum_k p_k S(A|| B_k)$ is a special feature for $\rho_{AB}$ we consider up to ${\cal O}(g^2)$, in which all the measurement-basis dependence happens to vanish. Otherwise, for a generic $X$-state, $\sum_k p_k S(A|| B_k)$ will generally depend on the choice of ${\{ B_k \}}$ as discussed in \cite{ali2010quantum, chen2011quantum, PhysRevA.88.014302, yurischev2015quantum}. }. The resultant quantum discord is 
\bea
D(A,B)&=& {g^2 \over 2 \ln 2} \Bigg[ (P_A+ P_B) \ln (P_A P_B - C^2)- 2 P_A \ln P_A - 2 P_B \ln P_B \ \qquad \nn \\
&& \qquad  + \sqrt{(P_A-P_B)^2 + 4 C^2} \ln {P_A + P_B + \sqrt{(P_A-P_B)^2 + 4 C^2} \over P_A + P_B - \sqrt{(P_A-P_B)^2 + 4 C^2}} \Bigg] \;.  \qquad \label{def_discord}
\eea
For the identical UDW detectors with $P_A=P_B:=P$, it can be reduced to
\be \label{QD_id}
D(A,B)= {g^2 \over \ln 2}\Big[ (P+|C|) \ln (P+|C|) + (P-|C|)\ln (P-|C|) -2 P \ln P  \Big]\;,
\ee
which can be further reduced to $D(A,B)=2 g^2 P$ if these two identical UDW detectors are placed at the same spatial position so that $C=P$. Thus, considering up to ${\cal O}(g^2)$ of two identical UDW detectors in the same position, their quantum discord is proportional to the transition probability.

\subsection{Euclidean vacuum and $\alpha$-vacua in de Sitter space}
As shown in the previous subsection, the final reduced density matrix of the UDW detectors is dictated by the Wightman function of the scalar field. As we aim to study the vacuum states of the scalar field in de Sitter space probed by the UDW detectors, we need to understand the basics of the corresponding Wightman function. In this subsection, we briefly review the necessary materials for our consideration. 

We will consider the invariant vacuum states under the full isometry group $O(1,4)$ of the de Sitter space, which includes the disconnected components related, for example, by the antipodal mapping. Such vacuum states are called $\alpha$-vacua \cite{PhysRevD.32.3136, PhysRevD.65.104039}, including the Euclidean vacuum state, also known as the Bunch-Davies vacuum. For this purpose, we consider the $4$-dimensional de Sitter space, denoted as dS$_4$, in the global coordinates, which can cover the patches connected by the disconnected components of $O(1,4)$. The $O(1,4)$ is also the full Lorentz global of 5-dimensional Minkowski space, in which we can consider dS$_4$ as the embedded hyperbola, with the embedding constraint
\begin{equation}\label{hyperbola}
-X_0^2+X_1^2+X_2^2+X_3^2+X_4^2=L^2.
\end{equation}
where $L$ is the radius of the cosmic horizon of dS$_4$. One can solve this embedding relation for $X^M(x)$ with $x^{\mu}$ the chosen 4-dimensional coordinates for dS$_4$. Then the $O(1,4)$-invariant length interval between two points $x$ and $x'$ is given by 
\be 
\begin{aligned}\label{O14_P}
    P(x,x')=&\frac{1}{L^2}\eta_{MN}X^M(x) X'^N(x')\;.
\end{aligned}
\ee

A dS$_4$ vacuum state $|\Lambda \rangle$ and the Wightman function can be defined by the mode decomposition of the scalar field operator $\phi(x)$ as follows:
\be
\phi(x)=\sum_n \big[ a_n \phi_n(x) + a_n^{\dagger} \phi^*_n(x)\big] 
\ee
where $a_n$ ($a^{\dagger}_n$) the annihilation (creation) operator of eigenmode $\phi_n$ of Klein-Gordon equation in dS$_4$ with appropriate boundary conditions so that
\be
a_n |\Lambda\rangle =0\;, \quad \forall{n}\;.
\ee
The corresponding Wightman function is then given by 
\be
W_{\Lambda}(x,x')= \langle \Lambda|\phi(x)\phi(x') |\Lambda\rangle = \sum_n \phi_n(x) \phi^*_n(x')\;.
\ee
It should be a function of $P(x,x')$ with an appropriate $ i\epsilon$ prescription with $\epsilon=0^+$ to take care of the pole due to light-like separation with  $P(x,x')=0$.  

By definition, a vacuum state should obey the isometry of the underlying spacetime. In dS$_4$, this is either $SO(1,4)$ for the static patch or $O(1,4)$ for the global patch. For the latter, the associate Hadamard Green function, i.e., symmetrized Wightman function or explicitly $H(x,x')=\sum_n \big(\phi_n(x) \phi^*_n(x') + \phi^*_n(x) \phi_n(x')\big)$ should be invariant under the CPT (charge-parity-time reversal) map, which sends a point $x$ into its CPT conjugate point $\bar{x}$ \cite{PhysRevD.32.3136}, i.e., $(t,\vec{x})\rightarrow (-t,-\vec{x})$ or $X^M \rightarrow -X^M$ on the embedding hyperbola under antipodal map. This implies that the vacuum state is {\rm CPT} invariant and the eigenmodes $\{\phi_n \}$ should satisfy 
\be\label{O14_c}
\phi_n(\bar{x})=\phi_n^*(x)\;, \quad \forall{n} \;.
\ee
In this work, we will consider the $O(1,4)$-invariant vacuum states, which include Euclidean vacuum and $\alpha$-vacua \cite{PhysRevD.32.3136}. For such a purpose, we need to consider dS$_4$ in the global coordinate. 

  The global coordinate system of dS$_4$ can be defined through the 
following embedding relations, which solve \eq{hyperbola},
\bea
&& X^0 = L \sinh{t\over L}\;, \qquad X^1=L\cosh{t\over L} \cos\chi\;, \nn \\
&& X^a = L \cosh{t\over L} \sin\chi \; \big(\cos\theta, \sin\theta \cos\phi, \sin\theta \sin\phi \big), \qquad a=2,3,4\;. \label{embed_1}
\eea
From the above, we can turn the 5-dimensional Minkowski metric into the following dS$_4$ metric in the global coordinate,
\begin{equation}\label{dS4_metric}
g_{\mu\nu} dx^{\mu} dx^{\nu}=-d t^2+L^2 \cosh ^2\left(\frac{t}{L}\right) \left(d \chi^2+\sin ^2\chi(d \theta^2+\sin ^2\theta d \phi^2)\right),
\end{equation}
where $t\in(-\infty, \infty)$,~ $(\chi, \theta) \in[0, \pi]$  and $\phi \in[0,2 \pi]$.
In this coordinate, the $P(x,x')$ of \eq{O14_P} in the global coordinate becomes 
\be\label{Pxx_g}
    P(x,x')= -\sinh \frac{t}{L} \sinh \frac{t^{\prime}}{L} + \cosh \frac{t}{L} \cosh \frac{t^{\prime}}{L} \cos\Delta \chi 
\ee
where, without loss of generality, we have chosen $x=(t,\Delta \chi,\theta,\phi)$ and $x=(t',0,\theta,\phi)$.

A particular vacuum state, called Euclidean vacuum or Bunch-Davies vacuum \cite{Chernikov:1968zm} and denoted by $|E\rangle$, of which the Wightman function is given by  \cite{PhysRevD.32.3136, PhysRevD.65.104039, PhysRevD.98.065014} 
\begin{equation}\label{Euclid_Wight_fn_d4}
W_E\left(x, x^{\prime}\right)=\frac{\Gamma\left(\frac{3}{2}-\nu\right) \Gamma\left(\frac{3}{2}+\nu\right)}{16 \pi^2 L^2} { }_2 F_1\left(\frac{3}{2}-\nu, \frac{3}{2}+\nu, 2 ; \frac{1+P(x,x')}{2}\right)\;, \quad  \nu=\sqrt{\frac{9}{4}- m^2 L^2 -12 \xi} 
\end{equation}
with $m$ the scalar's mass and $\xi$ the coupling constant of $\phi^2 R$. For simplicity, in this work, we will only consider the case with $\nu=1/2$, which can be identified as a conformally coupled massless scalar. In this case, \eq{Euclid_Wight_fn_d4} can be simplified to 
\be\label{Euclid_wightman2}
W_E\left(x, x^{\prime}\right)=\frac{1}{8 \pi^2 L^2} \frac{1}{\left(1- P\left(x,x^{\prime }\right)\right)}\;.
\ee
This Wightman function can also be obtained by using the embedding relation \eq{embed_1}  to replace $X^M$ in the Wightman function of the 5-dimensional Minkowski space for a massless scalar, that is, 
\be\label{W_E1}
W_E(x,x') := \langle E|\phi(x)\phi(x') |E\rangle  =-{1\over 4\pi^2} \frac{1}{(X^0(x)-X^0(x') - i \epsilon )^2 - |\vec{X}(x)-\vec{X}(x')|^2}\;. 
\ee
The associated Hadamard function will inherit the $O(1,4)$ invariance of its parent $5$-dimensional Hadamard function; thus, the Euclidean vacuum is $O(1,4)$ invariant.

The Euclidean vacuum is not the only $O(1,4)$-invariant state. To see this,  we first perform a global Bogoliubov transformation for all eigenmodes, i.e., 
\be\label{alpha_mode}
\Tilde{\phi}_{n}(x)=\cosh \alpha ~\phi_n(x)+e^{i\beta}\sinh \alpha ~\phi^{\star}_n(x),
\ee
where the parameters $\alpha \in [0,\infty)$ and $\beta \in (-\pi,\pi)$ are real. Use this set of modes to expand the scalar field operator $\phi(x)=\sum_n \big( \tilde{a}_n \tilde{\phi}_n(x) + {\rm h.c.} \big)$ to define the so-called $\alpha$-vacua $|\alpha, \beta\rangle$ with $\tilde{a}_n|\alpha,\beta\rangle =0$  \cite{PhysRevD.32.3136}. The corresponding Wightman function $W_{\alpha,\beta}=\langle \alpha,\beta| \phi(x) \phi(x') |\alpha, \beta\rangle =\sum_n \tilde{\phi}_n(x) \tilde{\phi}^*(x')_n$ is given by 
\begin{equation}\label{wightman_alpha}
\begin{aligned}
    W_{\alpha,\beta} \left(x, x^{\prime}\right)=\cosh ^2 \alpha W_E\left(x, x^{\prime}\right)+\sinh ^2 \alpha W_E\left(\bar{x}, \bar{x}^{\prime}\right)
    +\frac{1}{2} \sinh 2 \alpha\left(e^{-i\beta}W_E\left(x, \bar{x}^{\prime}\right)+e^{i\beta}W_E\left(\bar{x}, x^{\prime}\right)\right)\;.
\end{aligned}
\end{equation}
In the above, we have used the relation \eq{O14_c}. It is straightforward to see that the associated Hadamard Green function is invariant under the CPT map only if $\beta=0$ \cite{PhysRevD.32.3136}. Otherwise, the vacuum state is invariant under $SO(1,4)$ but not under ${\rm CPT}$ map. We will refer to these $O(1,4)$-invariant vacuum states as the $\alpha$-vacua, denoted as $|\alpha\rangle$, and the associated Wightman function by $W_{\alpha}(x,x')$. 

The vacuum states $|\alpha,\beta\rangle$ have been revived in \cite{PhysRevD.65.104039} to discuss the dS/CFT correspondence. In \cite{PhysRevD.65.104039}, these states are parameterized by a single complex number; we denote it as $\tilde{\alpha}$ with ${\rm Re}\; \tilde{\alpha}<0$ to distinguish from the $\alpha$ in $|\alpha,\beta\rangle$. Thus, we have $|\tilde{\alpha}\rangle=|\alpha,\beta\rangle$, which was shown in \cite{PhysRevD.65.104039} to be a squeezed state of the Euclidean vacuum, i.e., 
\be
|\tilde{\alpha}\rangle = \exp\Big[ \sum_n \big( c (a_n^{\dagger})^2 -  c^* (a_n)^2 \big) \Big] |E\rangle, \qquad c:={1\over 4} \Big(\ln \tanh ({-{\rm Re} \tilde{\alpha} \over 2}) \Big) e^{-i {\rm Im} \tilde{\alpha}}
\ee
and the relation between $\tilde{\alpha}$ and $\alpha, \beta$ are given by 
\be\label{ab_rel}
\alpha=\tanh^{-1}\left(\exp\left(\text{Re}\left(\tilde{\alpha}\right)\right)\right), \quad \text{and}\quad \beta=\text{Im}(\tilde{\alpha})
\ee
by noting that \eq{alpha_mode} can also be expressed as \cite{PhysRevD.65.104039}
\begin{equation}
\tilde{\phi}_n \equiv N_{\alpha^{\prime}}\left(\phi_n(x)+e^{\alpha^{\prime}} \phi^{\star}_n(x)\right), \quad N_{\alpha^{\prime}} \equiv \frac{1}{1-e^{2 \text{Re}(\alpha^\prime)}}\;.
\end{equation}

\subsection{Spectral representations of Wightman function and its antipodal counterparts}

As the reduced final state $\rho_{AB}$ of UDW detectors depends on the Wightman function $W_E(x,x')$ and its antipodal counterparts, we need to evaluate it either analytically or numerically. However, the required $i\epsilon$ prescription often makes numerical error difficult to control. Thus, we will evaluate it analytically by obtaining its spectral representation. However, as $W_E(x,x')$ depends on both $t$ and $t'$, we may not be able to have a spectral representation if $W_E(x,x')$ cannot be reduced to a function of a single variable of the linear combination of $t$ and $t'$. For the Wightman function  \eq{Euclid_wightman2} with $P(x,x')$ of \eq{Pxx_g} considered in this paper, we see this is the case if $\Delta \chi \ne 0, \pi$. Thus, in this work, we will only consider the two static UDW detectors separated by $\Delta \chi=0,\pi$, and denote the corresponding $W_E(x,x')$ as $W_E^-(x,x')$ and $W_E^+(x,x')$, respectively. These two cases are also the extreme cases for time-like and space-like separations. For simplicity, we will set the static detector's worldline time to coincide with the coordinate time. 

Using $P(x,x')$ of \eq{Pxx_g} for $\Delta \chi=0,\pi$ and the defining equation \eq{Euclid_wightman2} for the Wightman function, we have 
\be\label{W_a0}
W_E^-(x,x') = \frac{a_0}{1 - \cosh\big(s_--i\epsilon\big)} \qquad  {\rm and} \qquad W_E^+(x,x') = \frac{a_0}{1 + \cosh\big(s_+\big)} 
\ee
with
\be
s_{\mp}:=\frac{t\mp t'}{L}\;, \qquad a_0:={1\over 8 \pi^2 L^2}\;. 
\ee
The $i \epsilon$ prescription is inherited from \eq{W_E1}. However, there is no causality issue for the antipodal (or any space-like) separation, so there is no need to provide $i\epsilon$ prescription for $W_E^+(x,x')$. We can obtain the spectral representation of $W^{\mp}_E(x,x')$ by the Fourier transform over $s_{\mp}$. Restrict the spectral density to be bounded below, i.e., $\omega\ge 0$, and turn this Fourier integral into a contour integral on the complex $s_{\mp}$ plane. Using the residue theorem for the double poles \footnote{$W^{\mp}_E(x,x')$ has pure imaginary double poles at $s_-=i 2 n \pi$ with residue $2 i \omega e^{2 n \pi \omega}$, and at $s_+= i (2n+1) \pi$ with residue $2 i \omega e^{2 (n+1) \pi \omega}$ for $n\in \mathbb{Z}$, respectively.} in the lower half $s_{\mp}$ plane without including $n=0$ poles for $W_E^-(x,x')$, we obtain the spectral representation,  
\bea\label{Wxx_1}
W_E^-(x,x') &=&   \int_0^{\infty} d\omega ~ \rho_{0,0}(\omega)~ e^{i {\omega \over L} (t-t')} \;, \\
W_E^+(x,x') &=&    \int_0^{\infty} d\omega ~ \rho_{0,1}(\omega) ~ e^{i {\omega \over L} (t + t')} \;, \label{Wxx_2}
\eea
with the spectral densities
\be\label{rho_lk}
\rho_{\ell,k}(\omega) := 2 a_0 \Big({T \over L}\Big)^{\ell} \; {\omega^{\ell+1} e^{k \pi\omega} \over e^{2\pi \omega}-1}\;, \quad {\rm for} \quad \ell=0,2\;,
\ee
where $\ell=0$ for monopole coupling and $\ell=2$ for dipole coupling, which will be discussed later. Note that the Boltzmann factor ${1\over e^{2\pi \omega}-1}$ reflects the thermal nature of the de Sitter vacua with the temperature related to the Hubble scale, i.e., here $\omega$ is dimensionless with Hubble as the basic unit.


To evaluate the Wightman function for the $\alpha$-vacua, i.e., \eq{wightman_alpha}, we also need to obtain the spectral representation for the CPT-conjugate partners: $W_E^{\mp}(\bar{x},\bar{x}')$, $W_E^{\mp}(\bar{x},x')$ and $W_E^{\mp}(x,\bar{x}')$. Recall that the CPT map sends a point $x=(t,\vec{x})$ to $\bar{x}=(-t,-\vec{x})$ \footnote{Thus, based on \eq{Pxx_g}, the two sitautions considered with $\Delta \chi=0,\pi$ will be swapped, i.e., $W_E^+ \Leftrightarrow W_E^-$, when either $x$ or $x'$ is CPT-conjugated.}. By applying the CTP map to \eq{W_a0} and obtaining the spectral density by Fourier transform via the residue theorem,  we have, 
\be
W^-_E(\bar{x},x')=W^-_E(x,\bar{x}')= \frac{a_0}{1 + \cosh\big(\mp s_-\big)} =   \int_0^{\infty} d\omega ~ \rho_{0,1}(\omega)~ e^{i {\omega \over L} (t - t')}  \label{Wxx_3}
\ee
and 
\bea 
W^+_E(x, \bar{x}') &=& \frac{a_0}{1 - \cosh\big(s_+ -i \epsilon \big)}=   \int_0^{\infty} d\omega ~ \rho_{0,0}(\omega)~ e^{i {\omega \over L} (t+t')} \;, \label{Wxx_4}
\\
W^+_E(\bar{x},x') &=& \frac{a_0}{1 - \cosh\big(-s_+ - i\epsilon \big)}=    \int_0^{\infty} d\omega ~ \rho_{0,2}(\omega)~ e^{i {\omega \over L} (t+t')}  \label{Wxx_5}\;.
\eea
Two coordinate arguments in \eq{Wxx_3} are in antipodal separation, so there is no need for $i\epsilon$ prescription. In contrast, the two coordinate arguments in \eq{Wxx_4} or \eq{Wxx_5} are in zero separation, so we must reinstall the $i\epsilon$ prescription \footnote{In \eq{Wxx_4}, we have chosen the $-i\epsilon$ for $W^+_E(x, \bar{x}')$ by not changing the original $i\epsilon$ prescription under the CPT map. However, it seems the alternative is also possible. When defining the Wightman function for the $(\alpha,\beta)$-vacua by \eq{wightman_alpha}, this ambiguity can be absorbed by changing the sign of $\beta$.}. Also, in \eq{Wxx_5} it is in the $s_+ + i\epsilon$ prescription, so that the $n=0$ poles are included so that $\rho_{0,0}(\omega)$ is \eq{Wxx_4} is changed to $\rho_{0,2}(\omega)$.

Finally, if we perform the CPT map on both coordinate arguments of $W^{\mp}(x,x')$, it will not change the spatial separation but reverse their time order. This is the same as swapping the two coordinate arguments, changing the $s_--i\epsilon$ to $s_-+i\epsilon$. Thus, we have 
\bea 
W^-_E(\bar{x}, \bar{x}') &=& W^-_E(x',x) = \frac{a_0}{1 - \cosh\big(-s_- -i \epsilon \big)}=  \int_0^{\infty} d\omega ~ \rho_{0,2}(\omega)~ e^{i {\omega \over L} (t-t')} \;, \label{Wxx_6}
\\
W^+_E(\bar{x},\bar{x}') &=& W^+_E(x',x) = \frac{a_0}{1 + \cosh\big(-s_+ \big)}=     \int_0^{\infty} d\omega ~ \rho_{0,1}(\omega)~ e^{i {\omega \over L} (t+t')} = W_E^+(x,x')  \label{Wxx_7}\;.
\eea
Similarly, we have
\be\label{swap_r_1}
W^+_E(\bar{x}',x)=W_E^+(\bar{x},x')\;, \qquad W^+_E(x',\bar{x})=W_E^+(x, \bar{x}') \;,
\ee
and 
\be\label{swap_r-2}
W^-_E(\bar{x}',x)=W^-_E(x',\bar{x})=W^-_E(\bar{x},x')=W^-_E(x,\bar{x}')\;.
\ee
The above results for the CPT map of argument swapping of $W^-_E(x,x')$ agree with the ones in \cite{Mottola:1984ar, PhysRevD.32.3136, PhysRevD.65.104039}.

The above are the spectral representations for the Wightman functions used for the monopole coupling of static UDW detectors to a massless conformal scalar. We can generalize to the cases for dipole coupling. Recall the defining equation \eq{dipole_W} of the corresponding windowed Wightman function for dipole coupling; we will only use the scalar dipole for simplicity so that $\Phi_I(x)=e^{\mu}_I(x)\partial_{\mu}\phi(x)$. For static UDW detectors at $\chi=0,\pi$, the worldline tetrads are $e^{\mu}_0=(1,0,0,0)$, $e^{\mu}_{1}=(0,L\cosh{t \over L},0,0)$, and $e^{\mu}_{2,3}=(0,0,0,0)$. If the window function $\chi^1(\tau)$ is nonzero, the $\cosh{t\over L}$ factor in $e^{\mu}_2$ will prevent from obtaining the spectral representation with the same reason discussed before. Therefore, we will only consider the case with the window functions $\chi^1=0$ and $\chi^0$ given by \eq{chi_w}. Then, the corresponding Wightman functions denoted by ${\cal W}_{\mp}(x,x')$ for zero and antipodal separation are give by
\be\label{W_a2}
{\cal W}_E^{\mp}(x,x')= T^2 \partial_t \partial_{t'} W_E^{\mp}(x,x')={T^2 \over  L^2} { a_0  \; (2 \pm \cosh s_{\mp} ) \over \big(1\mp \cosh(s_{\mp}-i\epsilon) \big)^2 }\;. 
\ee
Here, we have compensated the dimension of $\partial_t$ by the resolution time scale $T$ given in \eq{chi_w} when defining the window function.
It turns out that ${\cal W}_E^{\mp}$ has pure imaginary double poles at $s_-=i 2 n \pi$ with residue $2 i \omega^3 e^{2 n \pi \omega}$, and at $s_+= i (2n+1) \pi$ with residue $2 i \omega^3 e^{2 (n+1) \pi \omega}$ for $n\in \mathbb{Z}$, respectively. Based on this, the spectral representations of the Wightman functions for the dipole coupling are just to replace the spectral density $\rho_{0,k}(\omega)$ in their monopole coupling counterparts by $\rho_{2,k}(\omega)$.

Based on the above results for the Euclidean vacuum, we can now obtain the Wightman function for the $|\alpha,\beta\rangle$ vacua. For the zero separation, we denote the Wightman function by $W^{(\ell,-)}_{\alpha, \beta}(x,x')$, and for the antipodal separation by $W^{(\ell,+)}_{\alpha, \beta}(x,x')$ with $\ell=0$ for monopole coupling and $\ell=2$ for dipole coupling.  Based on 
\eq{wightman_alpha}, we can write them in a more unified way as follows:
\be\label{W_f_1}
W^{(\ell,\mp)}_{\alpha,\beta}(x,x') := \sum^2_{k=0} f^{\mp}_k  w_k^{\mp} \;,  \qquad \ell=0,2\;,
\ee
with the component spectral representations 
\be
w^{\mp}_k:=\int_0^{\infty} d\omega \; \rho_{\ell,k}(\omega) \; e^{i {\omega \over L} (t \mp t')}\;.
\ee
The coefficients $f^{\mp}_k$ encode the information about $(\alpha,\beta)$-vacua. Explicitly,
\bea\label{W_f_3}
&& f_0^-= \cosh^2\alpha\;, \qquad f_1^-= \sinh 2\alpha \cos\beta\;, \qquad f_2^-= \sinh^2\alpha  \;; 
\\
&& f_0^+={1\over 2} \sinh 2\alpha \; e^{-i \beta}\;,  \qquad f_1^+= \cosh 2\alpha\;, \qquad f_2^+={1\over 2} \sinh 2\alpha \; e^{i \beta}\;. \label{W_f_4}
\eea
In order to calculate $X$ of \eq{defX}, we also need to have $W^{(\ell,\mp)}_{\alpha,\beta}(x',x)$. Using the swapping rules \eq{Wxx_6}, \eq{Wxx_7}, \eq{swap_r_1} and \eq{swap_r-2}, we obtain
\be\label{W_f_swap}
W^{(\ell,\mp)}_{\alpha,\beta}(x',x) = f^{\mp}_0 w^{\mp}_2 + f^{\mp}_1 w^{\mp}_1 + f^{\mp}_2 w^{\mp}_0 \;.
\ee

\section{Final states of UDW detectors of zero and antipodal separations in de Sitter vacua}\label{sec3}

Given the spectral representation of $W^{(\ell,\mp)}_{\alpha,\beta}(x,x')$ in \eq{W_f_1}-\eq{W_f_4}, we can calculate the corresponding reduced density matrix $\rho_{AB}$ defined in \eq{dtec_den_mat}-\eq{defX} for the zero and antipodal separations with monopole and dipole couplings to a massless conformal scalar field. 

Inspecting \eq{PJ} and \eq{defC},  we note that $P_{D=A,B}$ can be thought of as a special case of $C$ with zero separation and the same $\Omega$ for the pair of UDW detectors. Thus, we will first evaluate $C$, from which we can obtain $P_D$ straightforwardly. According to \eq{W_f_1}, $C$ for zero and antipodal separation denoted respectively by $C^-$ and $C^+$ will take the following form (we sometimes omit the $\ell$ labeling without confusion for simplicity)
\be
C^{\mp}= g^2 \sum_{k=0}^2 f^{\mp}_k c^{\mp}_k\;,
\ee
where $f_k^{\mp}$ as given in \eq{W_f_3} and \eq{W_f_4} encodes the information of $(\alpha,\beta)$-vacua, and 
\be
c^{\mp}_k := \int_0^{\infty} d\omega \; \rho_{\ell,k}(\omega) \int_{-\infty}^{\infty} dt_A \int_{-\infty}^{\infty} dt_B \chi(t_A) \chi(t_B) e^{-i (\Omega_A t_A- \Omega_B t_B)} e^{i {\omega \over L} (t_A \mp t_B)}\;.
\ee
Introduce the following new variables
\be
t_{\pm}={t_A \pm t_B}\;, \qquad {\rm and} \qquad \Omega_{\pm}={\Omega_A \pm \Omega_B \over 2}\;,
\ee
so that
\bea\label{temp_1}
&& \int_{-\infty}^{\infty} \int_{-\infty}^{\infty} dt_A  dt_B \; \chi(t_A) \chi(t_B) e^{-i (\Omega_A t _A \mp \Omega_B t_B)}\;,\nn
\\
&& \qquad = {1\over 2} \int_{-\infty}^{\infty} \int_{-\infty}^{\infty} dt_- dt_+ \chi\big({t_- \over \sqrt{2}}\big) \chi\big({t_+ \over \sqrt{2}}\big) e^{-i (\Omega_{\mp} t _+ + \Omega_{\pm} t_-)}:=\big[\cdots \big]_{\mp}  \;.
\eea
Using \eq{temp_1} and $\int_{-\infty}^{\infty} dt \; \chi({t\over \sqrt{2}}) e^{- i \Omega t}=2(2\pi)^{1/4} T e^{-2 T^2 \Omega^2}$ to carry out the double time integrals, we get 
\be\label{saddle_c_1}
c^{\mp}_k = 2\sqrt{2\pi} T^2 \int_0^{\infty} d\omega \; \rho_{\ell,k}(\omega) \exp\Big\{- 2 \Big( {T\over L} \Big)^2 \Big[ \Big(\omega -\Omega_{\pm} L \Big)^2  + \Big(\Omega_{\mp} L \Big)^2 \Big] \Big\} \;.
\ee
There is no closed form for the above integral; instead, we perform it using the saddle point approximation for large $T/L$, similar to \cite{Niermann:2024fvi}. That is, using
\be\label{saddle_c_2}
\int_0^{\infty} d\omega R(\omega) e^{-\big({T\over L} \big)^2 Q(\omega)} \simeq \sqrt{2 \pi \over Q''(\omega_0)} \Big({L \over T} \Big) R(\omega_0) e^{-\big({T\over L} \big)^2 Q(\omega_0)}\;, \qquad {\rm for \; large } \quad T/L\;,
\ee
with $\omega_0 \ge 0$ a strict minimum such that $g'(\omega_0)=0$ and $R(\omega_0)\ne 0$. Casting \eq{saddle_c_1} into the form of \eq{saddle_c_2}, we find 
\be
\omega_0=\Omega_{\pm} L\;, \qquad Q(\omega_0)= 2 \big(\Omega_{\mp} L\big)^2\;, \qquad Q''(\omega_0)= 4 \;.
\ee
Thus, the resultant $C^{\mp}$ is 
\be\label{C_f_1}
C^{\mp}={g^2 \over 2\pi} \Big({T\over L}\Big)^{\ell+1} \sum_k f_k^{\mp}  e^{-2 \big({T\over L} \big)^2 \big(\Omega_{\mp} L\big)^2} \bar{\rho}_{\ell,k}[\Omega_{\pm}L]\;,
\ee
where the dimensionless spectral density is defined by 
\be\label{rhobar}
\bar{\rho}_{\ell,k}[\omega]:= {\omega^{\ell+1} e^{k \pi\omega} \over e^{2\pi \omega}-1}\;.
\ee
It is related to the spectral density of \eq{rho_lk} by $\rho_{\ell,k}(\omega) =2 a_0 \Big({T \over L}\Big)^{\ell}\bar{\rho}_{\ell,k}[\omega]$.

From the fact that  $\lim_{\omega\rightarrow 0}\bar{\rho}_{\ell,k}[\omega] = {1\over 2\pi} \delta_{\ell,0}$ and  \eq{C_f_1}, we can obtain $P_{D=A,B}$ by taking $\Omega_-\rightarrow 0^+$ and $\Omega_+=\Omega_D$ of $C^-$, and the result is
\be\label{P_D_fab}
P_D = {g^2 \over 2\pi} \Big({T\over L}\Big)^{\ell+1}  \sum_k f_k^- \bar{\rho}_{\ell,k}[\Omega_D L]\;.
\ee
The above results for $C^{\mp}$ and $P_D$ hold for the $(\alpha,\beta)$-vacua, which include the Euclidean vacuum with only nonzero $f_0^-=f_0^+=1$. Moreover, the transition probability $P_D$ of \eq{P_D_fab} agrees with the one in  \cite{PhysRevD.65.104039} by using \eq{ab_rel} for a comparison. 

In the case with identical UDW detectors, by construction $C^-=P_D$, but
\be
C^+=\delta_{\ell,0}  {g^2 \over 4\pi^2} \Big({T\over L}\Big)^{\ell+1} \sum_k f_k^+  e^{-2 \big({T\over L} \big)^2 \big(\Omega_D L\big)^2} \;.
\ee
Note that $C^+=0$ for the identical UDW detectors with dipole coupling.

Finally, we calculate the matrix element $X$ of $\rho_{AB}$. First, note that the factor $e^{-i(\Omega_A t_A -\Omega_B)}$ in $C$ is replace by $e^{-i(\Omega_A t_A + \Omega_B)}$. There are two terms with different time orderings and the arguments of the Wightman function swapped.  
Denote $X$ for zero and antipodal separations by $X^-$ and $X^+$, respectively, then from \eq{W_f_1} and \eq{W_f_swap} they will take the following form
\bea
X^{\mp} &=& - g^2 \Big[ f_0^{\mp} \big(x^{\mp}_{<,0} +x^{\mp}_{>,2} \big) + f_1^{\mp} \big(x^{\mp}_{<,1} +x^{\mp}_{>,1} \big) + f_2^{\mp} \big(x^{\mp}_{<,2} +x^{\mp}_{>,0} \big) \Big] \;,
\\
&:=& - g^2 \sum_{k=0}^2 f_k^{\mp} \tilde{x}^{\mp}_k\;,
\eea
with
\be
x^{\mp}_{s,k} :=\int_0^{\infty} d\omega \; \rho_{\ell,k}(\omega) \big[\cdots \big]_+  e^{i{\omega \over L} t_{\mp}} \; \theta(t_s)  \;,
\ee
where $s=<$ or $>$ with $t_<=-t_-$ and $t_>=t_-$, and  $\big[\cdots \big]_+ \sim  {1\over 2}\int dt_- dt_+ \cdots e^{-i (\Omega_+ t _+ + \Omega_- t_-)}$ is defined in \eq{temp_1}. First, the two theta functions in $\tilde{x}_1=x_{<,1}+x_{>,1}$ can be combined into unity. Carrying out the Gaussian integrals over $t_{\pm}$ and the saddle point approximation for the integral over $\omega$, we arrive
\be
\tilde{x}^{\mp}_1 := 2\pi T^2  e^{-2 \big({T\over L} \big)^2 \big(\Omega_{\pm} L\big)^2} \rho_{\ell,1}[\Omega_{\mp} L] \;.
\ee
On the other hand, the two theta functions in $\tilde{x}_{k=0,2}$ cannot be combined into unity so that the integral over $t_-$ will yield the imaginary error function $\mathrm{erfi}[z]=-\mathrm{erfi}[-z]$. Carrying out the integrals over $t_{\pm}$ yields
\be
x^{\mp}_{s,k=0,2}=\sqrt{2\pi} T^2 \int_0^{\infty} d\omega \;  \rho_{\ell,k=0,2}(\omega) e^{-2 \big({T\over L}\big)^2 \big((\Omega_{\pm} L)^2 + (\omega - \Omega_{\mp} L)^2 \big)} \Bigg[1\mp i \; {\rm erfi}\Big[\sqrt{2}\Big({T\over L}\Big) \big(\omega-\Omega_{\mp} L\big) \Big] \Bigg] \;.
\ee
Perform the saddle point approximation for the above integral. The imaginary error function vanishes at the saddle point $\omega_0=\Omega_- L$. This then results in 
\be
\tilde{x}^{\mp}_0=\tilde{x}^{\mp}_2 = \pi T^2  e^{-2 \big({T\over L}\big)^2 \big(\Omega_{\pm} L \big)^2} \Big(\rho_{\ell,0}[\Omega_{\mp} L] + \rho_{\ell,2}[\Omega_{\mp} L] \Big)\; \;.
\ee

Combine all the above results, we obtain $X^{\mp}$ for the zero and antipodal separations in the $(\alpha,\beta)$-vacua are
\be\label{X_f_0}
X^{\mp} = -{g^2 \over 4\pi} \Big({T\over L}\Big)^{\ell+1} e^{-2 \big({T\over L}\big)^2 \big(\Omega_{\pm} L \big)^2}  \Bigg[ \Big( \bar{\rho}_{\ell,0}[\Omega_{\mp} L] + \bar{\rho}_{\ell,2}[\Omega_{\mp} L] \Big)\; (f_0^{\mp} + f_2^{\mp}) + 2 \bar{\rho}_{\ell,1}[\Omega_{\mp} L]  f_1^{\mp}  \Bigg]\;.
\ee
Again, this result for $X^{\mp}$ holds for the $(\alpha,\beta)$-vacua, including the Euclidean one with only nonzero  $f_0^-=f_0^+=1$. For the identical UDW detectors with $\Omega_-\rightarrow 0$ and $\Omega_+=\Omega_D$, we have
\be\label{XmD}
X^-_D=-\delta_{\ell,0} {g^2 \over 4\pi^2} \Big({T\over L}\Big)^{\ell+1} e^{-2 \big({T\over L}\big)^2 \big(\Omega_D L \big)^2} \sum_{k=0}^2 f^-_k \;,
\ee
and
\be
X^+_D=-{g^2 \over 4\pi} \Big({T\over L}\Big)^{\ell+1}  \Bigg[ \Big( \bar{\rho}_{\ell,0}[\Omega_D L] + \bar{\rho}_{\ell,2}[\Omega_D L] \Big)\; (f_0^+ + f_2^+) + 2 \bar{\rho}_{\ell,1}[\Omega_D L]  f_1^+  \Bigg] \;.
\ee
We note that $X^-_D=0$ for the dipole coupling.

In summary, the analytical forms of the elements of the reduced density matrix for a pair of UDW detectors at zero and antipodal separation with monopole/dipole coupling in the $(\alpha,\beta)$-vacua of de Sitter space are given in \eq{C_f_1}, \eq{P_D_fab} and \eq{X_f_0} supplemented with \eq{rhobar}. These are the key results of this paper.
So far, we have written the expressions of our key results by adopting $L$ as the unit for measuring $T$, and $1/L$ to measure $\Omega$ \footnote{This is different from the usual convention in the entanglement harvesting, e.g., \cite{Henderson:2018lcy, Perche:2022ykt, Maeso-Garcia:2022uzf} by using $1/T$ as the frequency unit and $T$ as the length unit (by setting light speed $c=1$) because $T$ is the overall measuring time. However, in de Sitter space, $L$ is a universal infrared cutoff. It is more natural to adopt it as the basic unit to measure other physical quantities.}. For simplicity, when presenting our results for entanglement harvesting and quantum discord in the next two sections, we will omit $L$ (or setting $L=1$) and treat both $\Omega$ and $T$ as dimensionless quantities with respect to the basic units defined by $L$. Thus, the superhorizon scale means $T$ is larger than ${\cal O}(L)$ or $\Omega$ is smaller than ${\cal O}(1/L)$.  

Finally, we emphasize that our analytical results are obtained by the saddle point approximation for the $\omega$-integral. The saddle point approximation is valid in the large $T$ limit. However, for all cases considered, only one dominant saddle exists.  It is then possible for the results to remain approximately valid even beyond the large $T$ regime. In the numerical plots shown below, we will sometimes also plot the regimes with small $T$. As a consistency check for the small $L$ regime, we will ensure the purity $\Tr \rho^2_{AB}$ is less than unity for all the numerical plots presented below.

\section{Entanglement harvesting from de-Sitter vacua} \label{sec4}

Based on the analytical results of the reduced density matrix given in the last section, we will apply them to calculating the concurrence of \eq{def_concurrence}. This represents the entanglement harvesting from the de sitter vacuum states by the UDW detectors. We then present the results in the numerical plots to demonstrate the dependence of entanglement harvesting on the energy gaps and the measuring time scales of the detectors. Due to the variety of dependent factors, we first need to clarify the logic of our presentation. 

We will start with the results of the Euclidean vacuum in the first subsection and the $(\alpha,\beta)$-vacua in the second one. The simplicity of the Euclidean vacuum helps to capture the essential features of the gravitating quantum information by their energy gap and time scale dependencies. Then, we will examine the effect of the variety of de Sitter vacuums and present the corresponding results by also showing the dependence on the values of $(\alpha,\beta)$. In each subsection, we will first consider the monopole coupling and then the dipole coupling. As the quantum information is usually nonlocal, it is important to observe the effect of the separation between two UDW detectors. Therefore, we will juxtapose the results of zero and antipodal separations for all the numerical plots. For simplicity, we will only present the plots for the identical UDW detectors.

\subsection{de Sitter Euclidean vacuum}
In this subsection, we analyze concurrence for two identical UDW detectors, i.e., $\Omega_A=\Omega_B=\Omega$
as a function of the detector energy gap and interaction time in the Euclidean vacuum in de Sitter space. We start with the scenario of monopole coupling \eq{Hamil_int_mono} and subsequently examine the case of dipole coupling \eq{dipole_H}.

\subsubsection{Monopole coupling in Euclidean vacuum}
\begin{figure}[h]
	\centering
	\begin{subfigure}{.4\textwidth}
		\centering
		\includegraphics[width=.9\linewidth]{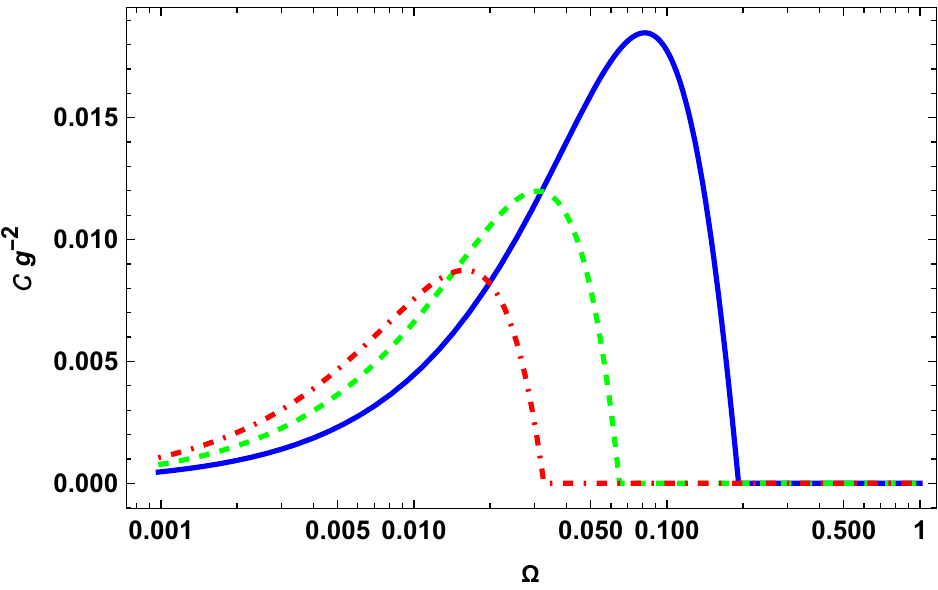}
		\caption{}
		\label{fig_con_BD_mono_N_vs_w}
	\end{subfigure}\hspace{.45cm}
	\begin{subfigure}{.4\textwidth}
		\centering
		\includegraphics[width=.9\linewidth]{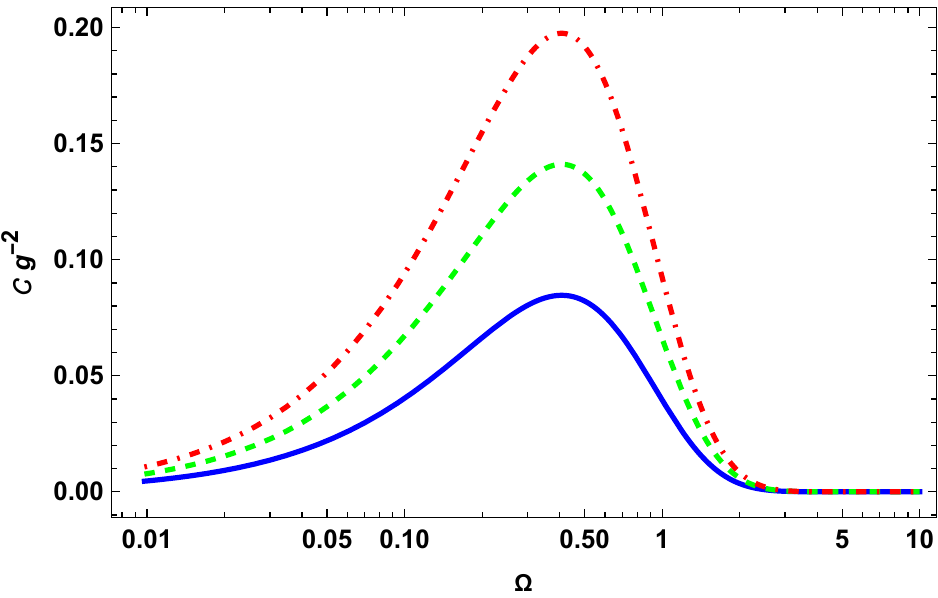}
		\caption{}
		\label{fig_con_BD_mono_NS_vs_w}
	\end{subfigure}
	\caption{The concurrence $\mathcal{C}$ of two monopole-coupling UDW identical detectors in Euclidean vacuum as a function of $\Omega$ for different values of $T=3$ (solid-blue), $T=5$ (green-dashed) and $T=7$ (red-dot-dashed) with (a) zero separation and (b) antipodal separation. Note that ``sudden death" occurs for (a), not for (b). }
		\label{fig_con_BD_mono_vs_w}
\end{figure}

\begin{figure}[h]
	\centering
	\begin{subfigure}{.4\textwidth}
		\centering
		\includegraphics[width=.9\linewidth]{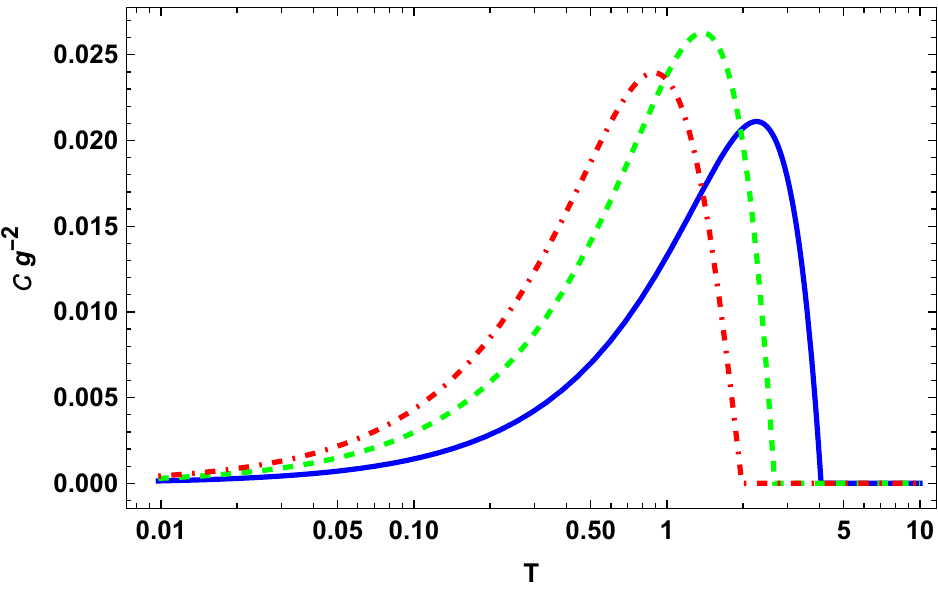}
		\caption{}
		\label{fig_con_BD_mono_N_vs_T}
	\end{subfigure}\hspace{.5cm}
	\begin{subfigure}{.4\textwidth}
		\centering
		\includegraphics[width=.9\linewidth]{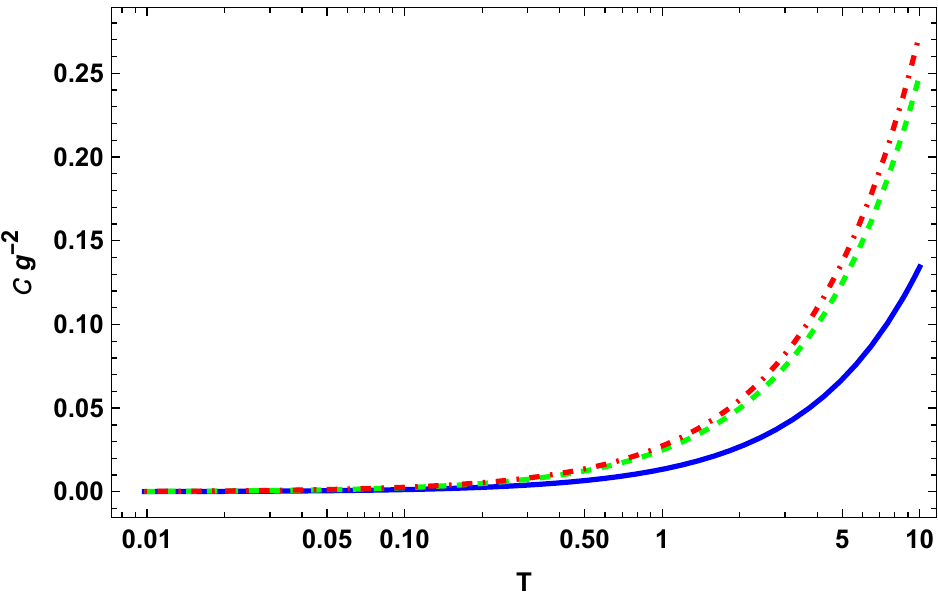}
		\caption{}
		\label{fig_con_BD_mono_NS_vs_T}
	\end{subfigure}
	\caption{The concurrence $\mathcal{C}$ of two monopole-coupling identical UDW detectors in Euclidean vacuum as a function of $T$ for different values of $\Omega=.1$ (solid-blue), $\Omega=.25$ (green-dashed) and $\Omega=.5$ (red-dot-dashed) with (a) zero separation and (b) antipodal separation. Note that ``sudden death" occurs for (a), not for (b). Moreover, in (b), $\mathcal{C}$ grows over time, sharply contrasted with decaying behavior for the other cases.}
 \label{fig_con_BD_mono_vs_T}
 \end{figure}

In \cref{fig_con_BD_mono_N_vs_w} and \cref{fig_con_BD_mono_NS_vs_w}, we present the concurrence as a function of the detector energy gap $\Omega$ for different measuring time $T$, for zero spatial separation and antipodal separations, respectively.   We observe for both cases that concurrence attains a maximum value before monotonically diminishing to zero as the detector energy becomes large. Interestingly, in the case of zero spatial, there is a phenomenon akin to the “sudden death” of concurrence occurring at a large detector energy gap. Furthermore, we depict the concurrence as a function of $T$ for various $\Omega$ in \cref{fig_con_BD_mono_N_vs_T} and \cref{fig_con_BD_mono_NS_vs_T} for zero and antipodal separations, respectively. We also notice the phenomena of “sudden death” of the concurrence when we plot it as a function of $T$ for the case of zero spatial separation, while for antipodal separation, the concurrence increases monotonically with time.
\FloatBarrier

\FloatBarrier

In the current setup, we use the UDW detectors to probe the entanglement structure of the underlying scalar vacuum. Thus, the zero and antipodal separations of UDW detectors probe the short-range and long-range quantum entanglement, respectively. From the above results, we can see that the short-range and long-range entanglements behave differently as a function of the energy gaps and the measuring time scale. In particular, ``sudden death'' only occurs for the short-range entanglement, not the long-range one. Besides, the \cref{fig_con_BD_mono_NS_vs_T} implies that the Euclidean vacuum generates more long-range entanglement over time, characterized by the growing concurrence for the antipodal separation. This contrasts with the zero separation cases where the concurrence decays to zero at large $T$.    
\begin{figure}[h]
	\centering
	\begin{subfigure}{.45\textwidth}
		\centering
		\includegraphics[width=.9\linewidth]{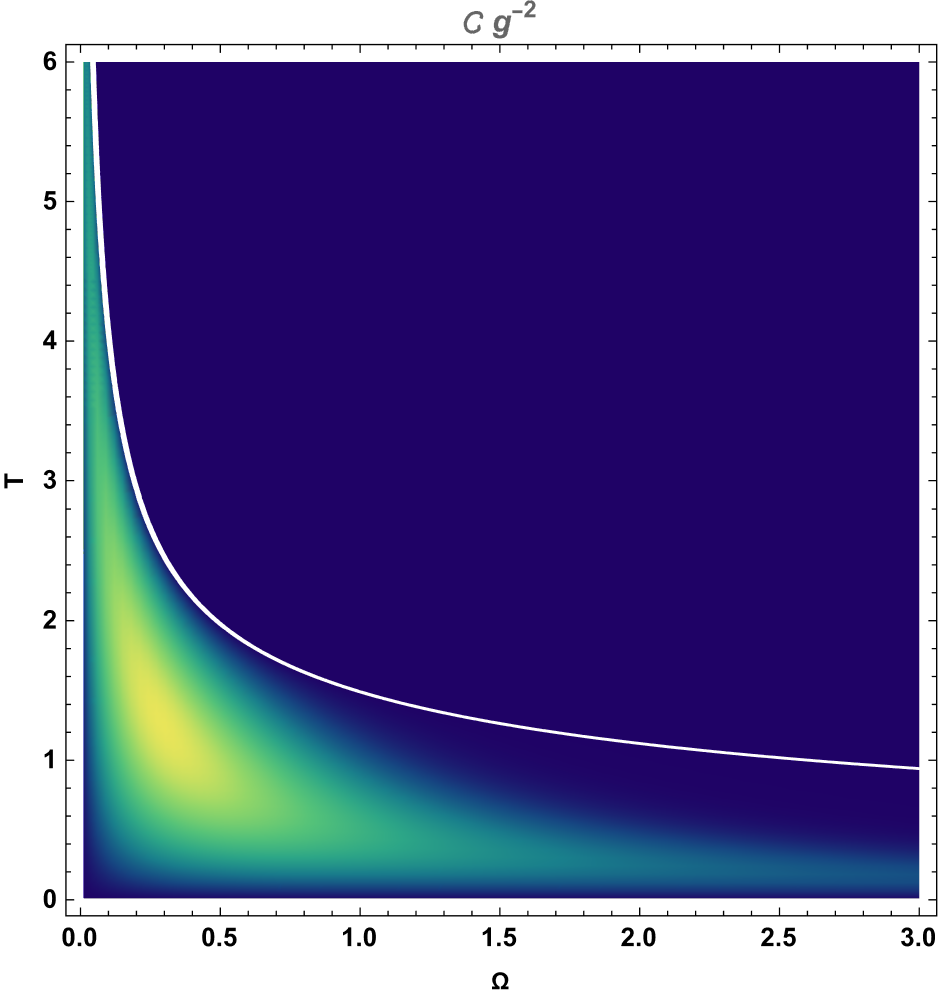}
		\caption{}
		\label{con_BD_mono_N_vs_wT}
	\end{subfigure}\hspace{.5cm}
	\begin{subfigure}{.45\textwidth}
		\centering
		\includegraphics[width=.9\linewidth]{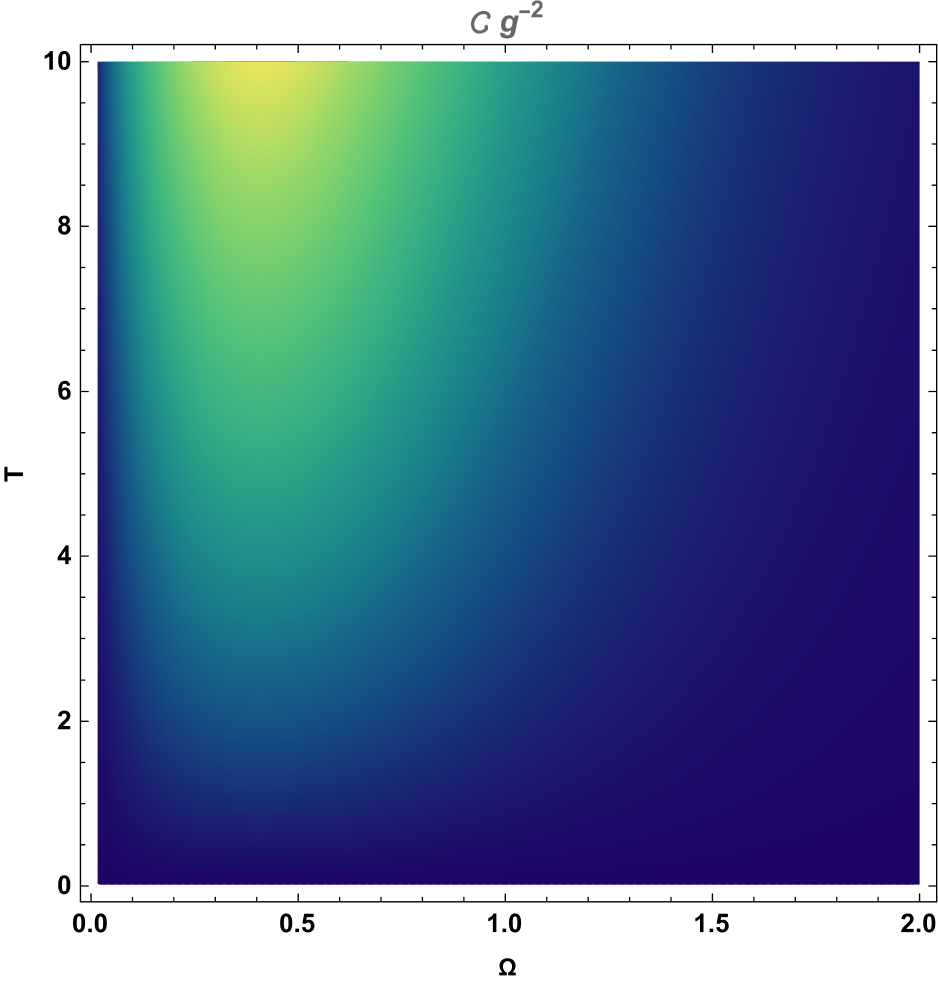}
		\caption{}
		\label{con_BD_mono_NS_vs_wT}
	\end{subfigure}
	\caption{The density plots of the concurrence $\mathcal{C}$ of two monopole-coupling identical UDW detectors in Euclidean vacuum as a function of  $\Omega$ and  $T$ with (a) zero separation and (b) antipodal separation. In (a), the ``sudden death" is indicated more precisely by a solid white curve.  Note that the silent and active regions of entanglement harvesting, with zero and high concurrence, are located quite differently in (a) and (b). }
		\label{con_BD_mono_vs_wT}
\end{figure}

 We think the reason for the sudden death shown in \cref{fig_con_BD_mono_N_vs_w}  and \cref{fig_con_BD_mono_N_vs_T}  is the same as mentioned in \cite{Henderson:2017yuv} due to the competing effects between entangling by probe-environment interactions and the unentangling due to thermal radiation/fluctuation. The first effect is characterized by $|X|$ and the latter by $P$, and the concurrence by ${\rm max}\big[ 0,2(|X|-P \big)]$. The cross-correlation $X$ will be suppressed by a large energy gap to result in the sudden death shown in \cref{fig_con_BD_mono_N_vs_w}, and the transition rate $P$ will increase as the detector receives more thermal quanta with increasing time, which results in the sudden death shown in \cref{fig_con_BD_mono_N_vs_T}.  In contrast, the concurrence of the antipodal separation shown in \cref{fig_con_BD_mono_NS_vs_T}  increases for large interaction time $T$. This could be due to no coherent long-ranged thermal fluctuation to enhance $P$ for antipodal separation.


To better comprehend the dependence of concurrence on various parameter spaces, we provide a density plot of concurrence as a function of $\Omega$ and $T$ in \cref{con_BD_mono_vs_wT}. The white curve in \cref{con_BD_mono_N_vs_wT} indicates the ``sudden death" of concurrence for zero spatial separation. On the other hand, this special feature is absent for antipodal separation in \cref{con_BD_mono_NS_vs_wT}. With this overview picture, we see that the silent and active regions of entanglement harvesting, with zero and high concurrence, are located quite differently for zero and antipodal separations.  The active region is located at low but nonzero $T$ and $\Omega$ part in \cref{con_BD_mono_N_vs_wT} but in the low $\Omega$ and high $T$ part in \cref{con_BD_mono_NS_vs_wT}. The last feature implies that the long-range entanglements in the de Sitter Euclidean vacuum grow over time, but the short-range ones decay.

\subsubsection{Dipole coupling in Euclidean vacuum}
\begin{figure}[h]
            \centering
		\includegraphics[width=.5\linewidth]{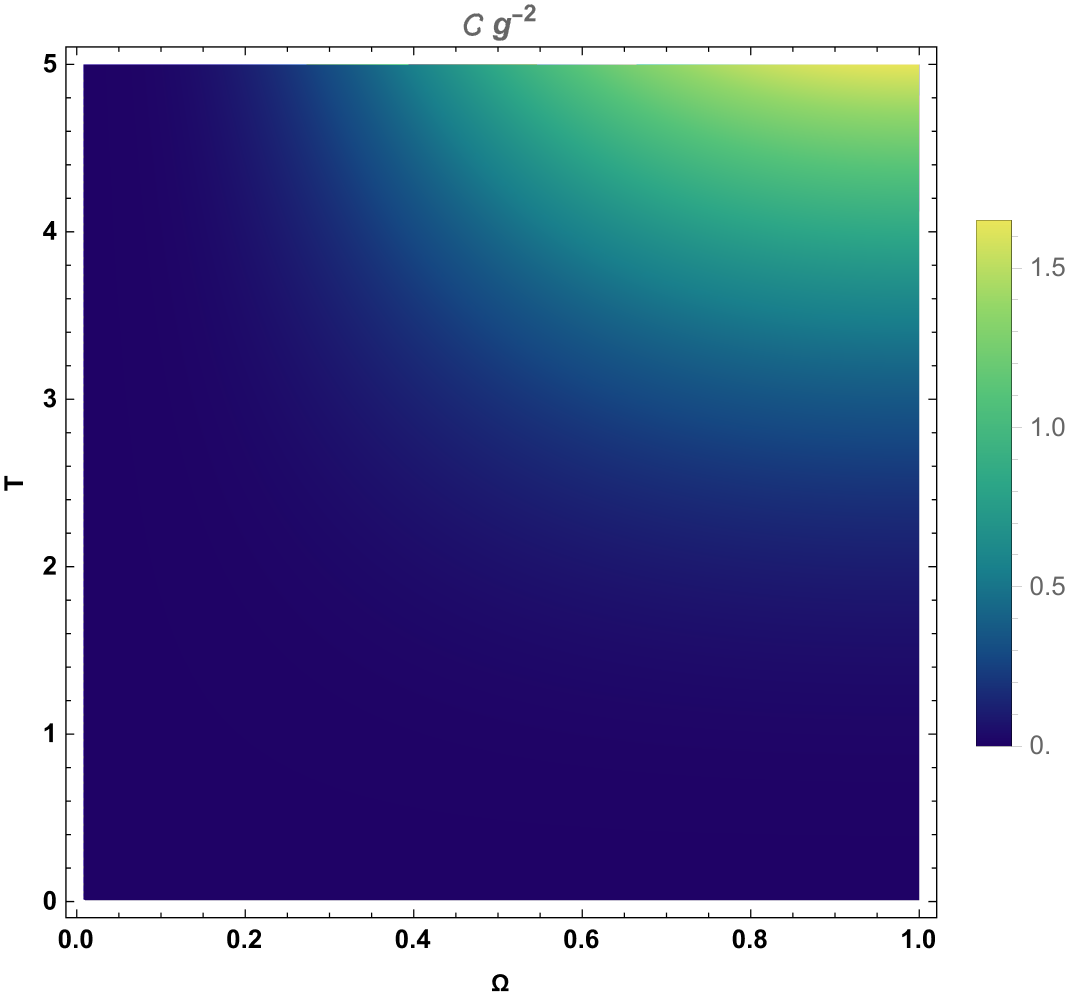}
	\caption{The density plots of the concurrence $\mathcal{C}$ of two dipole-coupling identical UDW detectors in Euclidean vacuum as a function of  $\Omega$ and $T$ with antipodal separation. In contrast, the concurrence is exactly zero for the zero separation, thus there is no need to show. Compared to \cref{con_BD_mono_NS_vs_wT} of monopole-coupling, the active region of entanglement harvesting to higher $T$ and $\Omega$ part with large value by a factor of $100$.}
 \label{con_BD_dipole_NS_vs_wT}
\end{figure}

We now examine entanglement harvesting for two identical dipole-coupling UDW detectors in Euclidean vacuum in de Sitter space. As noted in \eq{XmD}, $X^-_{\ell=2}=0$ for zero separation, this yields zero concurrence.  On the other hand, there is nonzero concurrence for the antipodal separation. This implies that the UDW detectors cannot explore the short-range but the long-range quantum entanglement through the dipole coupling. Here, we just present the density plot of the concurrence for the antipodal separation, as shown in \cref{con_BD_dipole_NS_vs_wT}. Compared to \cref{con_BD_mono_NS_vs_wT} of monopole-coupling, the active region of the concurrence shifts to higher $T$ and $\Omega$ part with larger values by a factor of $100$.


%

\FloatBarrier

\subsection{de Sitter $\alpha$-vacua}
We now consider the entanglement harvesting for the $\alpha$-vacua (with $\beta$=0) or the generic $(\alpha,\beta)$-vacua. The parameter $\beta$ is periodic with a period of $2\pi$, and the parameter $\alpha$ is a non-negative real number. It is important to mention that the purity $\Tr \rho_{AB}^2$ can be expressed as $1\pm \mathcal{O}(g^2)$, in the peturbation expansion of $g$. We notice that purity decreases with increasing $\alpha$ and for large $\alpha$ it becomes negative. The exact critical value of $\alpha$ where the purity becomes ill-defined depends on $\Omega$ and $T$. We suspect that this behavior may be attributed to the breakdown of the saddle point approximation. 
 Consequently, we will restrict our consideration for the $\alpha$-vacua with $\alpha \leq 1.5$ to ensure a well-defined density matrix $\rho_{AB}$.  Again, we start our analysis with the monopole coupling and, subsequently, the dipole coupling. 

\subsubsection{Monopole coupling in $\alpha$-vacua}

\begin{figure}[htb]
	\centering
	\begin{subfigure}{.45\textwidth}
		\centering
		\includegraphics[width=.9\linewidth]{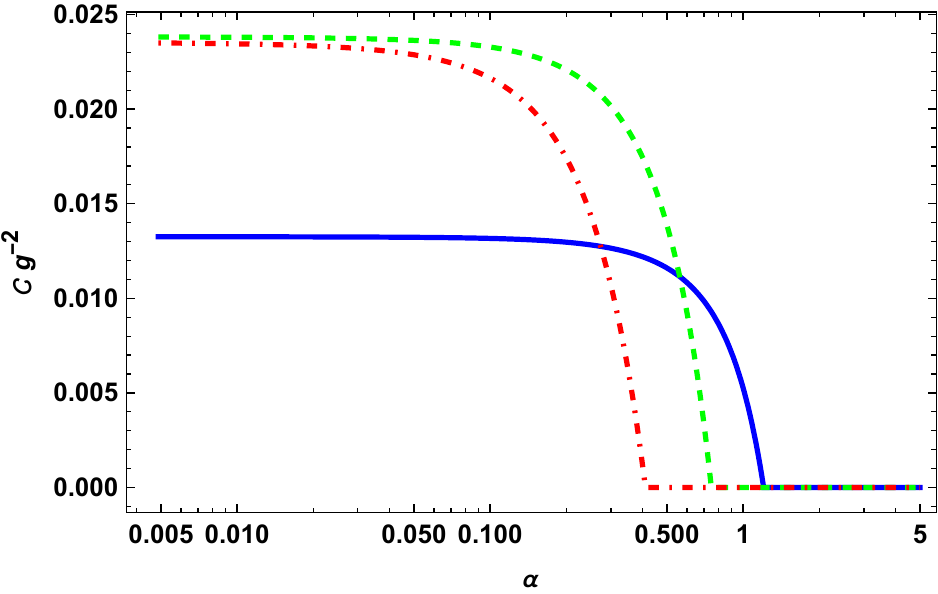}
		\caption{}
		\label{con_alpha_mono_N_vs_alpha_fixed_T}
	\end{subfigure}\hspace{.5cm}
	\begin{subfigure}{.45\textwidth}
		\centering
		\includegraphics[width=.9\linewidth]{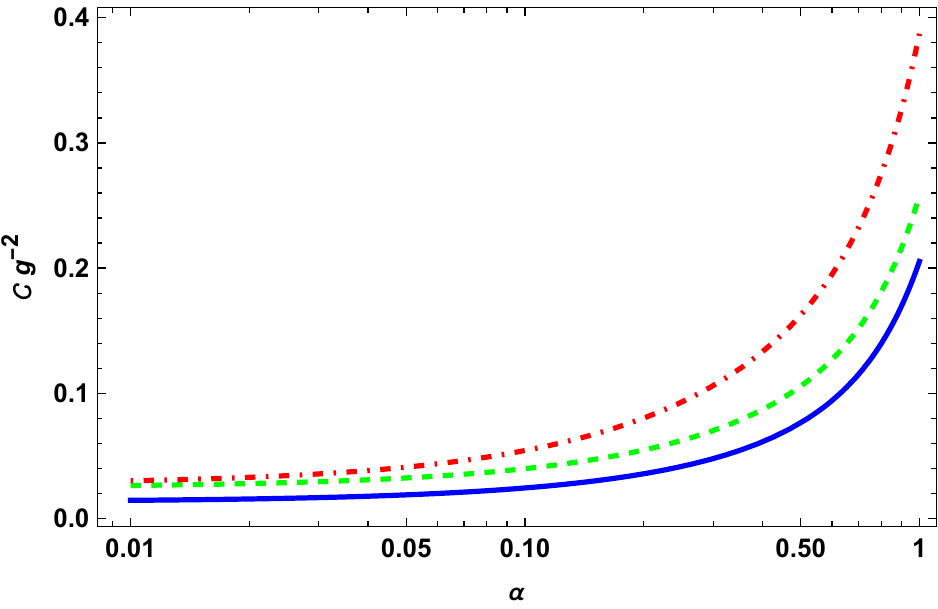}
		\caption{}
		\label{con_alpha_mono_NS_vs_alpha_fixed_T}
	\end{subfigure}
	\caption{The concurrence $\mathcal{C}$ of two identical monopole-coupling UDW detectors in the $\alpha$-vacua as a function of  $\alpha$ for $T=1.0$ and different values of $\Omega=.1$ (solid-blue), $\Omega=.25$ (green-dashed) and $\Omega=.5$ (red-dot-dashed) with (a) zero separation and (b) antipodal separation. The ``sudden death" occurs for (a) but not (b).}
		\label{con_alpha_mono_vs_alpha_fixed_T}
\end{figure}

We first consider the $\alpha$-vacua (i.e., $\beta$=0). For a small $\alpha$ value, we find the dependences of the concurrence on $\Omega$ and $T$ are similar to the ones in \cref{fig_con_BD_mono_vs_w} and \cref{fig_con_BD_mono_vs_T} of the Euclidean vacuum. However, this may not be the case for larger $\alpha$. In \cref{con_alpha_mono_vs_alpha_fixed_T}, we show the dependence of the concurrence on $\alpha$ for a fixed value of $T$ and a few different values of $\Omega$ for the zero and antipodal separation. Interestingly, we notice that the ``sudden death" of the entanglement occurs at some value of $\alpha$ for the zero separation scenario only. This can be understood by comparing \eq{P_D_fab} and \eq{XmD} as follows: an increase of $\alpha$ will enhance the transition rate $P$ to result in sudden death. On the other hand, in the case of antipodal separation, the concurrence grows with $\alpha$ monotonically. Similar $\alpha$ dependence can be obtained by fixing $\Omega$ and varying $T$. The above $\alpha$ dependence of concurrence implies that increasing $\alpha$ suppresses the short-range entanglement harvested by the UDW detectors but enhances the long-range entanglement for a given $\Omega$ and $T$.

\FloatBarrier

\FloatBarrier

\begin{figure}[htt]
	\centering
	\begin{subfigure}{.45\textwidth}
		\centering
		\includegraphics[width=.9\linewidth]{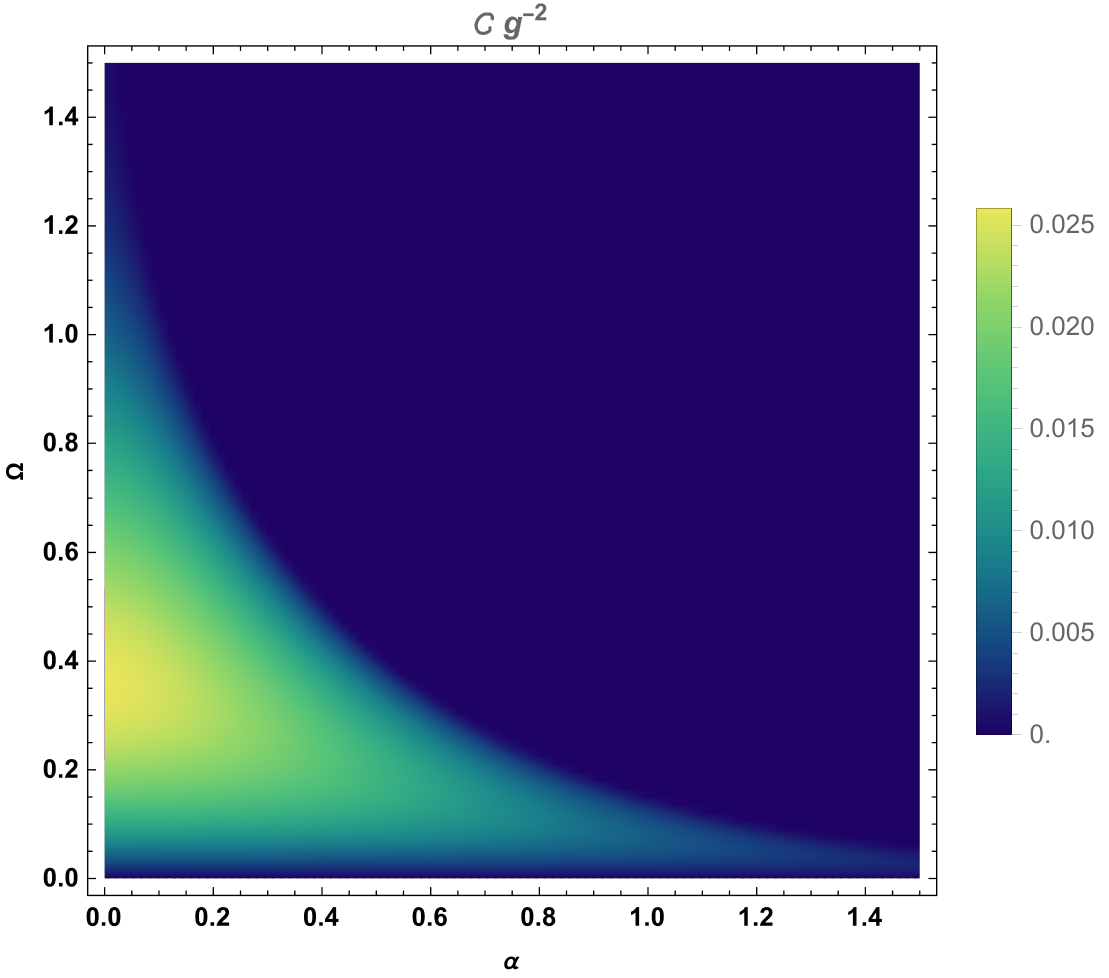}
		\caption{}
		\label{con_alpha_mono_N_vs_aw}
	\end{subfigure}\hspace{.5cm}
	\begin{subfigure}{.45\textwidth}
		\centering
		\includegraphics[width=.9\linewidth]{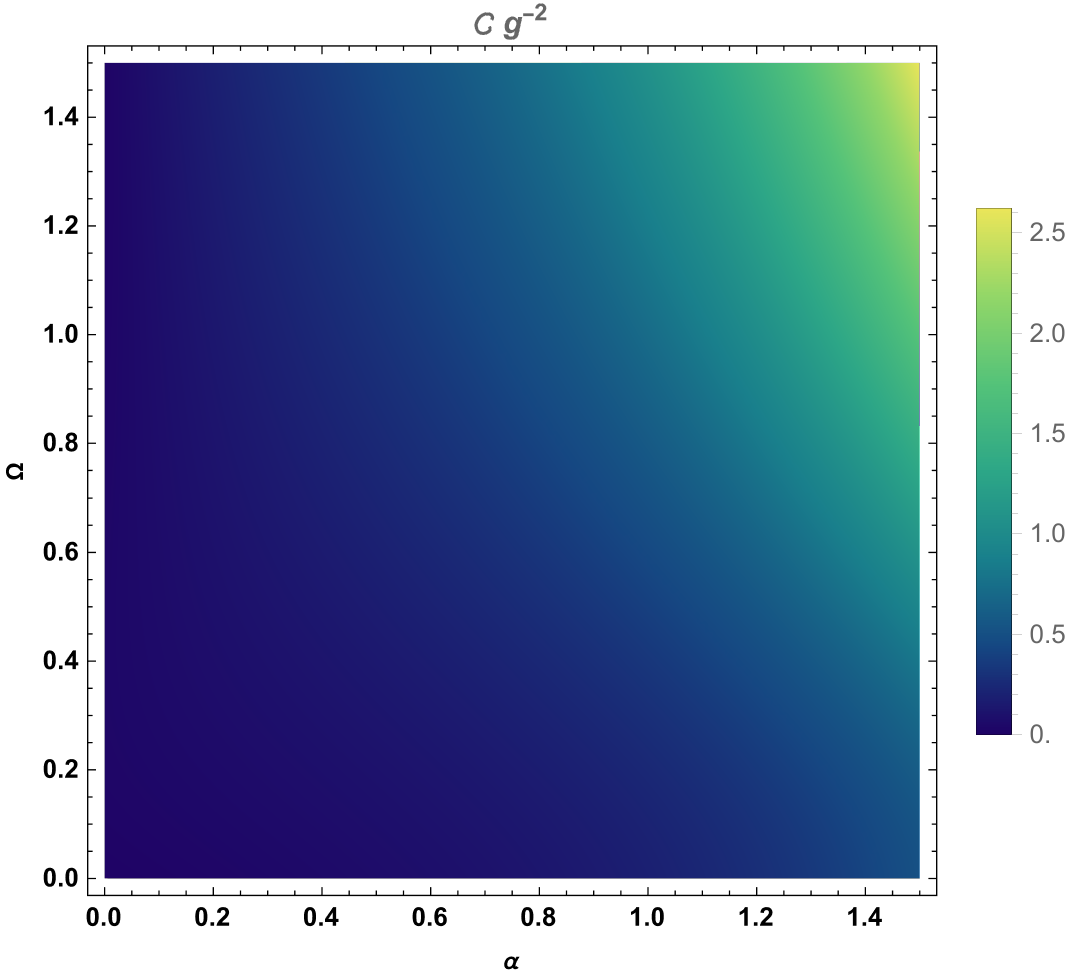}
		\caption{}
		\label{con_alpha_mono_NS_vs_aw}
	\end{subfigure}
	\caption{The density plots of the concurrence $\mathcal{C}$ of two monopole-coupling identical UDW detectors in $\alpha$-vacua with $T$=1 and $\beta=0$ as a function of  $\alpha$ and $\Omega$ with (a) zero separation and (b) antipodal separation. The plot implies that the $\alpha$-vacua prefer long-range entanglement harvesting.}
       \label{con_alpha_mono_vs_aW}
\end{figure}
\FloatBarrier

\begin{figure}[htt]
	\centering
	\begin{subfigure}{.45\textwidth}
		\centering
		\includegraphics[width=.9\linewidth]{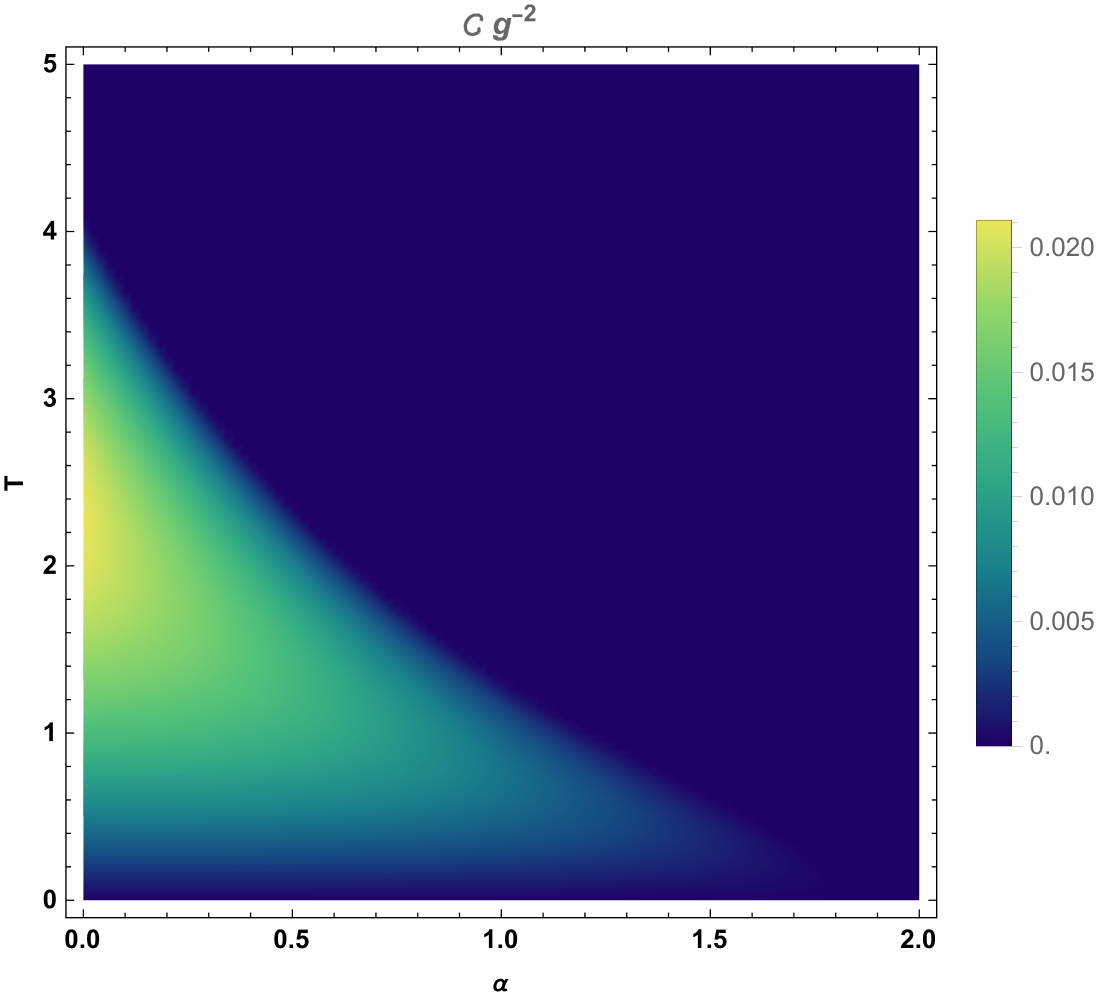}
		\caption{}
		\label{con_alpha_mono_N_vs_aT}
	\end{subfigure}\hspace{.5cm}
	\begin{subfigure}{.45\textwidth}
		\centering
		\includegraphics[width=.9\linewidth]{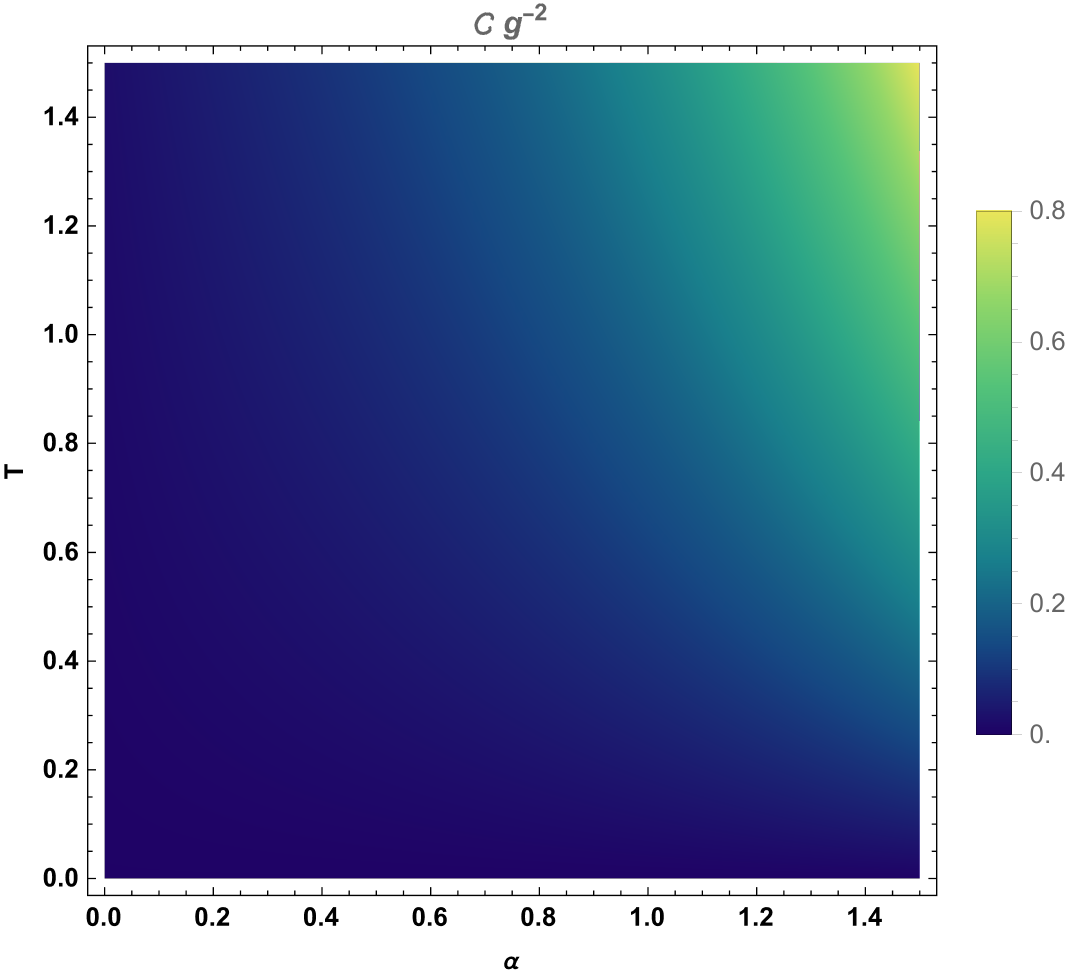}
		\caption{}
		\label{con_alpha_mono_NS_vs_aT}
	\end{subfigure}
	\caption{The density plots of the concurrence $\mathcal{C}$ of two monopole-coupling identical UDW detectors in $\alpha$-vacua with $\Omega=0.1$ and $\beta=0$ as a function of  $\alpha$ and $T$ with (a) zero separation and (b) antipodal separation. The plot again implies that the $\alpha$-vacua prefer long-range entanglement harvesting.}
	\label{con_alpha_mono_vs_aT}
\end{figure}

From the previous discussions, it is clear that concurrence exhibits interesting behavior as a function of the detector energy gap, measuring time scale and the parameter $\alpha$. To elucidate this behavior more comprehensively, we provide the following density plots. If we fix $\alpha$, the density plot for the concurrence as a function of $\Omega$ and $T$ shows an analogous pattern as illustrated in \cref{con_BD_mono_vs_wT} of Euclidean vacuum. Thus we do not include it for brevity. Instead, we present the density plots of the concurrence as a function of $\alpha$ and $\Omega$ with a fixed $T$ in \cref{con_alpha_mono_vs_aW}, and a function of $\alpha$ and $T$ with a fixed $\Omega$ in \cref{con_alpha_mono_vs_aT}. These density plots reconfirm the implication that the $\alpha$-vacua prefer long-range entanglement harvesting.

\begin{figure}[htbp]
	\centering
	\begin{subfigure}{.45\textwidth}
		\centering
		\includegraphics[width=.9\linewidth]{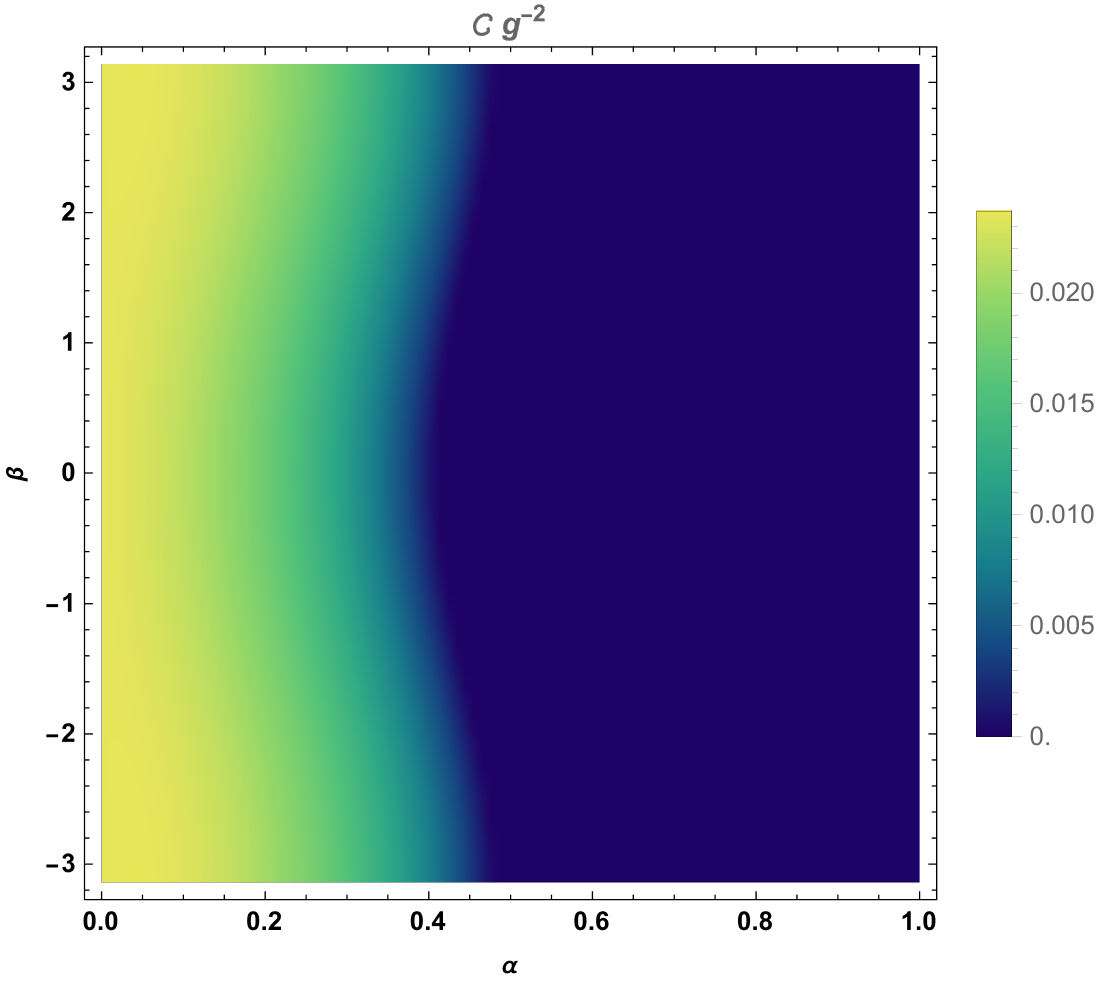}
		\caption{}
		\label{con_alpha_mono_N_vs_ab}
	\end{subfigure}\hspace{.5cm}
	\begin{subfigure}{.45\textwidth}
		\centering
		\includegraphics[width=.9\linewidth]{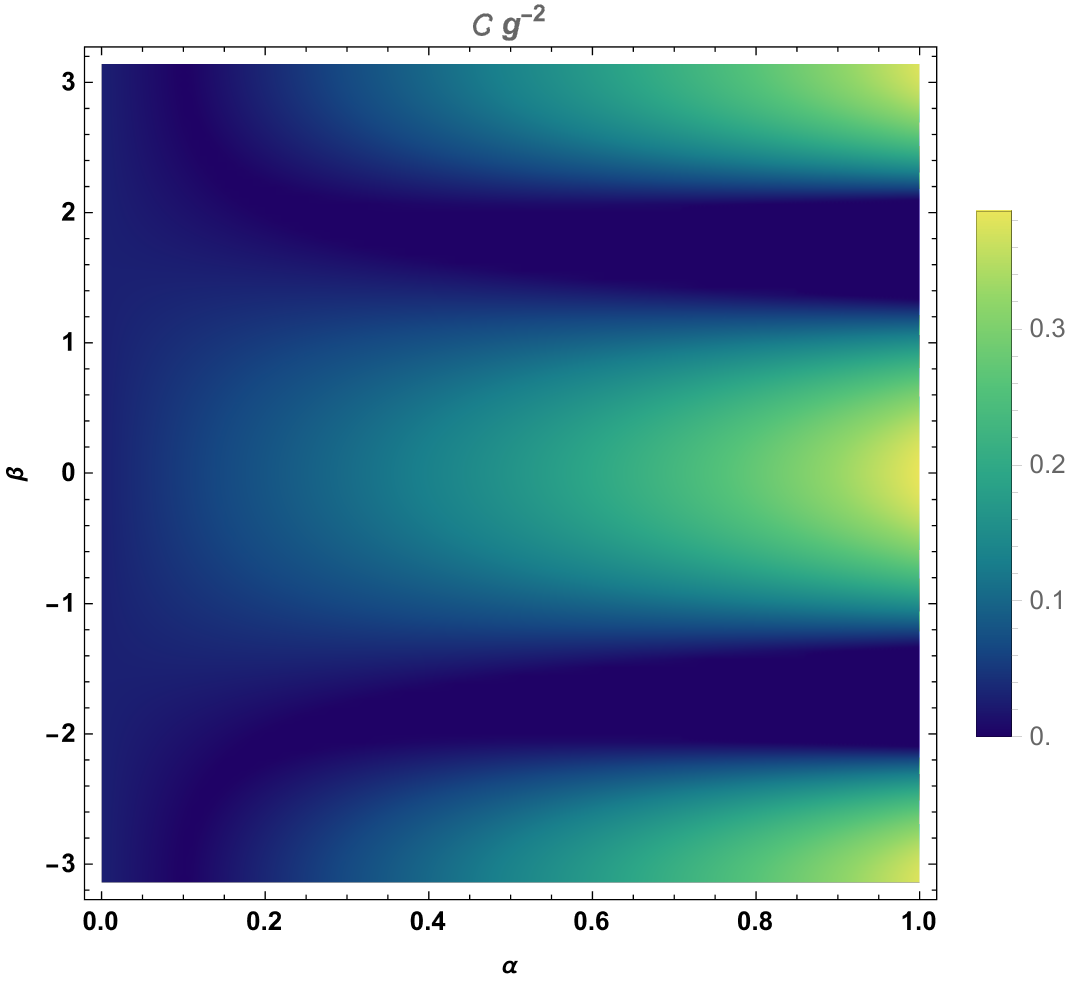}
		\caption{}
		\label{con_alpha_mono_NS_vs_ab}
	\end{subfigure}
	\caption{The density plots of the concurrence $\mathcal{C}$ of two monopole-coupling identical UDW detectors in $(\alpha, \beta)$-vacua with $\Omega=0.5$ and $T=1.0$ as a function of  $\alpha$ and $\beta$ with (a) zero separation and (b) antipodal separation. In (b), the novel phenomena of ``sudden death and revival" of the entanglement appear when tuning $\beta$ for the antipodal separation. }
		\label{con_alpha_mono_vs_ab}
\end{figure}
\FloatBarrier
Finally, we show concurrence for the generic $(\alpha,\beta)$-vacua in \cref{con_alpha_mono_N_vs_ab} and \cref{con_alpha_mono_NS_vs_ab} for zero and antipodal separations, respectively. Interestingly, we observe the novel phenomena of ``sudden death and revival" of the entanglement when tuning $\beta$ for the antipodal separation. Otherwise, the $\alpha$-dependence of concurrence that we notice from these density plots confirms the implication previously discussed.

\begin{figure}[h]
	\centering
         \begin{subfigure}{.45\textwidth}
		\centering
		\includegraphics[width=.8\linewidth]{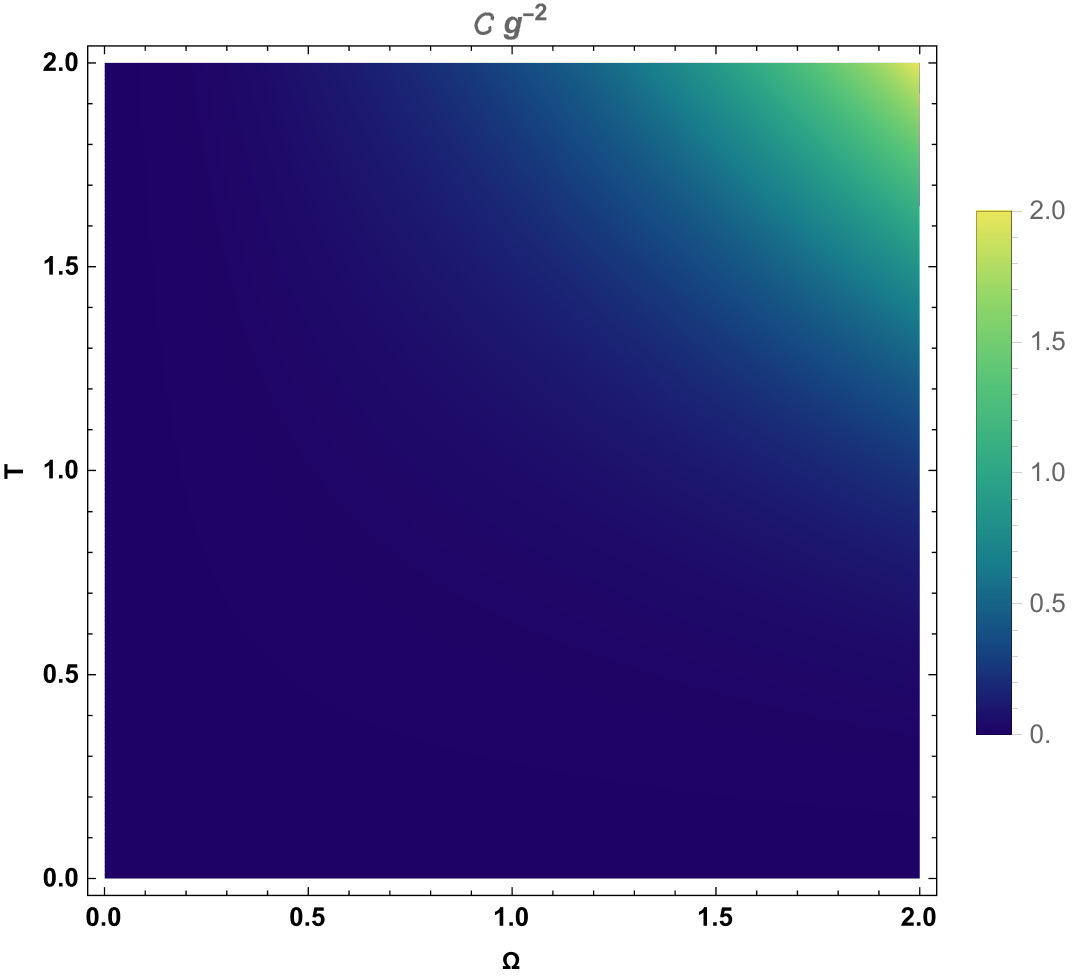}
		\caption{}
		\label{con_alpha_di_NS_vs_wT}
	\end{subfigure}\hspace{.5cm}
        \begin{subfigure}{.45\textwidth}
		\centering
		\includegraphics[width=.8\linewidth]{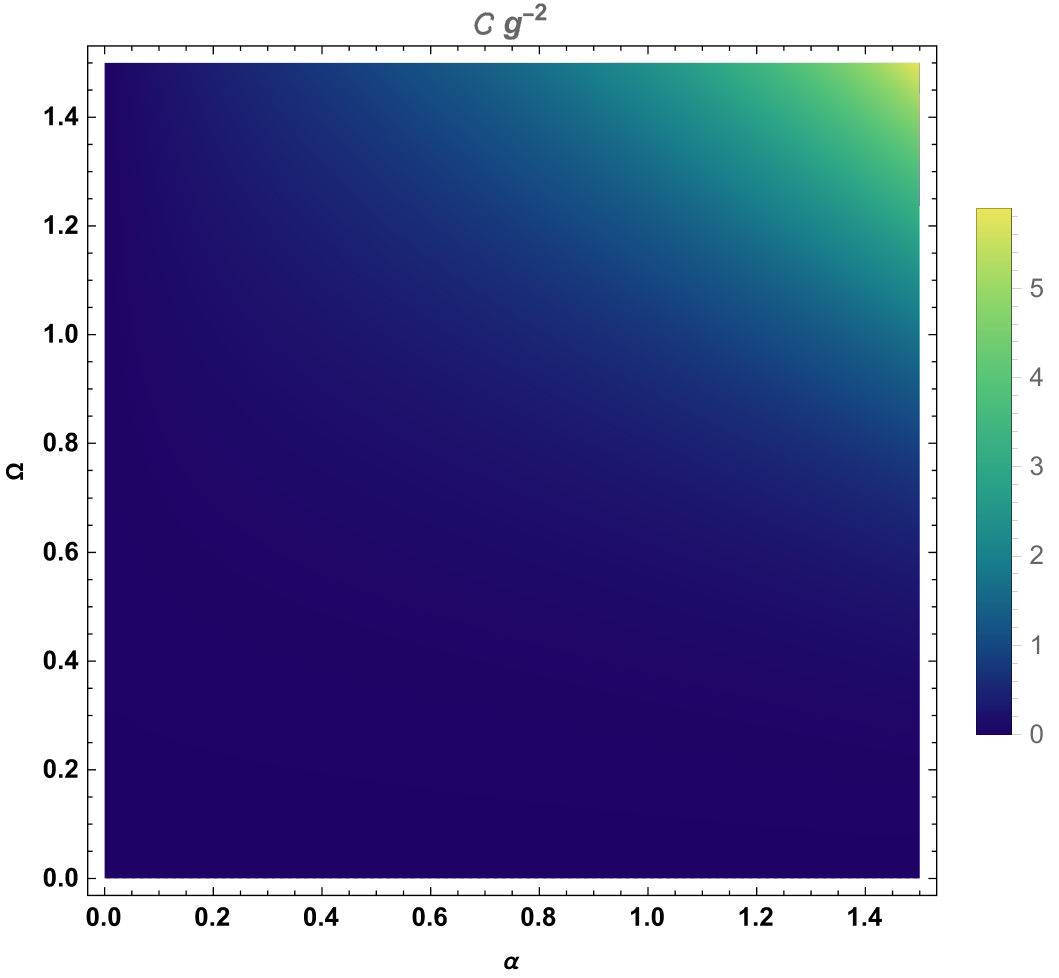}
		\caption{}
		\label{con_alpha_di_NS_vs_aw}
	\end{subfigure}\hspace{.5cm}
      \begin{subfigure}{.45\textwidth}
		\centering
		\includegraphics[width=.8\linewidth]{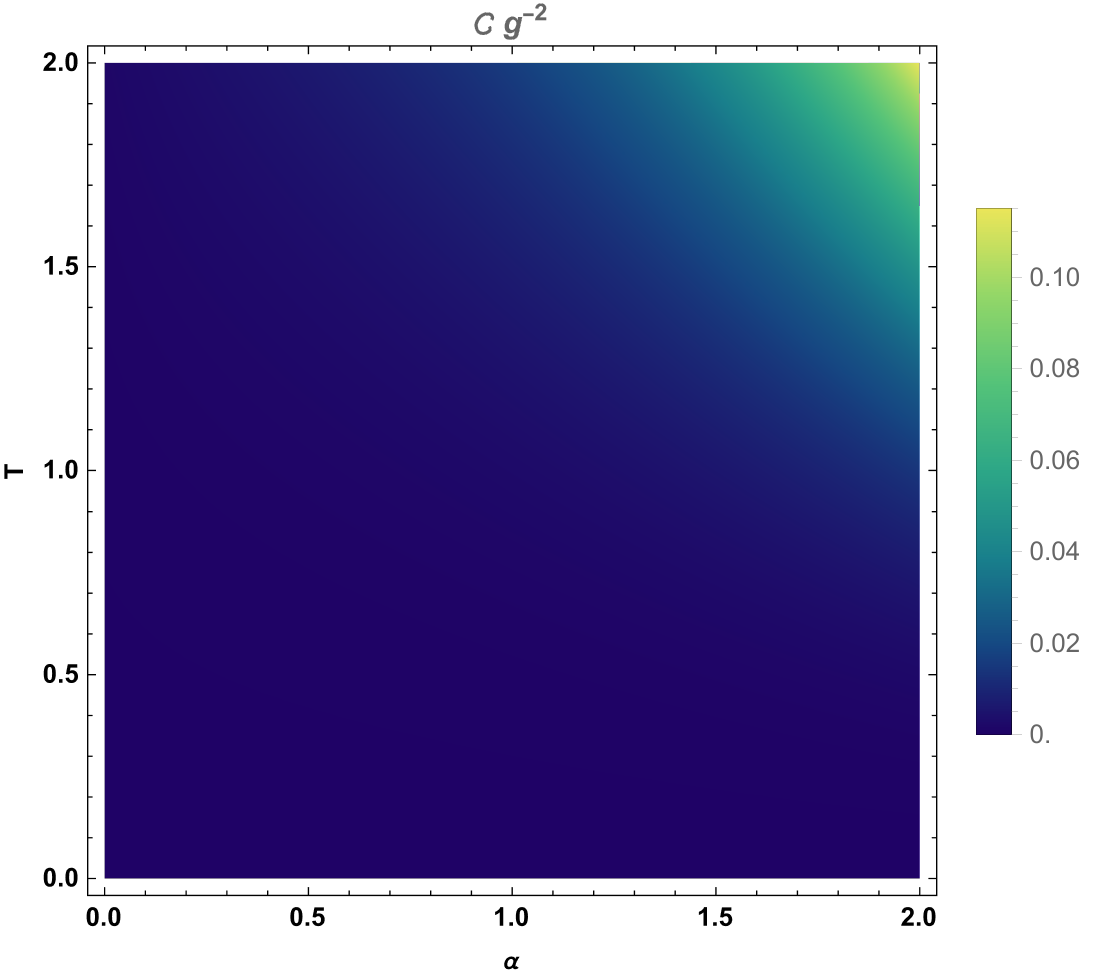}
		\caption{}
		\label{con_alpha_di_NS_vs_aT}
	\end{subfigure}\hspace{.5cm}
      \begin{subfigure}{.45\textwidth}
		\centering
		\includegraphics[width=.8\linewidth]{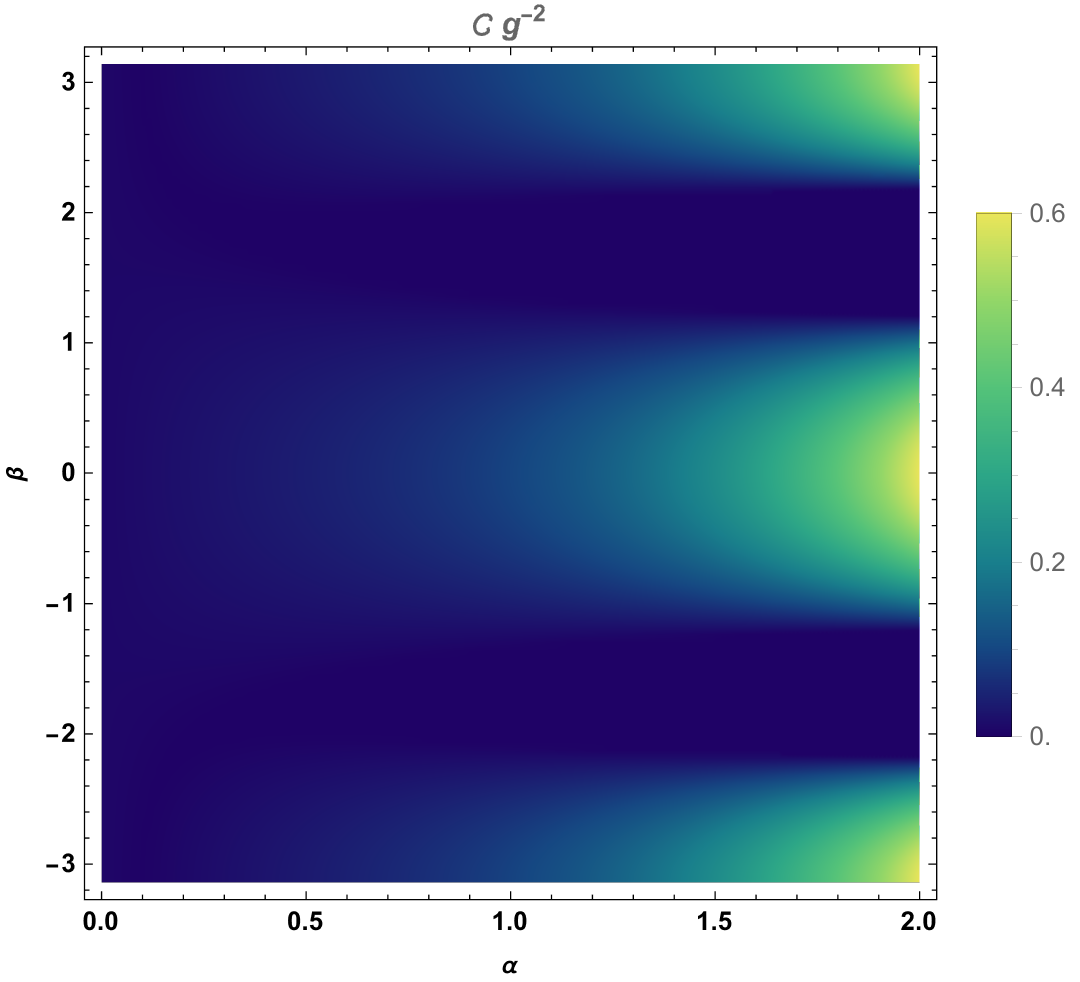}
		\caption{}
		\label{con_alpha_di_NS_vs_ab}
	\end{subfigure}\hspace{.5cm}
	\caption{The density plots of the concurrence $\mathcal{C}$ of two dipole-coupling identical UDW detectors with antipodal separation in $\alpha$-vacua as a function of  (a) $\Omega$ and $T$ for $\alpha=0.05$ and $\beta=0$,  (b) $\alpha$ and $\Omega$ for $T=1.0$  and $\beta=0$, (c) $\alpha$ and $T$ for $\Omega=0.1$  and $\beta=0$, and in $(\alpha,\beta)$-vacua as a function of (d) $\alpha$ and $\beta$ for $T=1.0$ and $\Omega=0.5$. They show similar features  to their monopole-coupling counterparts, as shown respectively in 
 \cref{con_BD_mono_NS_vs_wT},  \cref{con_alpha_mono_NS_vs_aw}, \cref{con_alpha_mono_NS_vs_aT}, and \cref{con_alpha_mono_NS_vs_ab} but with different overall magnitudes.}
 \label{fig_con_BD_mono_vs_wT}
\end{figure}
\FloatBarrier
\subsubsection{Dipole coupling in $\alpha$-vacua}
We now proceed to discuss entanglement harvesting by two dipole-coupling distinct detectors in the $\alpha$-vacua.
As noted in \eq{XmD}, $X^-_D=0$ for the dipole coupling for two identical detectors, resulting in vanishing concurrence for the case of zero separation. Therefore, we only need to consider the antipodal separation scenario. We first depict a density plot in \cref{con_alpha_di_NS_vs_wT} to present an overview of $T$ and $\Omega$ dependence of the concurrence for an $\alpha$-vacuum with a typical value of $\alpha=0.05$. It can be compared with the antipodal separation results of Euclidean vacuum for either the monopole-coupling one in \cref{con_BD_mono_NS_vs_wT} or the dipole-coupling one in \cref{con_BD_dipole_NS_vs_wT}. Moreover, in \cref{con_alpha_di_NS_vs_aw}, \cref{con_alpha_di_NS_vs_aT}, and \cref{con_alpha_di_NS_vs_ab}, we provide the density plot of concurrence as a function of $\alpha$ and $\Omega$, $\alpha$ and $T$, and $\alpha$ and $\beta$ respectively. We then compare the plots with the monopole-coupling counterparts illustrated in \cref{con_alpha_mono_NS_vs_aw}, \cref{con_alpha_mono_NS_vs_aT}, and \cref{con_alpha_mono_NS_vs_ab}. They exhibit similar features, including the growth of concurrence with $\alpha$ and the phenomena of ``sudden death and revival" upon tuning $\beta$, but with different overall magnitudes compared to their monopole-coupling counterparts.
%

\section{Quantum discord of de Sitter vacua}\label{sec5}

To characterize the non-classical quantum correlation produced by the gravitating scalar vacuum states, in this section, we will apply the analytical results of the reduced density matrix of the UDW detectors provided in section \ref{sec3} to obtain the quantum discord $D$ of \eq{def_discord}. We subsequently present the results in numerical plots to demonstrate the dependence of gravitating quantum correlation on the measuring time scale and energy gaps of the detectors. The logic of the presentation is similar to that adopted for entanglement harvesting in section \ref{sec4}.

As noted in \eq{def_discord}, the quantum discord $D=D(A,B)$ depends only on $P_{D=A,B}$ and $C$ but not on $X$. For the identical UDW detectors, it can be further reduced to \eq{QD_id}. The latter leads to $D=2 g^2 P_D$ for the zero separation. It also leads to $D=0$ for the dipole-coupling identical UDW detectors of antipodal separation since $C^+_{\ell=2}=0$. Due to these two special trivial cases for identical UDW detectors, we will consider the quantum discord for the non-identical UDW detectors that can be characterized by $\Omega_B$ and the parameter of energy gap difference
\be
\delta :=\Omega_A - \Omega_B\;.
\ee
Given that the quantum discord $D(A,B)$ \eq{def_discord} is symmetric when exchanging $A$ and $B$, we simply assume $\delta\ge 0$ without loss of generality. Moreover, $D(A,B)$ is defined only for real $P_D$ \cite{yurischev2015quantum}, which is only true if $\beta=0$ or $\delta=0$. Since we will mostly investigate scenarios with nonzero $\delta$, our analysis will focus on non-identical UDW detectors in $\alpha$-vacua only, i.e., $\beta=0$, when considering the dependence of $D$ on $\delta$, $\Omega_B$, $T$ and $\alpha$ for both monopole- and dipole-coupling with zero and antipodal separations. 

\subsection{de Sitter Euclidean vacuum}
In this subsection, we study quantum discord between two distinct UDW detectors in the Euclidean vacuum. We first consider the monopole-coupling and, subsequently, the dipole-coupling.  

\subsubsection{Monopole coupling in Euclidean vacuum}
\begin{figure}[ht]
	\centering
	\begin{subfigure}{.45\textwidth}
		\centering
		\includegraphics[width=.9\linewidth]{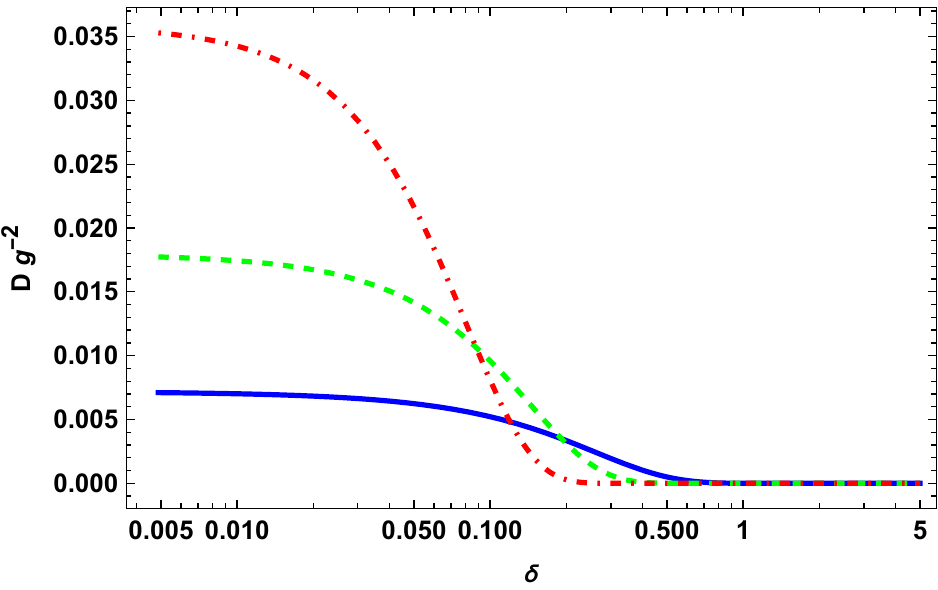}
		\caption{  }
		\label{QD_BD_mono_N_vs_r}
	\end{subfigure}\hspace{.5cm}
	\begin{subfigure}{.45\textwidth}
		\centering
		\includegraphics[width=.9\linewidth]{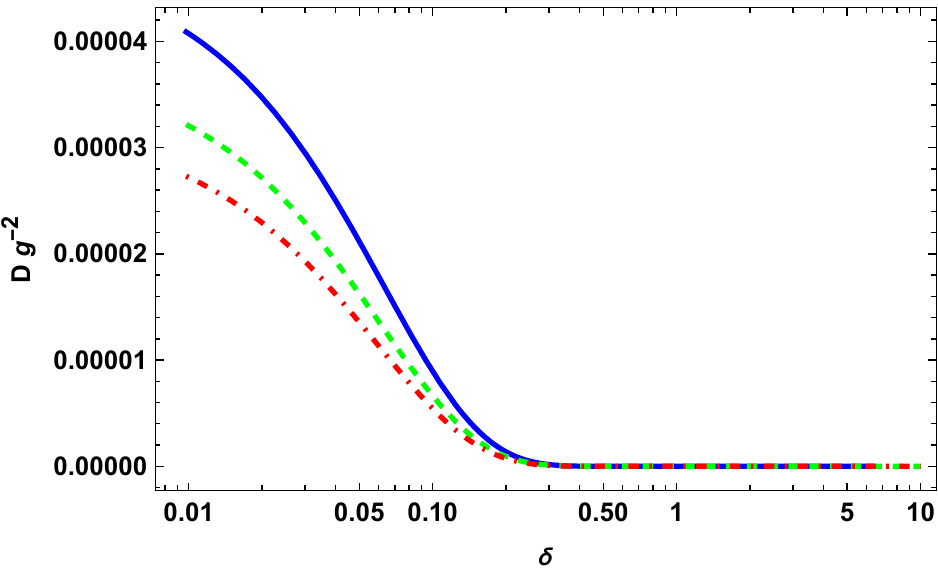}
		\caption{}
		\label{QD_BD_mono_NS_vs_r}
	\end{subfigure}
	\caption{Quantum discord $D$ of two monopole-coupling UDW detectors in Euclidean vacuum as a function of the energy gap difference $\delta$ for $\Omega_B=0.5$ and $T=1$ (solid-blue), $T=2.5$ (green-dashed) and $T=5$ (red-dot-dashed) with (a) zero separation and (b) antipodal separation. This implies that quantum correlations are suppressed for incompatible detectors.}
	\label{QD_BD_mono_vs_r}
\end{figure}

In \cref{QD_BD_mono_N_vs_r} and \cref{QD_BD_mono_NS_vs_r}, we show the behavior of quantum discord $D$ as a function of the energy gap difference $\delta$ for a given set of $(\Omega_B, T)$ for zero and antipodal separations respectively. It shows that $D$ decays to zero when $\delta$ becomes an order of $\Omega_B$ for both cases,  although their detailed decay patterns differ.  This implies the quantum correlations are heavily suppressed by the incompatibility of spectral gaps between the UDW detectors.

\begin{figure}[ht]
	\centering
	\begin{subfigure}{.45\textwidth}
		\centering
		\includegraphics[width=.9\linewidth]{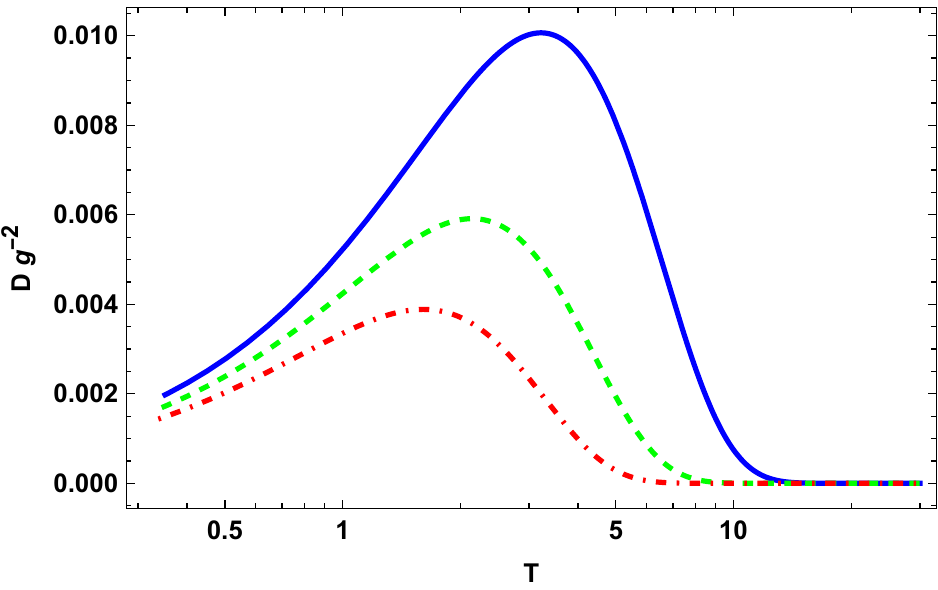}
		\caption{ }
		\label{QD_BD_mono_N_vs_T}
	\end{subfigure}\hspace{.5cm}
	\begin{subfigure}{.45\textwidth}
		\centering
		\includegraphics[width=.9\linewidth]{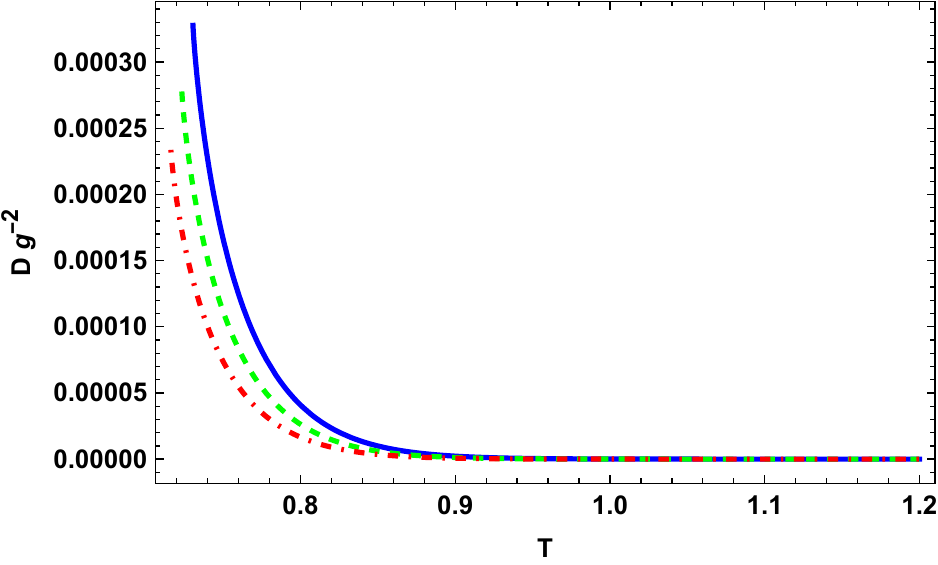}
		\caption{}
		\label{QD_BD_mono_NS_vs_T}
	\end{subfigure}
	\caption{Quantum discord $D$ of two of two monopole-coupling UDW detectors in Euclidean vacuum as a function of $T$ for $\Omega_B=0.5$ and $\delta=.1$ (solid-blue), $\delta=.15$ (green-dashed) and $\delta=.2$ (red-dot-dashed) with (a) zero separation and (b) antipodal separation. The results imply the difficulty in maintaining the coherence of long-time quantum correlations and that the short-range quantum correlations are more vibrant than the long-range ones, as expected. } 
		\label{QD_BD_mono_vs_T}
\end{figure}

To further explore the interplay between the $T$ and separation dependences of $D$, we present the plots  \cref{QD_BD_mono_N_vs_T} and \cref{QD_BD_mono_NS_vs_T}, for zero and antipodal separations respectively. As it can be observed from these plots \cref{QD_BD_mono_N_vs_T} and \cref{QD_BD_mono_NS_vs_T}, unlike the entanglement harvesting there is no ``sudden death" behavior for quantum discord. For both zero and antipodal separations, the corresponding discords decay to zero for sufficiently large $T$.
 However, in  \cref{QD_BD_mono_N_vs_T}, $D$ reaches a maximum before decaying to zero, whereas in \cref{QD_BD_mono_NS_vs_T}, it decreases monotonically to zero.  This implies that the coherence of long-time correlations is hard to maintain. In terms of length scale, the above results imply that there is no quantum correlation beyond an order of a few Hubble scales, i.e., no super-horizon quantum correlation, as clearly shown in \cref{QD_BD_mono_NS_vs_T} for antipodal separation.  This corresponds to the decoherence of superhorizon quantum fluctuation in the inflationary universe scenario. On the other hand, we see in \cref{con_BD_mono_NS_vs_wT} that the quantum entanglements of the superhorizon scales are not suppressed. This highlights an interesting contrast between quantum entanglement and quantum correlation at superhorizon scales.
\FloatBarrier
\FloatBarrier

Moreover, the magnitude and the range of $T$ for nonzero $D$ are larger in the zero separation than in the antipodal separation. This agrees with the expectation that the short-range quantum correlations are more vibrant than the long-range ones.

\begin{figure}[htp]
	\centering
	\begin{subfigure}{.45\textwidth}
		\centering
		\includegraphics[width=.9\linewidth]{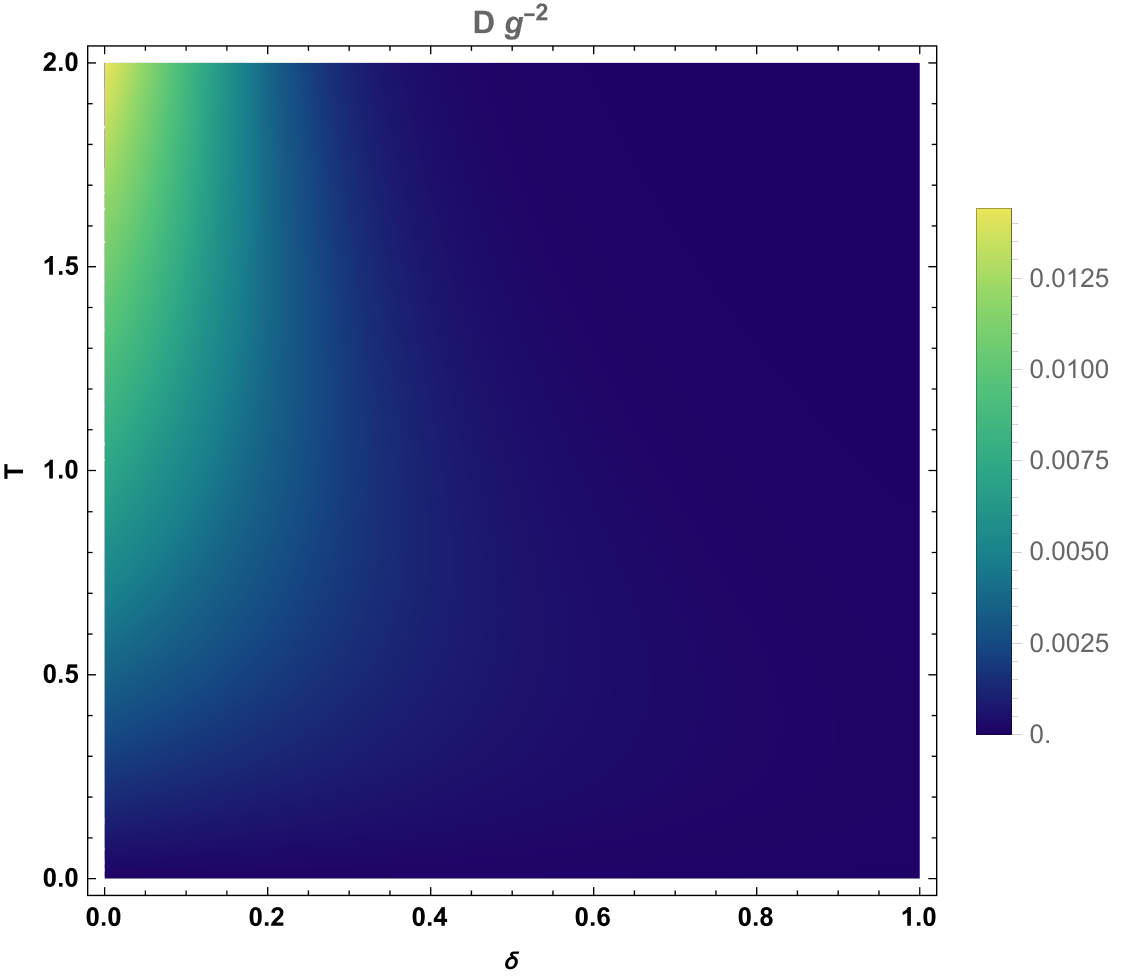}
		\caption{ }
		\label{QD_BD_mono_N_vs_rT}
	\end{subfigure}\hspace{.5cm}
	\begin{subfigure}{.45\textwidth}
		\centering
		\includegraphics[width=.9\linewidth]{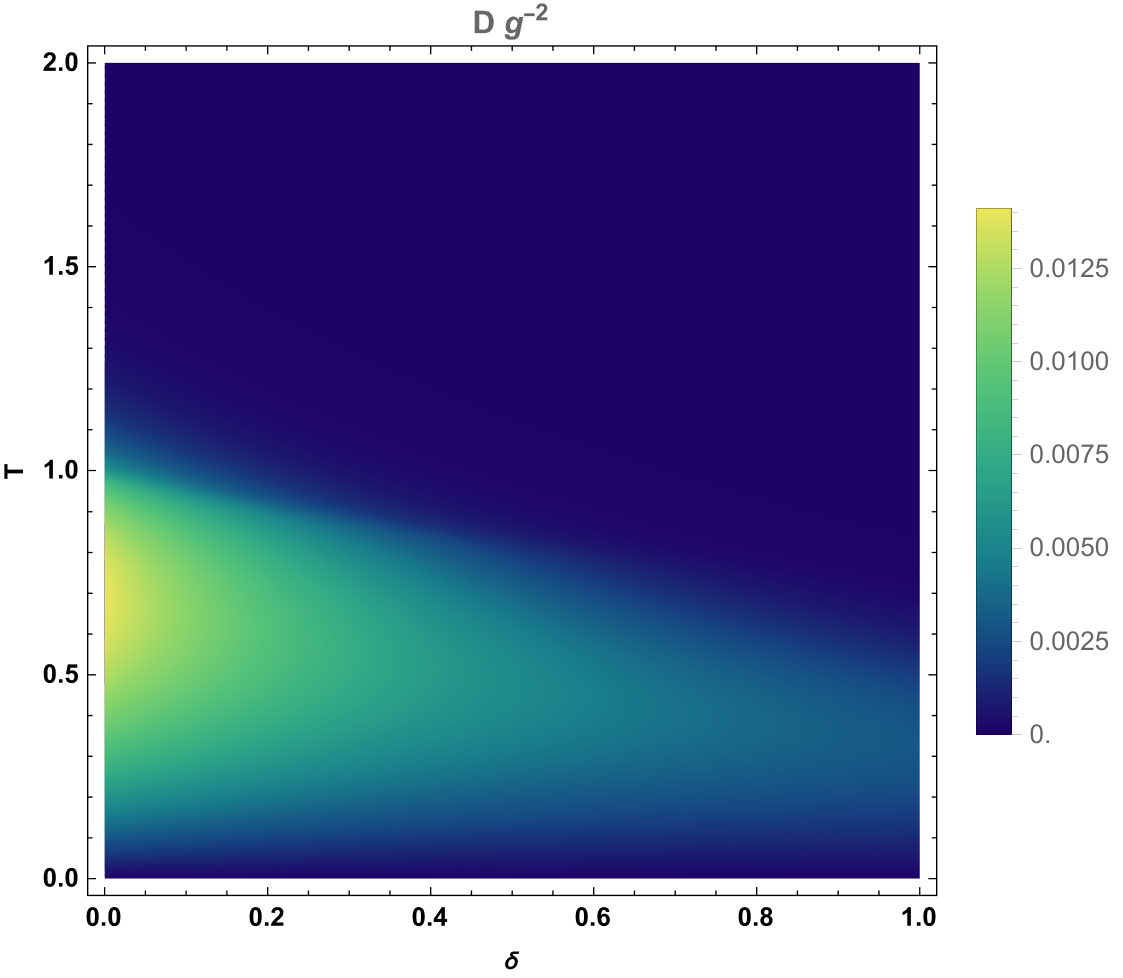}
		\caption{}
		\label{QD_BD_mono_NS_vs_rT}
	\end{subfigure}
	\caption{The density plots of quantum discord $D$ of two monopole-coupling UDW detectors in Euclidean vacuum as a function of $\delta$ and $T$ given $\Omega_B=0.5$ with (a) zero separation and (b) antipodal separation. In (b), the superhorizon-scale quantum correlations are suppressed.}
		\label{QD_BD_mono_vs_rT}
\end{figure}

Finally, to obtain an overview of the $\delta$ and $T$ dependence of $D$, we present the corresponding density plots in \cref{QD_BD_mono_vs_rT} for both zero and antipodal cases for a fixed value of $\Omega_B$. We notice that the quantum discord extends more along the $T$-direction in \cref{QD_BD_mono_N_vs_rT} for zero separation, but more along the $\delta$-direction in \cref{QD_BD_mono_NS_vs_rT} for antipodal separation. The former is more constrained by the decoherence due to the incompatibility of spectral gaps, and the latter is more by the decoherence of superhorizon quantum correlations. 

\FloatBarrier

\FloatBarrier


\FloatBarrier
\begin{figure}[htp]
	\centering
	\begin{subfigure}{.45\textwidth}
		\centering
		\includegraphics[width=.9\linewidth]{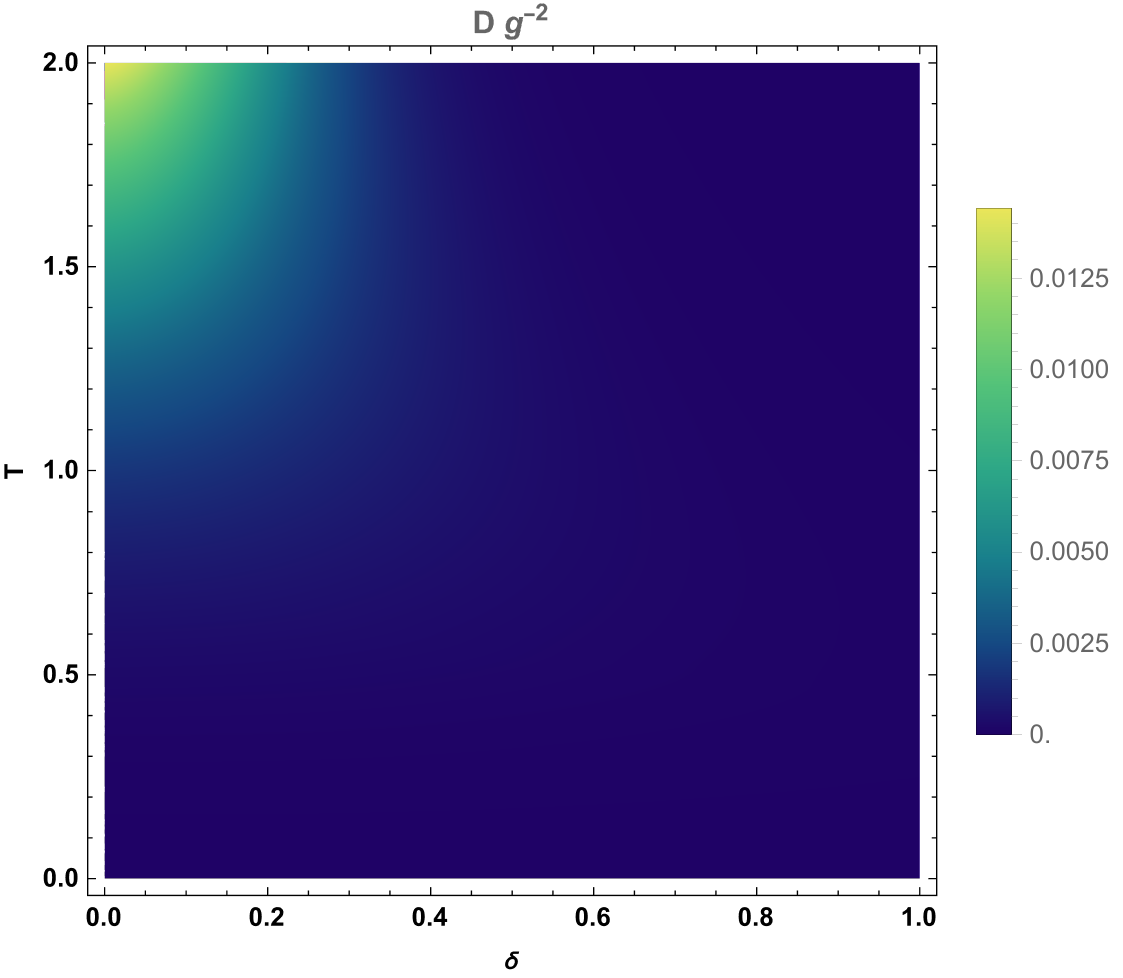}
		\caption{ }
		\label{QD_BD_di_N_vs_rT}
	\end{subfigure}\hspace{.5cm}
	\begin{subfigure}{.45\textwidth}
		\centering
		\includegraphics[width=.9\linewidth]{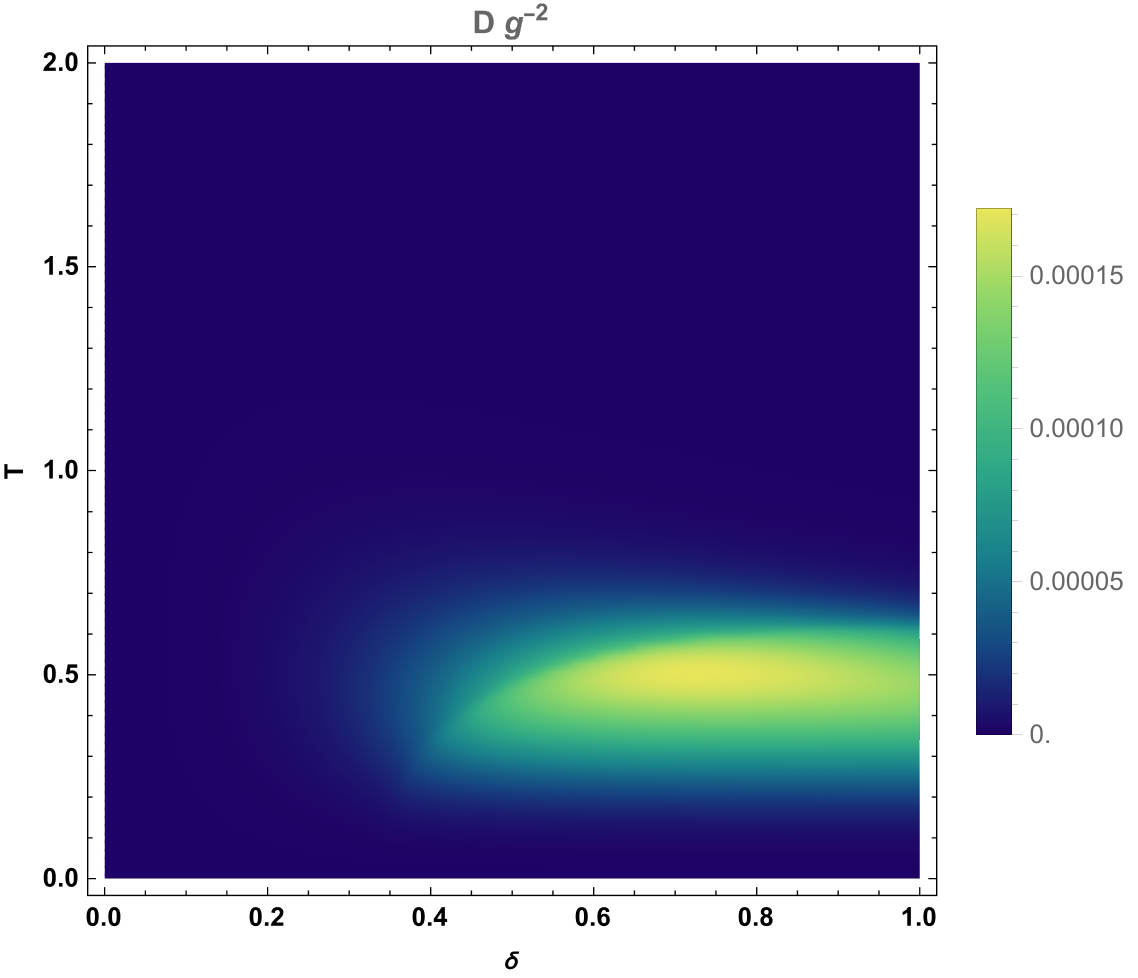}
		\caption{}
		\label{QD_BD_di_NS_vs_rT}
	\end{subfigure}
	\caption{The density plots of quantum discord $D$ of two dipole-coupling UDW detectors in Euclidean vacuum as a function of $r$ and $T$ given $\Omega_B=0.5$ with (a) zero separation and (b) antipodal separation. For (b), the novel feature is that the active region of $D$ is located at the part of large spectral incompatibility.  }
		\label{QD_BD_di_vs_rT}
\end{figure}
	
\FloatBarrier
\subsubsection{Dipole coupling in Euclidean vacuum}
To compare with the monopole-coupling counterparts illustrated in \cref{QD_BD_mono_vs_rT}, we present the density plots of quantum discord as a function of $\delta$ and $T$ for dipole-coupling detectors
in \cref{QD_BD_di_N_vs_rT} and \cref{QD_BD_di_NS_vs_rT}, for zero and antipodal separations respectively. By comparison, the active region of the zero separation shrinks, with the lower $T$ region becoming silent. On the other hand, the active region of the antipodal separation is now located at the large $\delta$ part instead of the smaller one, which is now silent. Despite the overall magnitude being down by two orders compared to the monopole-coupling counterpart, it is still novel to see that the spectral incompatibility of the UDW detectors will enhance the quantum correlation in the dipole-coupling cases. However, the superhorizon suppression remains.
%

\subsection{de Sitter $\alpha$-vacua}
We will now consider quantum discord for the non-identical UDW detectors coupled to the scalar field in the $\alpha$-vacua. As before, we first consider the monopole-coupling and then dipole-coupling. We have four model parameters $\Omega_B$, $\delta$, $T$ and $\alpha$. In what follows, we will present the density plots of quantum discord as a function of two model parameters, with the values of the other two being fixed. To ensure consistency and facilitate straightforward comparison, we will use the following fixed parameter values throughout our analysis: $\Omega_B=0.5$, $T=1.5$, $\delta=0.1$, and $\alpha=0.1$.

\subsubsection{Monopole coupling in $\alpha$-vacua}
\begin{figure}[htp]
	\centering
	\begin{subfigure}{.45\textwidth}
		\centering
		\includegraphics[width=.9\linewidth]{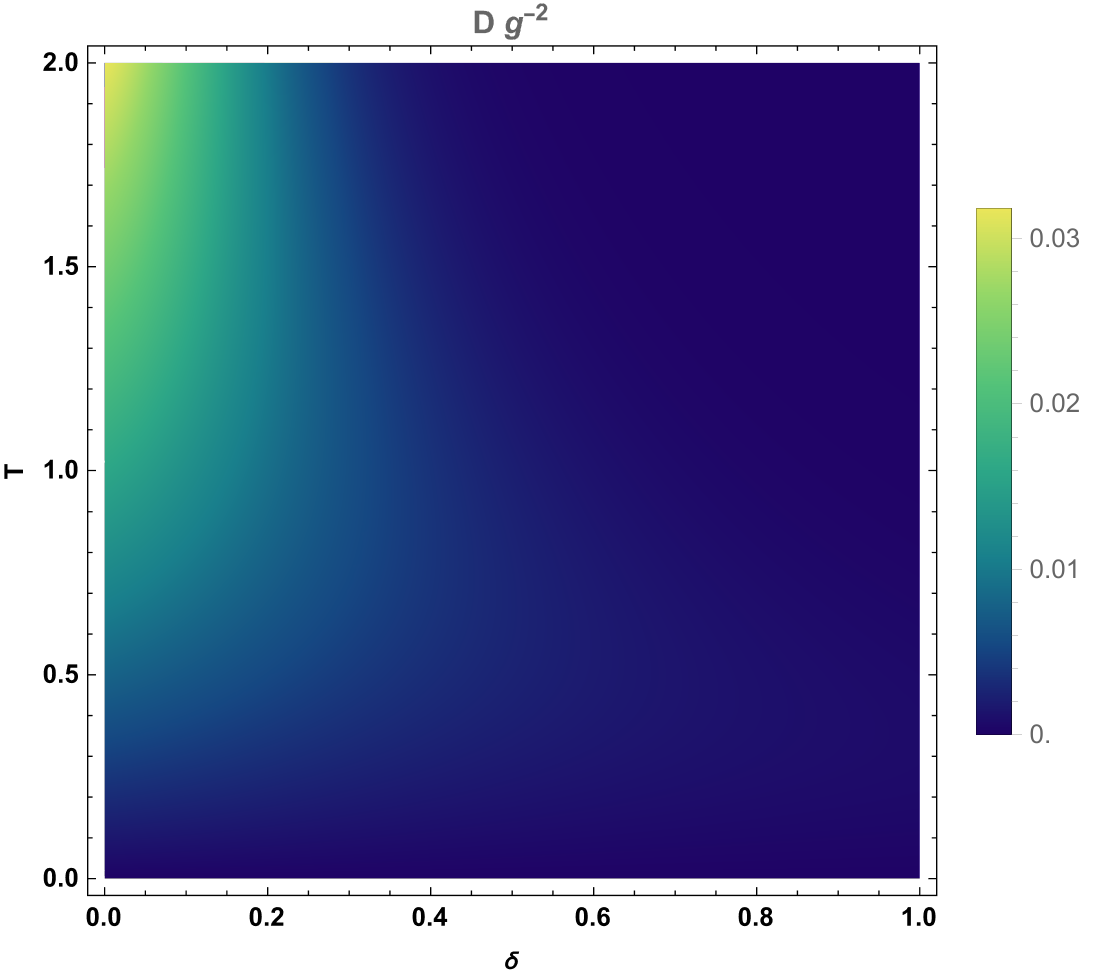}
		\caption{ }
		\label{QD_alpha_mono_N_vs_rT}
	\end{subfigure}\hspace{.5cm}
	\begin{subfigure}{.45\textwidth}
		\centering
		\includegraphics[width=.9\linewidth]{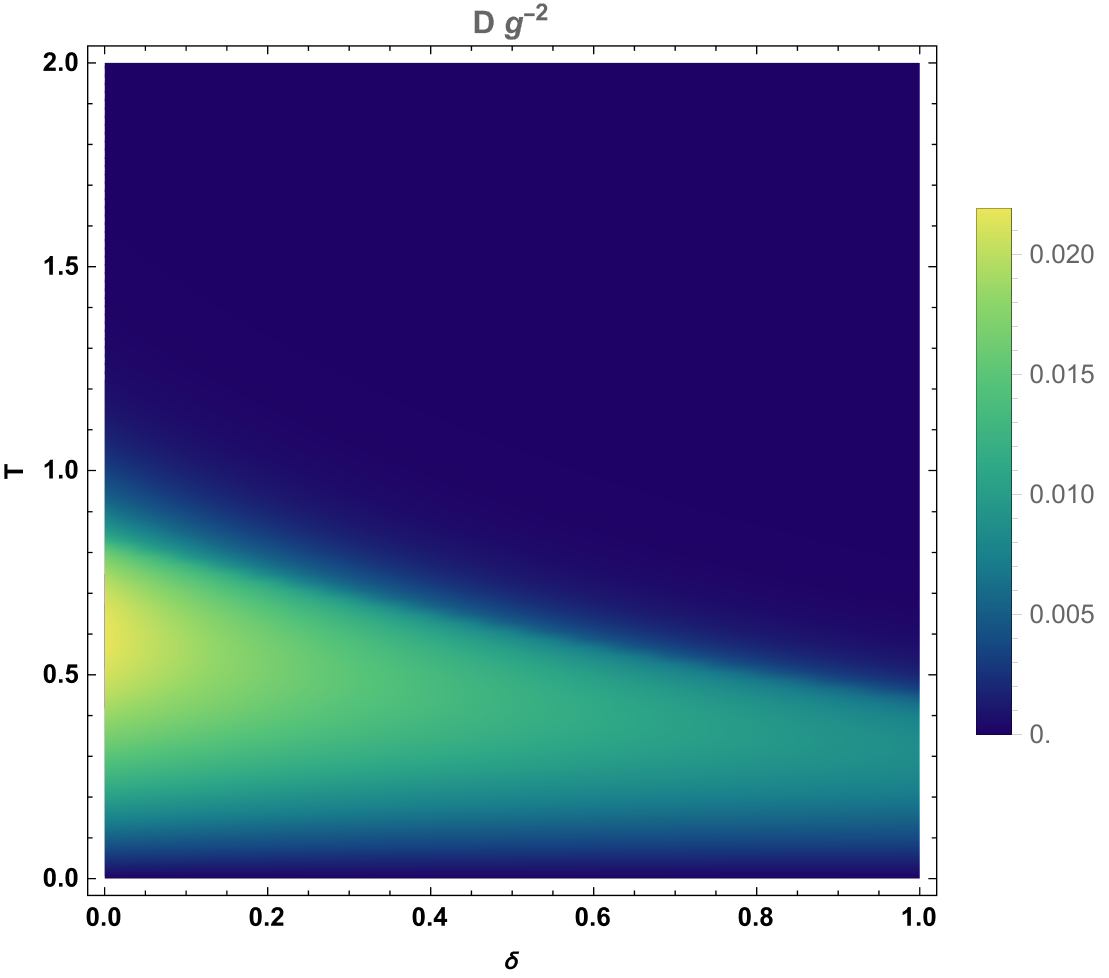}
		\caption{}
		\label{QD_alpha_mono_NS_vs_rT}
	\end{subfigure}
	\caption{The density plots of the quantum discord $D$ of two monopole-coupling identical UDW detectors in $\alpha$-vacua as a function of $\delta$ and $T$ given $\Omega_B=0.5$ and $\alpha=0.1$ with (a) zero separation and (b) antipodal separation.}
		\label{QD_alpha_mono_vs_rT}
\end{figure}

We present three sets of density plots for the monopole-coupling case, contrasting zero and antipodal separations. The first one is depicted in \cref{QD_alpha_mono_N_vs_rT} and \cref{QD_alpha_mono_NS_vs_rT} for zero and antipodal separation respectively, which shows the $\delta$ and $T$ dependence of $D$.  This can be compared to \cref{QD_BD_mono_N_vs_rT} and \cref{QD_BD_mono_NS_vs_rT} for the Euclidean vacuum scenario. They differ slightly and share all the relevant features, such as the suppressions of the quantum fluctuations due to superhorizon decoherence or spectral incompatibility. 

\FloatBarrier

\begin{figure}[htt]
	\centering
	\begin{subfigure}{.45\textwidth}
		\centering
		\includegraphics[width=.8\linewidth]{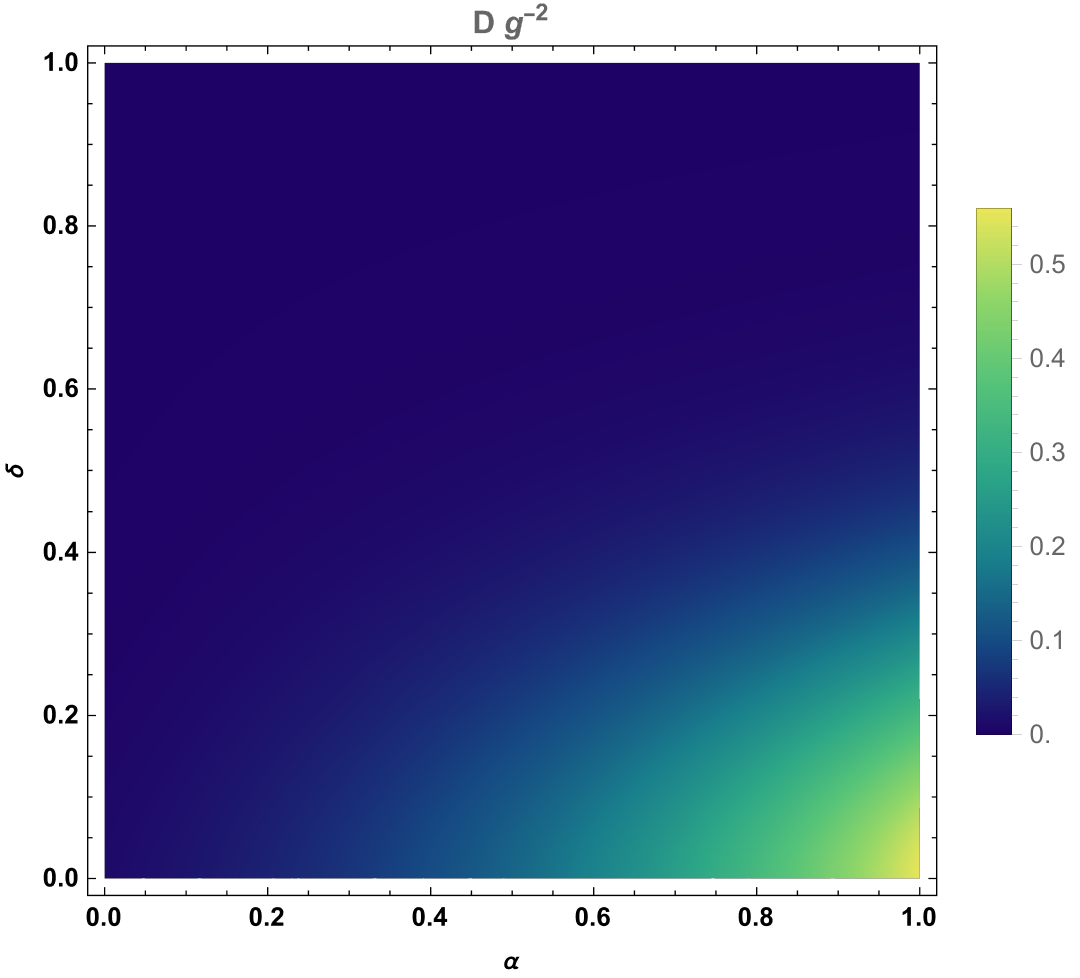}
		\caption{ }
		\label{QD_alpha_mono_N_vs_ar}
	\end{subfigure}\hspace{.5cm}
	\begin{subfigure}{.45\textwidth}
		\centering
		\includegraphics[width=.85\linewidth]{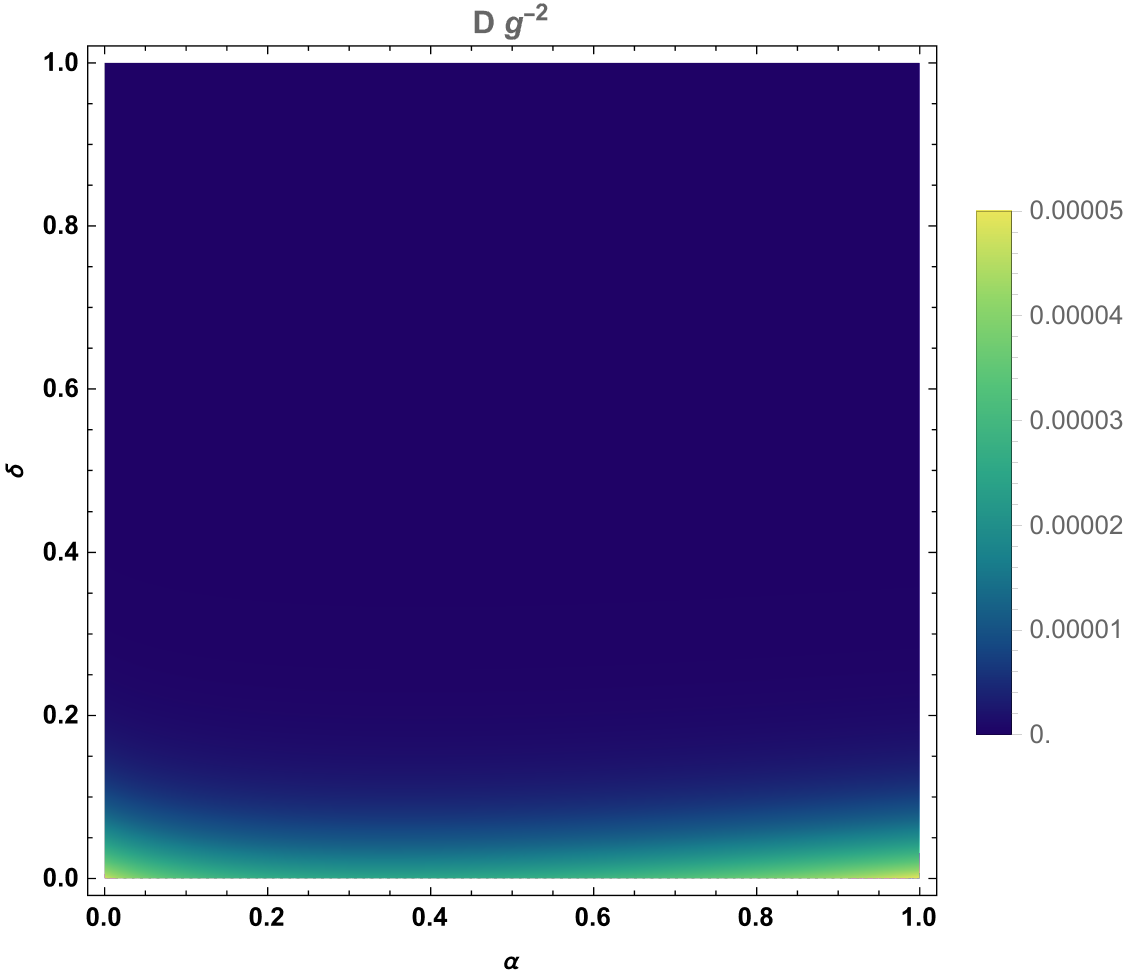}
		\caption{}
		\label{QD_alpha_mono_NS_vs_ar}
	\end{subfigure}\hspace{.5cm}
	\caption{The density plots of the quantum discord $D$ of two monopole-coupling identical UDW detectors in $\alpha$-vacua as a function of  $\alpha$ and $\delta$ given $\Omega_B=0.5$ and $T=1.5$ with (a) zero separation and (b) antipodal separation.}
	\label{QD_alpha_mono_vs_ar}
\end{figure}

\FloatBarrier
The second set of density plots is shown in \cref{QD_alpha_mono_N_vs_ar} and \cref{QD_alpha_mono_NS_vs_ar} for zero and antipodal separations, which exhibits the interplay between the $\alpha$ and $\delta$ dependence of $D$. We observe that increasing the value of $\alpha$  enhances the quantum discord for the zero separation cases significantly, while there is a minimal effect for the antipodal separation cases.  The overall magnitude for the zero separation scenario is about four orders larger than the one for the antipodal separation. In both cases, the spectral incompatibility diminishes the quantum discord.

\FloatBarrier

\FloatBarrier

\begin{figure}[htt]
	\centering
	\begin{subfigure}{.45\textwidth}
		\centering
		\includegraphics[width=.85\linewidth]{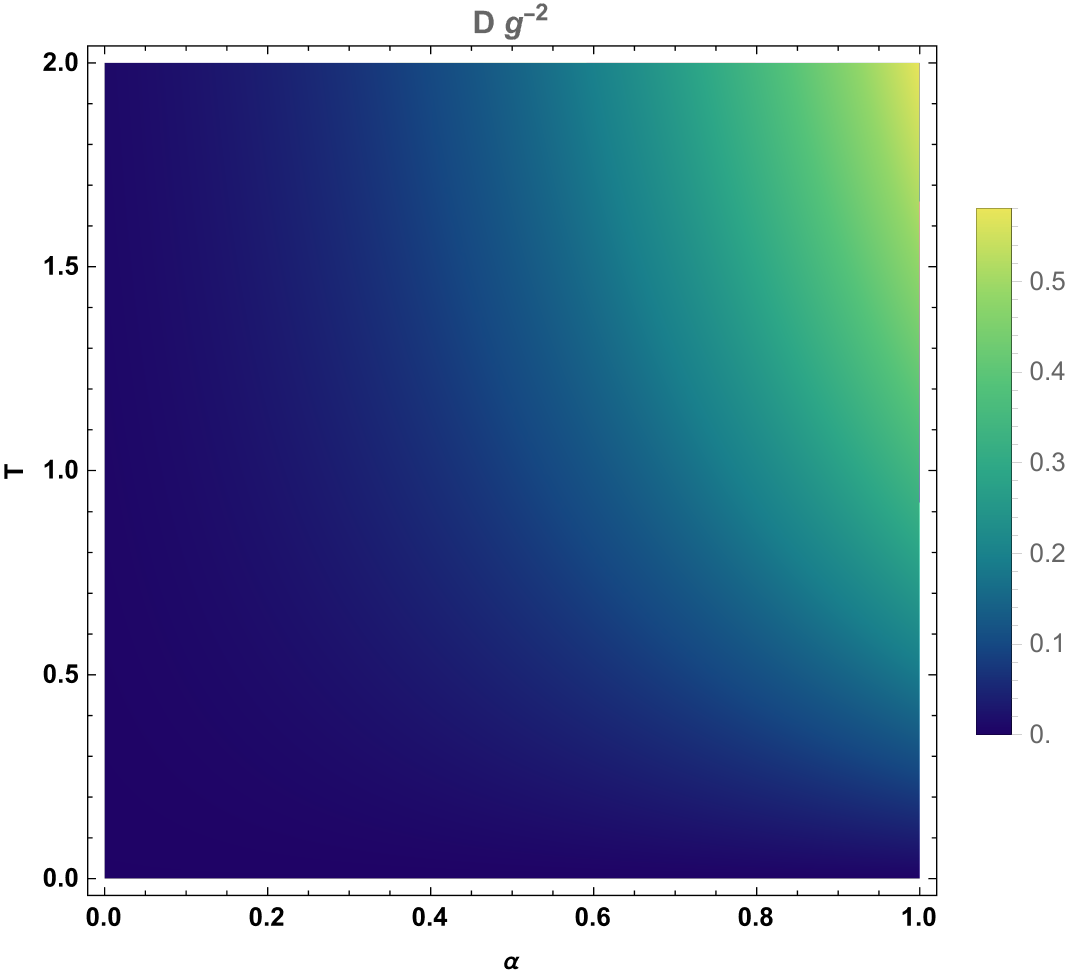}
		\caption{ }
		\label{QD_alpha_mono_N_vs_aT}
	\end{subfigure}\hspace{.5cm}
	\begin{subfigure}{.45\textwidth}
		\centering
		\includegraphics[width=.85\linewidth]{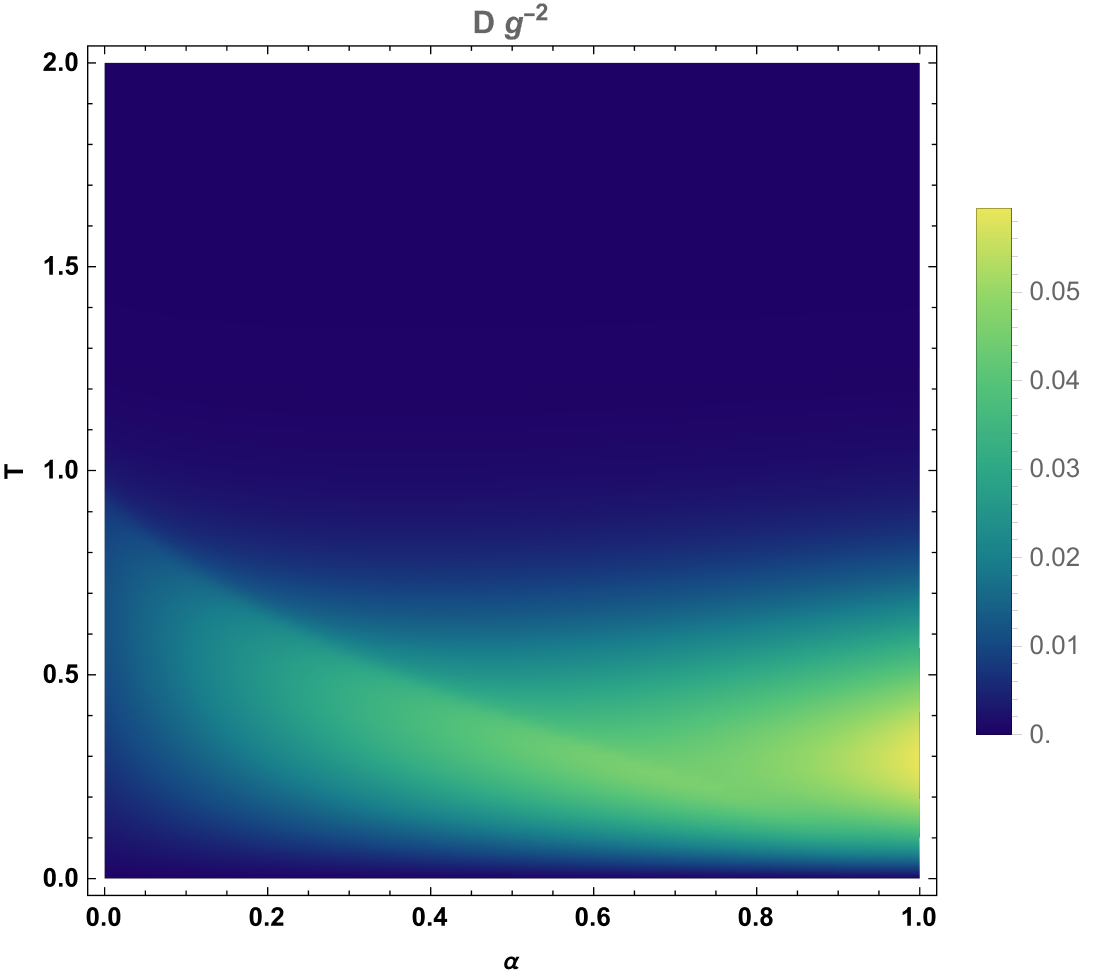}
		\caption{}
		\label{QD_alpha_mono_NS_vs_aT}
	\end{subfigure}
	\caption{The density plots of the quantum discord $D$ of two monopole-coupling identical UDW detectors in $\alpha$-vacua as a function of  $\alpha$ and $T$ given $\Omega_B=.5$ and $\delta=0.1$ with (a) zero separation and (b) antipodal separation.}
		\label{QD_alpha_mono_vs_aT}
\end{figure}

The third set of density plots is presented in \cref{QD_alpha_mono_N_vs_aT} and \cref{QD_alpha_mono_NS_vs_aT} for zero and antipodal separations, which exhibits the interplay between the $\alpha$ and $T$ dependence of $D$. We again see that increasing the value of $\alpha$ can enhance the quantum discord for both zero and antipodal separations. However, their ways of enhancement are different.  In \cref{QD_alpha_mono_N_vs_aT}, the quantum discord of larger $T$ gets more enhanced, which is in contrast with \cref{QD_alpha_mono_NS_vs_aT}, where the quantum discord of lower $T$ ($T\simeq 0.25$) gets more enhanced. 

In conclusion, our results demonstrate that the $\alpha$-vacua have a non-trivial effect on quantum discord, but the superhorizon decoherence remains.

\FloatBarrier
%

\subsubsection{Dipole coupling in $\alpha$-vacua}

Finally, we present the quantum discord for the dipole-coupling of UDW detectors in $\alpha$-vacua. We parallel what we have presented for the monopole coupling cases for comparison.

\begin{figure}[htt]
	\centering
	\begin{subfigure}{.45\textwidth}
		\centering
		\includegraphics[width=.9\linewidth]{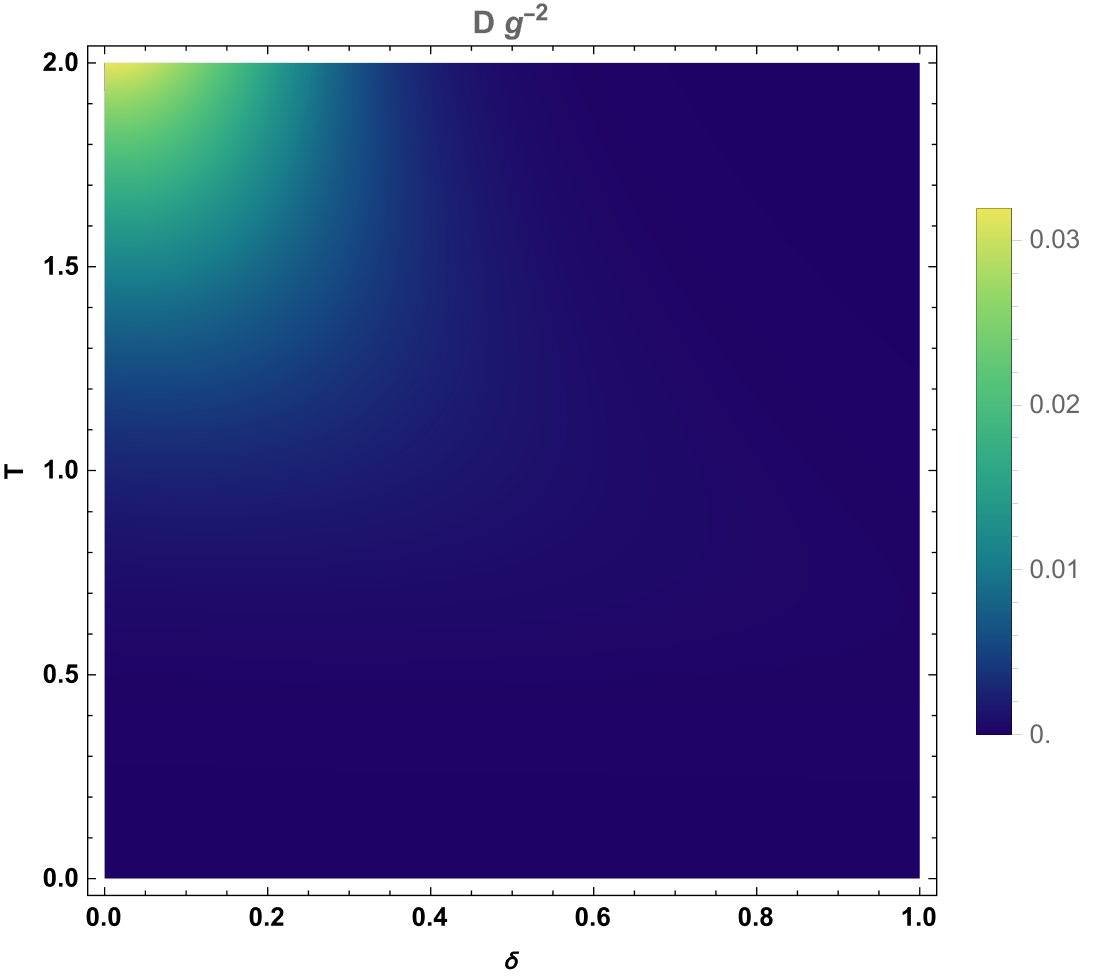}
		\caption{ }
		\label{QD_alpha_di_N_vs_rT}
	\end{subfigure}\hspace{.5cm}
	\begin{subfigure}{.45\textwidth}
		\centering
		\includegraphics[width=.9\linewidth]{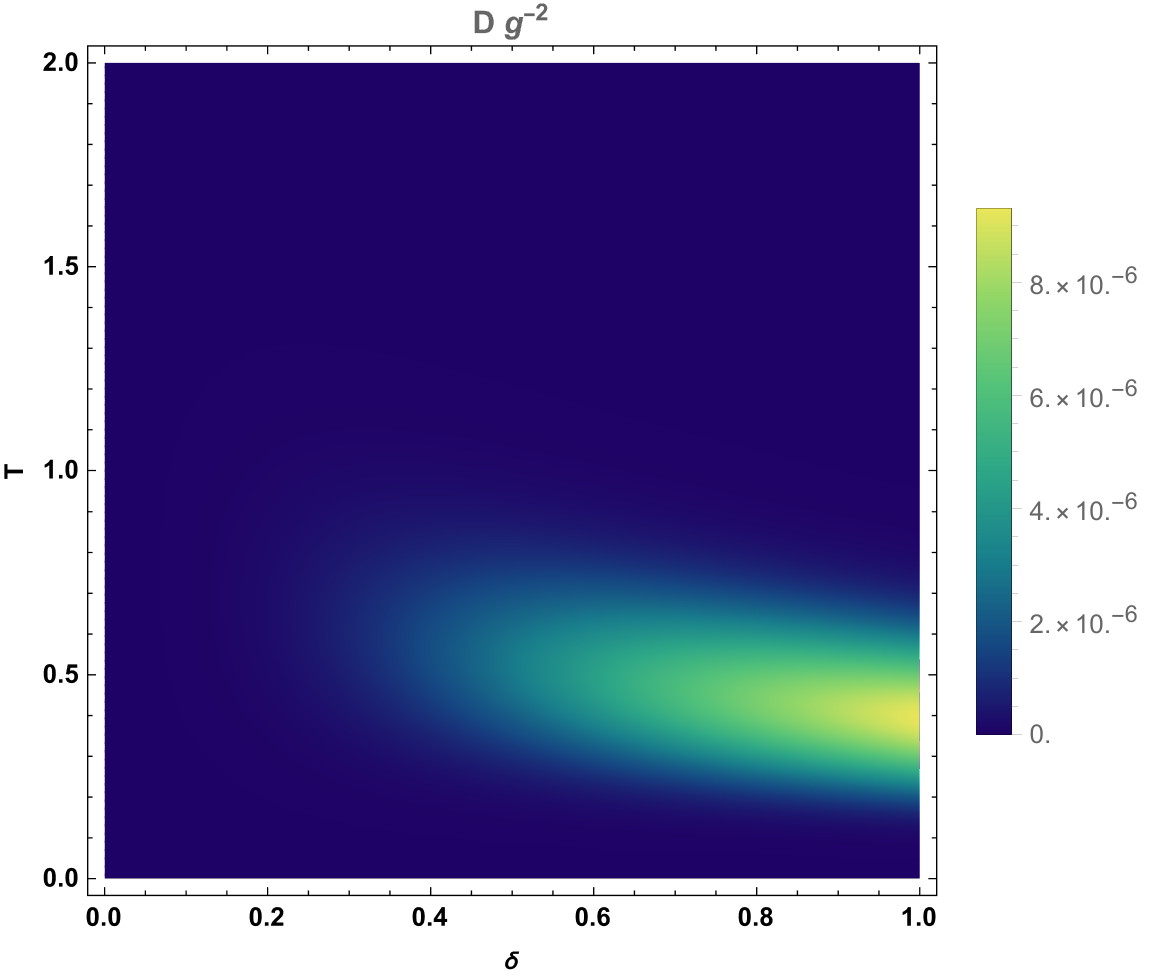}
		\caption{}
		\label{QD_alpha_di_NS_vs_rT}
	\end{subfigure}
	\caption{The density plots of the quantum discord $D$ of two dipole-coupling identical UDW detectors in $\alpha$-vacua as a function of $\delta$ and $T$ given $\Omega_B=0.5$ and $\alpha=0.1$ with (a) zero separation and (b) antipodal separation.}
		\label{QD_alpha_di_vs_rT}
\end{figure}
\FloatBarrier


\begin{figure}[H]
	\centering
	\begin{subfigure}{.45\textwidth}
		\centering
		\includegraphics[width=.85\linewidth]{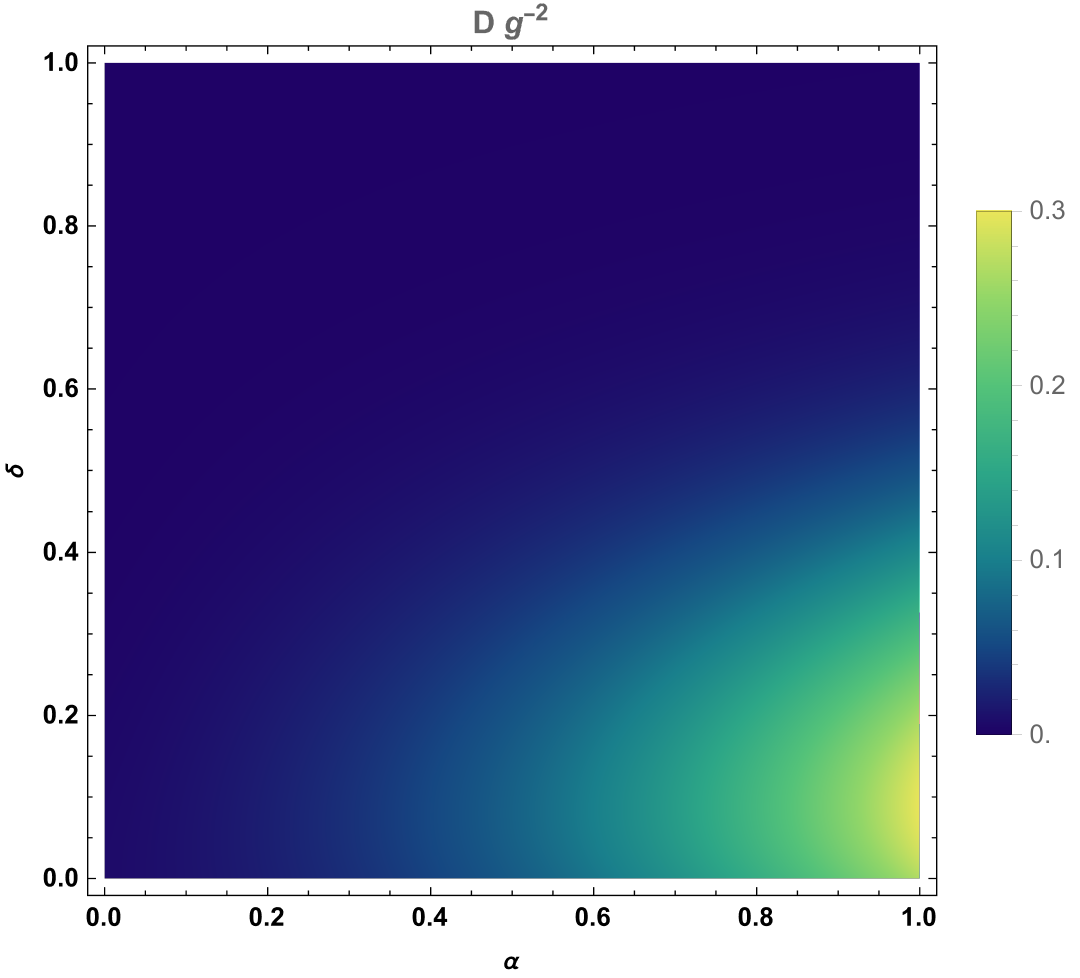}
		\caption{}
		\label{QD_alpha_di_N_vs_ar}
	\end{subfigure}\hspace{.5cm}
	\begin{subfigure}{.45\textwidth}
		\centering
		\includegraphics[width=.9\linewidth]{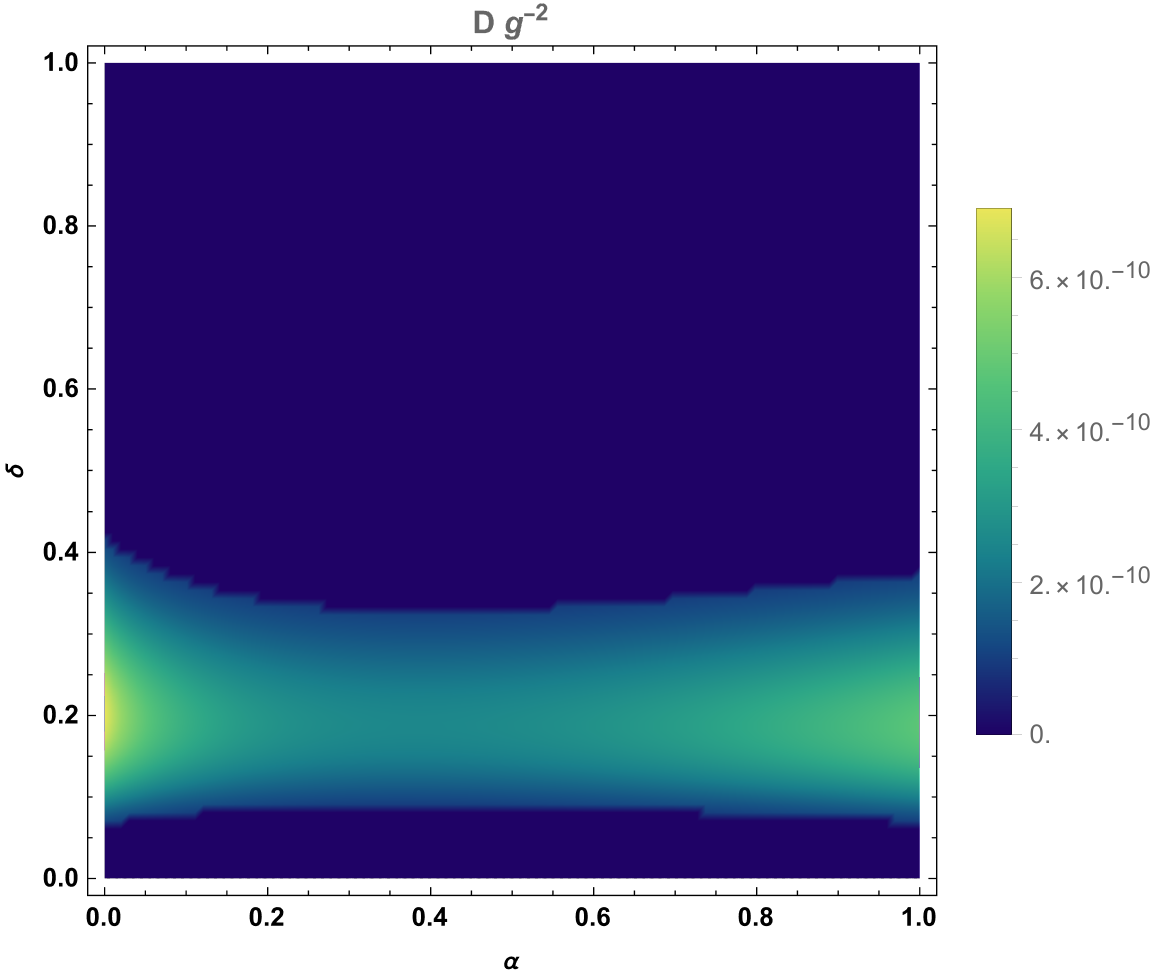}
		\caption{}
		\label{QD_alpha_di_NS_vs_ar}
	\end{subfigure}
\caption{The density plots of the quantum discord $D$ of two dipole-coupling identical UDW detectors in $\alpha$-vacua as a function of $\alpha$ and $\delta$ given $\Omega_B=0.5$, $T=1.5$ with (a) zero separation and (b) antipodal separation.}
		\label{QD_alpha_di_vs_ar}
\end{figure}
\FloatBarrier
The first set of density plots is shown in \cref{QD_alpha_di_N_vs_rT} and \cref{QD_alpha_di_NS_vs_rT} for zero and antipodal separations, to exhibit the $\delta$ and $T$ dependence of $D$. Its difference from its monopole-coupling counterpart of \cref{QD_alpha_mono_vs_rT} is quite similar to the corresponding difference for Euclidean vacuum, i.e., the difference of 
\cref{QD_BD_di_vs_rT} from \cref{QD_BD_mono_vs_rT}.  Thus, turning on $\alpha$ will not drastically change the overall patterns of the quantum discord of the Euclidean vacuum. 

The second set of density plots is shown in \cref{QD_alpha_di_N_vs_ar} and \cref{QD_alpha_di_NS_vs_ar} for zero and antipodal separations, to exhibit the interplay between the $\alpha$ and $\delta$ dependence of $D$.
Compared to its monopole coupling counterpart of \cref{QD_alpha_mono_vs_ar}, the overall magnitude and pattern of the active region of the zero separation cases do not change much. On the other hand, for the antipodal separation cases, the overall magnitude is reduced by more than four orders and can be considered zero.

\begin{figure}[htp]
    \hspace*{-0.5in}
    \centering
	\begin{subfigure}{.45\textwidth}
		\centering
		\includegraphics[width=.85\linewidth]{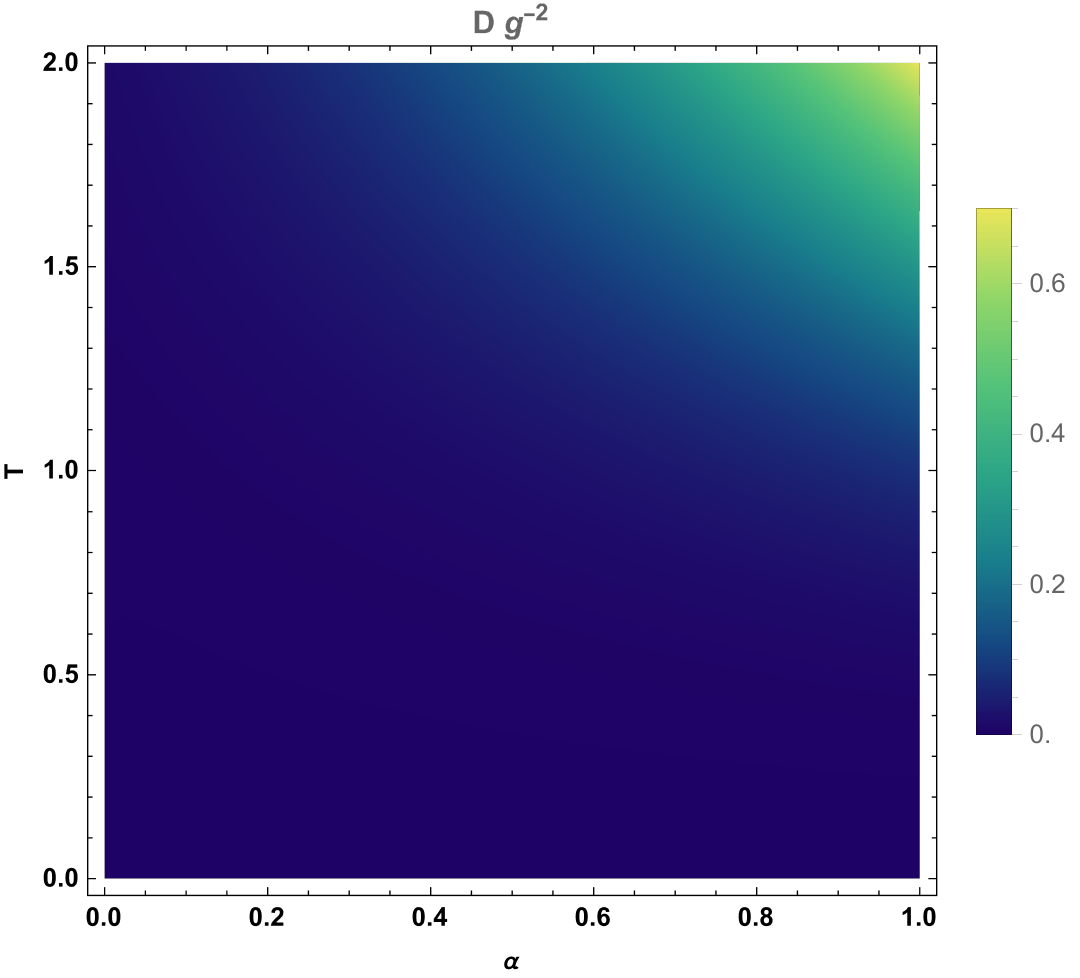}
		\caption{ }
		\label{QD_alpha_di_N_vs_aT}
	\end{subfigure}\hspace{.5cm}
	\begin{subfigure}{.45\textwidth}
		\centering
		\includegraphics[width=.9\linewidth]{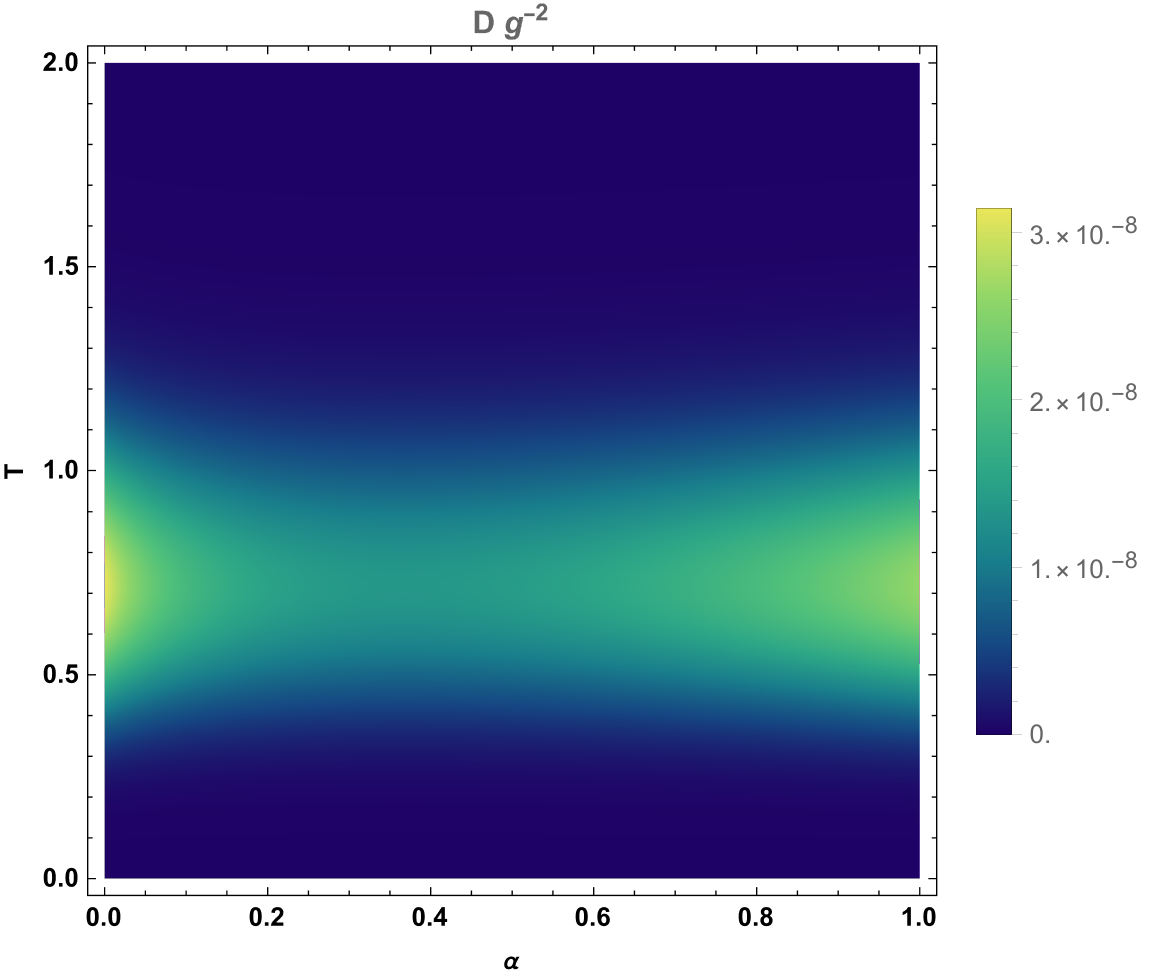}
		\caption{}
		\label{QD_alpha_di_NS_vs_aT}
	\end{subfigure}
 \caption{The density plots of the quantum discord $D$ of two dipole-coupling identical UDW detectors in $\alpha$-vacua as a function of $\delta$ and $T$ given $\Omega_B=0.5$, $\delta=0.1$ with (a) zero separation and (b) antipodal separation.}
	\label{QD_alpha_di_vs_aT}
\end{figure}

The third set of density plots is shown in \cref{QD_alpha_di_N_vs_aT} and \cref{QD_alpha_di_NS_vs_aT} for zero and antipodal separations, which is in parallel with its monopole-coupling counterpart of \cref{QD_alpha_mono_vs_aT}, to exhibit the interplay between the $\alpha$ and $T$ dependence of $D$. The change in the overall magnitude and patterns compared to the monopole-coupling counterpart is similar to that of the second set of density plots.

Overall, we see that the patterns in the cases of monopole and dipole couplings are quite similar. However, the overall magnitude remains the same for the zero separation but decreases by a few orders for the antipodal separation.


\FloatBarrier


\section{Conclusion}\label{sec6}

De Sitter space is the most simple dynamical spacetime and plays an important role in the inflationary universe scenario for initiating the primordial curvature perturbations from the fluctuations of quantum fields. Besides, the CPT invariant vacuum states of a quantum field in de Sitter space, called $\alpha$-vacua, are not unique. In this paper, we have explored the de Sitter space and the associated scalar $\alpha$-vacuum states by a pair of probe UDW detectors. Our results demonstrated how the de Sitter gravity affects the relativistic quantum information of the vacuum polarizations. In this work, we, in particular, have studied two quantum information quantities, concurrence and quantum discord of the reduced final state of the static UDW detectors. The concurrence characterizes the quantum entanglement harvested by the UDW detectors from the vacuum states of the environmental scalar. Notably, the quantum discord for the reduced state is obtained analytically in this paper for the first time in the literature. It characterizes the non-classical quantum correlations between UDW detectors, which reflect the gravitated quantum correlation of the scalar's vacuum states. These two quantities, though quantum, are intrinsically quite different, especially on the nonlocal features.  For this purpose, we considered a pair of static UDW detectors in either time-like zero separation or space-like antipodal separation. 

As the reduced state is obtained by tracing out the environmental scalar field in curved spacetime, it is usually hard to have an analytical form, and the calculations of the derived quantum information quantities are usually based on numerical analysis of the integrals of the Wightman functions. Due to the nontrivial $i\epsilon$ prescription of the Wightman function, the results will be subjected to numerical error for nontrivial background spacetime. The de Sitter space, though time-dependent, is maximally symmetric, so the Wightman function has a simple analytical spectral representation. Exploiting this and the saddle point approximation, we have obtained the analytical form of the reduced state in the $\alpha$-vacuum states of de Sitter space for the first time in the literature. 

Based on the analytical reduced states, we compute the 
corresponding concurrence and quantum discord. We then present the numerical density plots to understand how their patterns are affected by the interplays of four model parameters: two spectral gaps of UDW detectors, the measuring time, and the value of $\alpha$  labeling the $\alpha$-vacua. From the patterns of these density plots, we draw our main conclusions as follows. By increasing the measuring time or the value of $\alpha$, we observed ``sudden" death behavior for the short-range quantum entanglements probed by UDW detectors in zero (or time-like) separation, but not for the long-range ones probed by UDW detectors in antipodal (or space-like) separation. Moreover, the long-range quantum entanglements grow with the measuring time and the values of $\alpha$. This implies that the de Sitter gravity enhances the long-range entanglement. On the other hand, for the quantum discord, we found that there exists suppression of quantum discord at the superhorizon scale, which can be characterized by the measuring time scale. This conforms to the folklore about the decoherence of the quantum correlation at the superhorizon scales in the inflationary universe scenario. This is also consistent with the intuition that it is difficult to maintain long-time quantum correlations. Several minor points, as drawn from the density plots, such as the dependence on the spectral (in)compatibility of UDW detectors in time-like and space-like separations, have also been noted in the paper. Overall, our study and results have helped one gain more tools and insights to understand the de Sitter space from the perspective of relativistic quantum information.

\section*{Acknowledgement}
The work of FLL is supported by Taiwan's NSTC with grant numbers 109-2112-M-003-007-MY3 and 112-2112-M-003-006-MY3, and the work of SM is supported by Taiwan's NSTC with grant numbers 112-2811-M-003-014.

\providecommand{\href}[2]{#2}\begingroup\raggedright\endgroup


\begin{thebibliography}{10}

\bibitem{gibbons1983very}
G.~Gibbons, S.~Hawking and S.~Siklos, \emph{{The Very Early Universe: Proceedings of the Nuffield Workshop, Cambridge 21 June to 9 July, 1982}}, .

\bibitem{Mottola:1984ar}
E.~Mottola, \emph{{Particle Creation in de Sitter Space}}, \href{https://doi.org/10.1103/PhysRevD.31.754}{\emph{Phys. Rev. D} {\bfseries 31} (1985) 754}.

\bibitem{PhysRevD.32.3136}
B.~Allen, \emph{Vacuum states in de sitter space}, \href{https://doi.org/10.1103/PhysRevD.32.3136}{\emph{Phys. Rev. D} {\bfseries 32} (1985) 3136}.

\bibitem{PhysRevD.98.065014}
A.~Higuchi and K.~Yamamoto, \emph{Vacuum state in de sitter spacetime with static charts}, \href{https://doi.org/10.1103/PhysRevD.98.065014}{\emph{Phys. Rev. D} {\bfseries 98} (2018) 065014}.

\bibitem{PhysRevD.65.104039}
R.~Bousso, A.~Maloney and A.~Strominger, \emph{Conformal vacua and entropy in de sitter space}, \href{https://doi.org/10.1103/PhysRevD.65.104039}{\emph{Phys. Rev. D} {\bfseries 65} (2002) 104039}.

\bibitem{Srednicki:1993im}
M.~Srednicki, \emph{{Entropy and area}}, \href{https://doi.org/10.1103/PhysRevLett.71.666}{\emph{Phys. Rev. Lett.} {\bfseries 71} (1993) 666} [\href{https://arxiv.org/abs/hep-th/9303048}{{\ttfamily hep-th/9303048}}].

\bibitem{Callan:1994py}
C.G.~Callan, Jr. and F.~Wilczek, \emph{{On geometric entropy}}, \href{https://doi.org/10.1016/0370-2693(94)91007-3}{\emph{Phys. Lett. B} {\bfseries 333} (1994) 55} [\href{https://arxiv.org/abs/hep-th/9401072}{{\ttfamily hep-th/9401072}}].

\bibitem{Holzhey:1994we}
C.~Holzhey, F.~Larsen and F.~Wilczek, \emph{{Geometric and renormalized entropy in conformal field theory}}, \href{https://doi.org/10.1016/0550-3213(94)90402-2}{\emph{Nucl. Phys. B} {\bfseries 424} (1994) 443} [\href{https://arxiv.org/abs/hep-th/9403108}{{\ttfamily hep-th/9403108}}].

\bibitem{Schlieder1965SomeRA}
S.~Schlieder, \emph{Some remarks about the localization of states in a quantum field theory}, {\emph{Communications in Mathematical Physics} {\bfseries 1} (1965) 265}.

\bibitem{Witten:2018zxz}
E.~Witten, \emph{{APS Medal for Exceptional Achievement in Research: Invited article on entanglement properties of quantum field theory}}, \href{https://doi.org/10.1103/RevModPhys.90.045003}{\emph{Rev. Mod. Phys.} {\bfseries 90} (2018) 045003} [\href{https://arxiv.org/abs/1803.04993}{{\ttfamily 1803.04993}}].

\bibitem{Sanders:2008gs}
K.~Sanders, \emph{{On the Reeh-Schlieder Property in Curved Spacetime}}, \href{https://doi.org/10.1007/s00220-009-0734-3}{\emph{Commun. Math. Phys.} {\bfseries 288} (2009) 271} [\href{https://arxiv.org/abs/0801.4676}{{\ttfamily 0801.4676}}].

\bibitem{Maldacena:1997re}
J.M.~Maldacena, \emph{{The Large N limit of superconformal field theories and supergravity}}, \href{https://doi.org/10.4310/ATMP.1998.v2.n2.a1}{\emph{Adv. Theor. Math. Phys.} {\bfseries 2} (1998) 231} [\href{https://arxiv.org/abs/hep-th/9711200}{{\ttfamily hep-th/9711200}}].

\bibitem{Ryu:2006bv}
S.~Ryu and T.~Takayanagi, \emph{{Holographic derivation of entanglement entropy from AdS/CFT}}, \href{https://doi.org/10.1103/PhysRevLett.96.181602}{\emph{Phys. Rev. Lett.} {\bfseries 96} (2006) 181602} [\href{https://arxiv.org/abs/hep-th/0603001}{{\ttfamily hep-th/0603001}}].

\bibitem{Chen:2021lnq}
B.~Chen, B.~Czech and Z.-z.~Wang, \emph{{Quantum information in holographic duality}}, \href{https://doi.org/10.1088/1361-6633/ac51b5}{\emph{Rept. Prog. Phys.} {\bfseries 85} (2022) 046001} [\href{https://arxiv.org/abs/2108.09188}{{\ttfamily 2108.09188}}].

\bibitem{PhysRevD.14.870}
W.G.~Unruh, \emph{Notes on black-hole evaporation}, \href{https://doi.org/10.1103/PhysRevD.14.870}{\emph{Phys. Rev. D} {\bfseries 14} (1976) 870}.

\bibitem{DeWitt:1980hx}
B.S.~DeWitt, \emph{{Quantum gravity: The new synthesis }},  in \emph{{General Relativity}: {An Einstein Centenary Survey}}, pp.~680--745 (1980).

\bibitem{summers1985vacuum}
S.J.~Summers and R.~Werner, \emph{The vacuum violates bell's inequalities}, {\emph{Physics Letters A} {\bfseries 110} (1985) 257}.

\bibitem{Summers:1987ze}
S.J.~Summers and R.~Werner, \emph{{Maximal Violation of Bell's Inequalities Is Generic in Quantum Field Theory}}, \href{https://doi.org/10.1007/BF01207366}{\emph{Commun. Math. Phys.} {\bfseries 110} (1987) 247}.

\bibitem{VALENTINI1991321}
A.~Valentini, \emph{Non-local correlations in quantum electrodynamics}, \href{https://doi.org/https://doi.org/10.1016/0375-9601(91)90952-5}{\emph{Physics Letters A} {\bfseries 153} (1991) 321}.

\bibitem{Reznik:2002fz}
B.~Reznik, \emph{{Entanglement from the vacuum}}, \href{https://doi.org/10.1023/A:1022875910744}{\emph{Found. Phys.} {\bfseries 33} (2003) 167} [\href{https://arxiv.org/abs/quant-ph/0212044}{{\ttfamily quant-ph/0212044}}].

\bibitem{PhysRevA.71.042104}
B.~Reznik, A.~Retzker and J.~Silman, \emph{Violating bell's inequalities in vacuum}, \href{https://doi.org/10.1103/PhysRevA.71.042104}{\emph{Phys. Rev. A} {\bfseries 71} (2005) 042104}.

\bibitem{Wilson-Gerow:2024ljx}
J.~Wilson-Gerow, A.~Dugad and Y.~Chen, \emph{{Decoherence by warm horizons}},  \href{https://arxiv.org/abs/2405.00804}{{\ttfamily 2405.00804}}.

\bibitem{Danielson:2021egj}
D.L.~Danielson, G.~Satishchandran and R.M.~Wald, \emph{{Gravitationally mediated entanglement: Newtonian field versus gravitons}}, \href{https://doi.org/10.1103/PhysRevD.105.086001}{\emph{Phys. Rev. D} {\bfseries 105} (2022) 086001} [\href{https://arxiv.org/abs/2112.10798}{{\ttfamily 2112.10798}}].

\bibitem{Danielson:2022sga}
D.L.~Danielson, G.~Satishchandran and R.M.~Wald, \emph{{Killing horizons decohere quantum superpositions}}, \href{https://doi.org/10.1103/PhysRevD.108.025007}{\emph{Phys. Rev. D} {\bfseries 108} (2023) 025007} [\href{https://arxiv.org/abs/2301.00026}{{\ttfamily 2301.00026}}].

\bibitem{Danielson:2022tdw}
D.L.~Danielson, G.~Satishchandran and R.M.~Wald, \emph{{Black holes decohere quantum superpositions}}, \href{https://doi.org/10.1142/S0218271822410036}{\emph{Int. J. Mod. Phys. D} {\bfseries 31} (2022) 2241003} [\href{https://arxiv.org/abs/2205.06279}{{\ttfamily 2205.06279}}].

\bibitem{Dhanuka:2022ggi}
A.~Dhanuka and K.~Lochan, \emph{{Unruh DeWitt probe of late time revival of quantum correlations in Friedmann spacetimes}}, \href{https://doi.org/10.1103/PhysRevD.106.125006}{\emph{Phys. Rev. D} {\bfseries 106} (2022) 125006} [\href{https://arxiv.org/abs/2210.11186}{{\ttfamily 2210.11186}}].

\bibitem{Gralla:2023oya}
S.E.~Gralla and H.~Wei, \emph{{Decoherence from horizons: General formulation and rotating black holes}}, \href{https://doi.org/10.1103/PhysRevD.109.065031}{\emph{Phys. Rev. D} {\bfseries 109} (2024) 065031} [\href{https://arxiv.org/abs/2311.11461}{{\ttfamily 2311.11461}}].

\bibitem{Salton:2014jaa}
G.~Salton, R.B.~Mann and N.C.~Menicucci, \emph{{Acceleration-assisted entanglement harvesting and rangefinding}}, \href{https://doi.org/10.1088/1367-2630/17/3/035001}{\emph{New J. Phys.} {\bfseries 17} (2015) 035001} [\href{https://arxiv.org/abs/1408.1395}{{\ttfamily 1408.1395}}].

\bibitem{Martin-Martinez:2015eoa}
E.~Martin-Martinez and B.C.~Sanders, \emph{{Precise space\textendash{}time positioning for entanglement harvesting}}, \href{https://doi.org/10.1088/1367-2630/18/4/043031}{\emph{New J. Phys.} {\bfseries 18} (2016) 043031} [\href{https://arxiv.org/abs/1508.01209}{{\ttfamily 1508.01209}}].

\bibitem{PhysRevD.93.044001}
E.~Mart\'{\i}n-Mart\'{\i}nez, A.R.H.~Smith and D.R.~Terno, \emph{Spacetime structure and vacuum entanglement}, \href{https://doi.org/10.1103/PhysRevD.93.044001}{\emph{Phys. Rev. D} {\bfseries 93} (2016) 044001}.

\bibitem{Henderson:2017yuv}
L.J.~Henderson, R.A.~Hennigar, R.B.~Mann, A.R.H.~Smith and J.~Zhang, \emph{{Harvesting Entanglement from the Black Hole Vacuum}}, \href{https://doi.org/10.1088/1361-6382/aae27e}{\emph{Class. Quant. Grav.} {\bfseries 35} (2018) 21LT02} [\href{https://arxiv.org/abs/1712.10018}{{\ttfamily 1712.10018}}].

\bibitem{Kukita:2017etu}
S.~Kukita and Y.~Nambu, \emph{{Harvesting large scale entanglement in de Sitter space with multiple detectors}}, \href{https://doi.org/10.3390/e19090449}{\emph{Entropy} {\bfseries 19} (2017) 449} [\href{https://arxiv.org/abs/1708.01359}{{\ttfamily 1708.01359}}].

\bibitem{Koga:2019fqh}
J.-i.~Koga, K.~Maeda and G.~Kimura, \emph{{Entanglement extracted from vacuum into accelerated Unruh-DeWitt detectors and energy conservation}}, \href{https://doi.org/10.1103/PhysRevD.100.065013}{\emph{Phys. Rev. D} {\bfseries 100} (2019) 065013} [\href{https://arxiv.org/abs/1906.02843}{{\ttfamily 1906.02843}}].

\bibitem{Perche:2022ykt}
T.R.~Perche, B.~Ragula and E.~Mart\'\i{}n-Mart\'\i{}nez, \emph{{Harvesting entanglement from the gravitational vacuum}}, \href{https://doi.org/10.1103/PhysRevD.108.085025}{\emph{Phys. Rev. D} {\bfseries 108} (2023) 085025} [\href{https://arxiv.org/abs/2210.14921}{{\ttfamily 2210.14921}}].

\bibitem{Mendez-Avalos:2022obb}
D.~Mendez-Avalos, L.J.~Henderson, K.~Gallock-Yoshimura and R.B.~Mann, \emph{{Entanglement harvesting of three Unruh-DeWitt detectors}}, \href{https://doi.org/10.1007/s10714-022-02956-x}{\emph{Gen. Rel. Grav.} {\bfseries 54} (2022) 87} [\href{https://arxiv.org/abs/2206.11902}{{\ttfamily 2206.11902}}].

\bibitem{ollivier2001introducing}
H.~Ollivier and W.H.~Zurek, \emph{Introducing quantum discord}, {\emph{arXiv preprint quant-ph/0105072} (2001) }.

\bibitem{Henderson:2001wrr}
L.~Henderson and V.~Vedral, \emph{{Classical, quantum and total correlations}}, \href{https://doi.org/10.1088/0305-4470/34/35/315}{\emph{J. Phys. A} {\bfseries 34} (2001) 6899} [\href{https://arxiv.org/abs/quant-ph/0105028}{{\ttfamily quant-ph/0105028}}].

\bibitem{Barman:2021kwg}
S.~Barman, D.~Barman and B.R.~Majhi, \emph{{Entanglement harvesting from conformal vacuums between two Unruh-DeWitt detectors moving along null paths}}, \href{https://doi.org/10.1007/JHEP09(2022)106}{\emph{JHEP} {\bfseries 09} (2022) 106} [\href{https://arxiv.org/abs/2112.01308}{{\ttfamily 2112.01308}}].

\bibitem{Barman:2021bbw}
D.~Barman, S.~Barman and B.R.~Majhi, \emph{{Role of thermal field in entanglement harvesting between two accelerated Unruh-DeWitt detectors}}, \href{https://doi.org/10.1007/JHEP07(2021)124}{\emph{JHEP} {\bfseries 07} (2021) 124} [\href{https://arxiv.org/abs/2104.11269}{{\ttfamily 2104.11269}}].

\bibitem{Barman:2022xht}
D.~Barman, S.~Barman and B.R.~Majhi, \emph{{Entanglement harvesting between two inertial Unruh-DeWitt detectors from nonvacuum quantum fluctuations}}, \href{https://doi.org/10.1103/PhysRevD.106.045005}{\emph{Phys. Rev. D} {\bfseries 106} (2022) 045005} [\href{https://arxiv.org/abs/2205.08505}{{\ttfamily 2205.08505}}].

\bibitem{yu2005evolution}
T.~Yu and J.~Eberly, \emph{Evolution from entanglement to decoherence of bipartite mixed" x" states}, {\emph{arXiv preprint quant-ph/0503089} (2005) }.

\bibitem{rau2009algebraic}
A.~Rau, \emph{Algebraic characterization of x-states in quantum information}, {\emph{Journal of physics a: Mathematical and theoretical} {\bfseries 42} (2009) 412002}.

\bibitem{PhysRevA.40.4277}
R.F.~Werner, \emph{Quantum states with einstein-podolsky-rosen correlations admitting a hidden-variable model}, \href{https://doi.org/10.1103/PhysRevA.40.4277}{\emph{Phys. Rev. A} {\bfseries 40} (1989) 4277}.

\bibitem{ali2010quantum}
M.~Ali, A.~Rau and G.~Alber, \emph{Quantum discord for two-qubit x states}, {\emph{Physical Review A} {\bfseries 81} (2010) 042105}.

\bibitem{Koga:2018the}
J.-I.~Koga, G.~Kimura and K.~Maeda, \emph{{Quantum teleportation in vacuum using only Unruh-DeWitt detectors}}, \href{https://doi.org/10.1103/PhysRevA.97.062338}{\emph{Phys. Rev. A} {\bfseries 97} (2018) 062338} [\href{https://arxiv.org/abs/1804.01183}{{\ttfamily 1804.01183}}].

\bibitem{PhysRevLett.80.2245}
W.K.~Wootters, \emph{Entanglement of formation of an arbitrary state of two qubits}, \href{https://doi.org/10.1103/PhysRevLett.80.2245}{\emph{Phys. Rev. Lett.} {\bfseries 80} (1998) 2245}.

\bibitem{chen2011quantum}
Q.~Chen, C.~Zhang, S.~Yu, X.~Yi and C.~Oh, \emph{Quantum discord of two-qubit x states}, {\emph{Physical Review A} {\bfseries 84} (2011) 042313}.

\bibitem{PhysRevA.88.014302}
Y.~Huang, \emph{Quantum discord for two-qubit $x$ states: Analytical formula with very small worst-case error}, \href{https://doi.org/10.1103/PhysRevA.88.014302}{\emph{Phys. Rev. A} {\bfseries 88} (2013) 014302}.

\bibitem{yurischev2015quantum}
M.A.~Yurischev, \emph{On the quantum discord of general x states}, {\emph{Quantum Information Processing} {\bfseries 14} (2015) 3399}.

\bibitem{Chernikov:1968zm}
N.A.~Chernikov and E.A.~Tagirov, \emph{{Quantum theory of scalar fields in de Sitter space-time}}, {\emph{Ann. Inst. H. Poincare A Phys. Theor.} {\bfseries 9} (1968) 109}.

\bibitem{Niermann:2024fvi}
L.~Niermann and L.C.~Barbado, \emph{{Particle detectors in superposition in de Sitter spacetime}},  \href{https://arxiv.org/abs/2403.02087}{{\ttfamily 2403.02087}}.

\bibitem{Henderson:2018lcy}
L.J.~Henderson, R.A.~Hennigar, R.B.~Mann, A.R.H.~Smith and J.~Zhang, \emph{{Entangling detectors in anti-de Sitter space}}, \href{https://doi.org/10.1007/JHEP05(2019)178}{\emph{JHEP} {\bfseries 05} (2019) 178} [\href{https://arxiv.org/abs/1809.06862}{{\ttfamily 1809.06862}}].

\bibitem{Maeso-Garcia:2022uzf}
H.~Maeso-Garc\'\i{}a, J.~Polo-G\'omez and E.~Mart\'\i{}n-Mart\'\i{}nez, \emph{{How measuring a quantum field affects entanglement harvesting}}, \href{https://doi.org/10.1103/PhysRevD.107.045011}{\emph{Phys. Rev. D} {\bfseries 107} (2023) 045011} [\href{https://arxiv.org/abs/2210.05692}{{\ttfamily 2210.05692}}].

\end{thebibliography}
\end{document}